\documentclass[12pt]{article}

\usepackage{framed}
\usepackage{algorithm}
\usepackage[noend]{algpseudocode}
\usepackage[english]{babel}
\usepackage[utf8x]{inputenc}
\usepackage[T1]{fontenc}
\usepackage{apacite}
\usepackage{algorithm}
\usepackage{tikz}
\usetikzlibrary{positioning}
\usepackage{booktabs}
\usepackage{subfiles} 
\usepackage{multirow}
\usepackage{siunitx}
\usepackage{algpseudocode}
\usepackage{mathrsfs}
\usepackage{listings}
\usepackage{color}
\usepackage{bbm}
\usepackage{adjustbox}
\usepackage{epsfig, epsf, graphicx}
\usepackage{tikz}
\usepackage{animate}
\usepackage{pstricks, pst-node, psfrag}
\usepackage{amssymb, amsmath, amsthm, bm}
\usepackage{verbatim,enumerate}
\usepackage{rotating, lscape}
\usepackage{diagbox}
\usepackage[doublespacing]{setspace}

\usetikzlibrary{arrows,calc,tikzmark}
\usepackage[colorlinks=true,linkcolor=blue,citecolor=blue]{hyperref}%
\usepackage{float}
\usepackage{natbib}
\usepackage[small]{caption}
\usepackage{subcaption}
\usepackage{lipsum}
\usepackage{threeparttable}
\usepackage{multirow}
\usepackage{lmodern}
\usepackage{multicol}
\usepackage[toc,page]{appendix}
\graphicspath{{figures/}}

\makeatletter
\def\BState{\State\hskip-\ALG@thistlm}
\makeatother
\makeatletter
\newcounter{phase}[algorithm]
\newlength{\phaserulewidth}
\newcommand{\setphaserulewidth}{\setlength{\phaserulewidth}}

\makeatother
\setphaserulewidth{.7pt}
\algrenewcommand\algorithmicrequire{\textbf{Input:}}
\algrenewcommand\algorithmicensure{\textbf{Output:}}
\newcounter{case}[algorithm]
\newlength{\caserulewidth}
\newcommand{\setcaserulewidth}{\setlength{\caserulewidth}}

\makeatother
\setcaserulewidth{.7pt}
\algdef{SE}[SUBALG]{Indent}{EndIndent}{}{\algorithmicend\ }%
\algtext*{Indent}
\algtext*{EndIndent}

\def\boxit#1{\vbox{\hrule\hbox{\vrule\kern6pt
			\vbox{\kern6pt#1\kern6pt}\kern6pt\vrule}\hrule}}

\def\bse{\begin{eqnarray*}}
	\def\ese{\end{eqnarray*}}
\def\be{\begin{eqnarray}}
\def\ee{\end{eqnarray}}
\def\bq{\begin{equation}}
\def\eq{\end{equation}}
\def\bse{\begin{eqnarray*}}
	\def\ese{\end{eqnarray*}}

\DeclareMathOperator{\Tr}{tr}

\newcommand{\0}{\mathbf{0}}

\newcommand{\bI}{\mathbf{I}}
\newcommand{\bL}{\mathbf{L}}

\newcommand{\redcolor}{\color{black}}

\newcommand{\bs}{\mathbf{s}}
\newcommand{\bSigma}{\bm{\Sigma}}

\newcommand{\btheta}{\bm{\theta}}

\newcommand{\bx}{\mathbf{x}}

\newcommand{\by}{\mathbf{y}}

\newcommand{\bz}{\mathbf{z}}
\newcommand{\bZ}{\mathbf{Z}}

\newcommand{\R}{\mathbb{R}}

\theoremstyle{plain}

\theoremstyle{definition}

\begin{document}

\thispagestyle{empty} \baselineskip=28pt \vskip 5mm
\begin{center} {\Huge{\bf Which Parameterization of the Matérn Covariance Function?}}
	
\end{center}

\baselineskip=12pt \vskip 10mm

\begin{center}\large
Kesen Wang\footnote[1]{\baselineskip=10pt Statistics Program,
King Abdullah University of Science and Technology,
Thuwal 23955-6900, Saudi Arabia.\\
E-mail: kesen.wang@kaust.edu.sa, marc.genton@kaust.edu.sa, ying.sun@kaust.edu.sa
}, Sameh Abdulah\footnote[2]{Extreme Computing Research Center (ECRC), King Abdullah University of Science and Technology,
Thuwal 23955-6900, Saudi Arabia. E-mail: sameh.abdulah@kaust.edu.sa 
\\
This research was supported by the
King Abdullah University of Science and Technology (KAUST).}, Ying Sun\textcolor{blue}{$^{1,2}$}, Marc G. Genton\textcolor{blue}{$^{1,2}$}
\end{center}
\baselineskip=17pt \vskip 4mm \centerline{\today} \vskip 7mm

\begin{center}
{\large{\bf Abstract}}
\end{center}

 The Matérn family of covariance functions is currently the most popularly used model in spatial statistics, geostatistics, and machine learning to specify the correlation between two geographical locations based on spatial distance. Compared to existing covariance functions, the Matérn family has more flexibility in data fitting because it allows the control of the field smoothness through a dedicated parameter.  Moreover, it generalizes other popular covariance functions. However, fitting the smoothness parameter is computationally challenging since it complicates the optimization process. As a result, some practitioners set the smoothness parameter at an arbitrary value to reduce the optimization convergence time. In the literature, studies have used various parameterizations of the Matérn covariance function, assuming they are equivalent. This work aims at studying the effectiveness of different parameterizations under various  settings. We demonstrate the feasibility of inferring all parameters simultaneously and quantifying their uncertainties on large-scale data using the \textit{ExaGeoStat} parallel software. We also highlight the importance of the smoothness parameter by analyzing the Fisher information of the statistical parameters. We show  that the various parameterizations have different properties and differ from several perspectives. In particular, we study the three most popular parameterizations in terms of parameter estimation accuracy, modeling accuracy and efficiency, prediction efficiency, uncertainty quantification, and asymptotic properties. We further  demonstrate their differing performances under nugget effects and approximated covariance. Lastly, we give recommendations for parameterization selection based on our experimental results.

\baselineskip=14pt

\par\vfill\noindent

\clearpage\pagebreak\newpage \pagenumbering{arabic}
\baselineskip=26pt
\section{Introduction}

The Matérn covariance function has been a longstanding tool for statisticians in spatial data analysis. Its purpose is to model the correlation between measurements at geospatial locations, allowing for predictions of unobserved values at new locations within the study region. Originally proposed by \cite{Matrn1960SpatialV} and later popularized by \cite{handcock1993bayesian}, the Matérn covariance is stationary and isotropic, that is, distance-dependent and invariant under translations and rotations. It characterizes the spatial correlation between two random variables at locations $\bs_i$ and $\bs_j$, separated by a distance $h=\|\boldsymbol{h}\|$, where $\boldsymbol{h}=\bs_i - \bs_j$:
\begin{align*}
    \text{Cor}(\|\boldsymbol{h}\|;\btheta)=\frac{1}{2^{\nu-1}\Gamma(\nu)}(\kappa \|\boldsymbol{h}\|)^\nu {\cal K}_\nu(\kappa \|\boldsymbol{h}\|) \propto \int_{\R^{d}}e^{i\boldsymbol{h}^\top\bz}(\|\bz\|^2+\kappa^2)^{-(2\nu+d)/2}\text{d}\bz, \quad \btheta=(\kappa,\nu)^\top,
\end{align*}
where $d$ is the dimension of the spatial field, $\kappa > 0$ denotes the scale parameter, $\nu > 0$ represents the smoothness parameter, and ${\cal K}_\nu(\cdot)$ denotes the modified Bessel function of the second kind with order $\nu$. The Matérn correlation function is independent of the field dimension $d$, which is integrated out from the spectral density $(\|\bz\|^2+\kappa^2)^{-(2\nu+d)/2}$. In addition, ${\text{Cor}}(h;\btheta)$ is standardized so that ${\text{Cor}}(0;\btheta)=1$. To make it a legitimate covariance function, statisticians often scale $\text{Cor}(h;\btheta)$ by a constant factor to account for non-unit variance. The Matérn covariance function is well-known for its flexibility which includes numerous covariance functions as special cases such as the exponential $(\nu=1/2)$, Whittle $(\nu=1)$, and squared-exponential $(\nu=\infty)$ covariances. In the literature, variants of the Matérn covariance have been developed for different purposes and motivations. There are currently three ubiquitous parameterizations of the Matérn covariance function in the literature: 
\begin{align}
   {\cal M}_1(h;\boldsymbol{\theta}_1)&=\frac{\sigma^2}{2^{\nu-1}\Gamma(\nu)}\left(\frac{h}{\beta}\right)^\nu {\cal K}_\nu \left(\frac{h}{\beta} \right) + \mathbbm{1}_{h=0}\cdot \tau^2, \quad &&\boldsymbol{\theta}_1  = (\sigma^2, \beta, \nu, \tau^2)^\top,
   \label{eq:matern1}\\
    {\cal M}_2(h;\boldsymbol{\theta}_2)&=\frac{\sqrt{\pi}\phi}{2^{\nu-1}\Gamma(\nu+\frac{1}{2})\alpha^{2\nu}}(\alpha h)^\nu {\cal K}_\nu(\alpha h)+ \mathbbm{1}_{h=0}\cdot \tau^2, \quad&&\boldsymbol{\theta}_2 = (\phi, \alpha, \nu, \tau^2)^\top,
    \label{eq:matern2}\\
    {\cal M}_3(h;\boldsymbol{\theta}_3)&=\frac{\sigma^2}{2^{\nu-1}\Gamma(\nu)}\left(\frac{2\sqrt{\nu}h}{\rho}\right)^\nu {\cal K}_\nu \left(\frac{2\sqrt{\nu}h}{\rho}\right)+ \mathbbm{1}_{h=0}\cdot \tau^2, \quad&& \boldsymbol{\theta}_3 = (\sigma^2, \rho, \nu, \tau^2)^\top.
    \label{eq:matern3}
\end{align}
In variant ${\cal M}_1$, \(\sigma^2\), \(\beta\), and \(\nu\) are the variance, range, and smoothness parameters, respectively. As for variant ${\cal M}_2$,  $\alpha$ partially functions like an inverse range parameter and primarily affects low-frequency behaviors \citep{stein1999interpolation}. Additionally, $\phi$ is the overall scale parameter that pertains to the high-frequency behavior of the spatial process. In variant ${\cal M}_3$, $\sigma^2$  plays the same role as in variant ${\cal M}_1$, and $\rho$ resembles the functionality of $\beta$ as the range parameter. The roles of the parameter $\nu$, which controls the decay rate at high frequencies or, correspondingly, the field smoothness, is immutable in all three variants. Although these variants may seem quite different, their essence remains unchanged. There are scale or range parameters (which can be easily transformed from one to another as presented in Table~\ref{table1}) that measure the dependence range and smoothness parameters that govern the decay rate of the dependence. When $h=0$, the three variants result in $\text{Var}\{Z(\bs)\}$, where $Z(\bs)$ represents a Gaussian random field (GRF) with locations $\bs$, in differing forms. For instance, $\text{Var}\{Z(\bs)\}=\sigma^2+\tau^2$ in ${\cal M}_1$ and ${\cal M}_3$ but has a much more complicated expression in ${\cal M}_2$, as specified in the links to $\sigma^2$. Moreover, $\tau^2$  represents the nugget effect due to the measurement error. 

\begin{table}[H]
    \centering
    \caption{Link functions for transformations in the ${\cal M}_1$, ${\cal M}_2$, and ${\cal M}_3$ parameterizations of the Matérn covariance function.}
    \resizebox{1\linewidth}{!}{
    \begin{tabular}{|c|c|c|c|}
    \hline
           $\nearrow$    & \({\cal M}_1\) & \({\cal M}_2\) & \({\cal M}_3\)\\
                       \hline
      ${\cal M}_1$   & $\sigma^2$, $\beta$, $\nu$, $\tau^2$ & \(\phi=\frac{\sigma^2\Gamma(\nu+\frac{1}{2})}{\pi^{1/2}\Gamma(\nu)\beta^{2\nu}}\), \(\alpha=\frac{1}{\beta}\), $\tau^2$ & $\sigma^2$, $\rho=2\nu^{1/2}\beta$, $\nu$, $\tau^2$\\
      
      \hline
      ${\cal M}_2$   & \(\sigma^2=\frac{\pi^{1/2}\phi\Gamma(\nu)}{\Gamma(\nu+\frac{1}{2})\alpha^{2\nu}}\), $\beta=\frac{1}{\alpha}$, $\nu$, $\tau^2$ & $\phi$, $\alpha$, $\nu$, $\tau^2$ & \(\sigma^2=\frac{\pi^{1/2}\phi\Gamma(\nu)}{\Gamma(\nu+\frac{1}{2})\alpha^{2\nu}}\), $\rho=\frac{2\nu^{1/2}}{\alpha}$, $\nu$, $\tau^2$ \\
      \hline
      ${\cal M}_3$ & $\sigma^2$, $\beta=\frac{\rho}{2\nu^{1/2}}$, $\nu$, $\tau^2$ & \(\phi=\frac{\sigma^2\Gamma(\nu+\frac{1}{2})(2\nu^{1/2})^{2\nu}}{\pi^{1/2}\Gamma(\nu)\rho^{2\nu}}\),  $\alpha=\frac{2\nu^{1/2}}{\rho}$, $\nu$, $\tau^2$ & $\sigma^2$, $\rho$, $\nu$, $\tau^2$  \\
      \hline
    \end{tabular}
    }
    \label{table1}
\end{table}

Regarding the associated applications, for instance, the correlation form of variant ${\cal M}_1$ was introduced and applied by \cite{handcock1993bayesian} to illustrate the idea of kriging in the Bayesian context. \cite{abdulah2018parallel}  and \cite{abdulah2019exageostatr} have applied variant ${\cal M}_1$ to model large-scale geospatial data. It is also applied in the commonly used R package for spatial statistics \textit{fields} \citep{fields} and \textit{geoR} \citep{geor} as the Matérn model. \cite{hong2021efficiency} used variant ${\cal M}_1$ to assess prediction efficiency for large spatial datasets with approximated covariance functions. Variant ${\cal M}_2$ was employed in \cite{stein1999interpolation} to define the Matérn class, avoiding the problematic spectral density concentration at the origin happening in
$
    {\cal M}_4(h;\btheta_2) = \phi (\alpha h)^\nu K_\nu(\alpha h) + {\mathbbm{1}_{h=0}}\cdot \tau^2. \nonumber
$
 In addition, variant ${\cal M}_2$ was also applied by \cite{stein1999interpolation} and \cite{loh2005fixed} to study domain asymptotics for Matérn GRFs. Variant ${\cal M}_3$ was recommended by \cite{handcock1994approach} and \cite{stein1999interpolation} for a more independent interpretation of $\rho$ from $\nu$ compared to their counterparts in ${\cal M}_2$.  \cite{geoga2020scalable} applied ${\cal M}_3$ for the log-likelihood optimization. \cite{geoga2022fitting} also used ${\cal M}_3$ to compute the Fisher information matrix. Furthermore, \cite{de2022information} studied the Fisher information pattern of $\nu$ with ${\cal M}_3$.  In addition, ${\cal M}_3$ is employed in the R package \textit{RandomFields} \citep{randomfield}, another popularly used spatial statistics package, as one of the Matérn models. 


 As mentioned above, practitioners seem to use the three parameterizations rather arbitrarily and interchangeably with the assumption that they are identical. For instance, these variants can be easily transformed from one to another according to the link functions provided in Table~\ref{table1}. Thus, the three parameterizations can characterize the same field if the relationships provided in Table \ref{table1} are applied. In addition, the various parameterizations are redundant in the maximum likelihood estimation. Indeed, by the {\redcolor invariance} property of maximum likelihood estimators (MLEs), we can transform the MLEs of one parameterization to those of another without losing optimality. However, the general {\redcolor interchangeability} assumption of the three parameterizations is not quite right because numerous aspects require further consideration. 
\begin{figure}[t!]
\begin{subfigure}{0.33\textwidth}
  \centering
  \includegraphics[width=1\textwidth,]{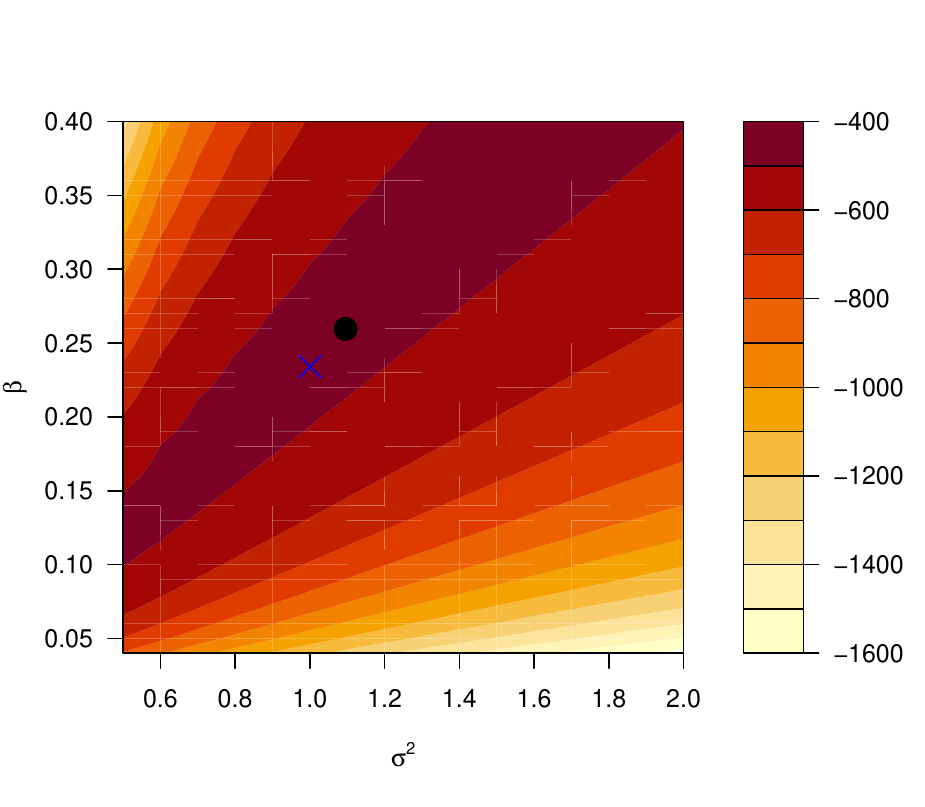}
  \caption*{${\cal M}_1$, strong}
\end{subfigure}
\begin{subfigure}{0.33\textwidth}
  \centering
  \includegraphics[width=1\textwidth,]{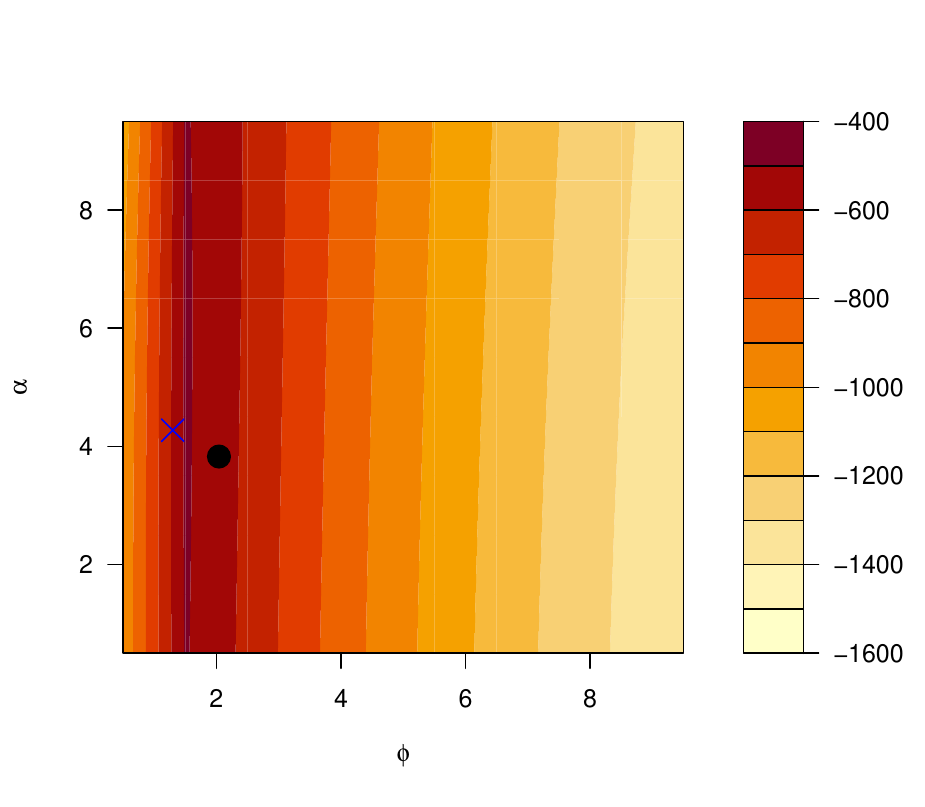}
  \caption*{${\cal M}_2$, strong}
\end{subfigure}
\begin{subfigure}{0.33\textwidth}
  \centering
  \includegraphics[width=1\textwidth,]{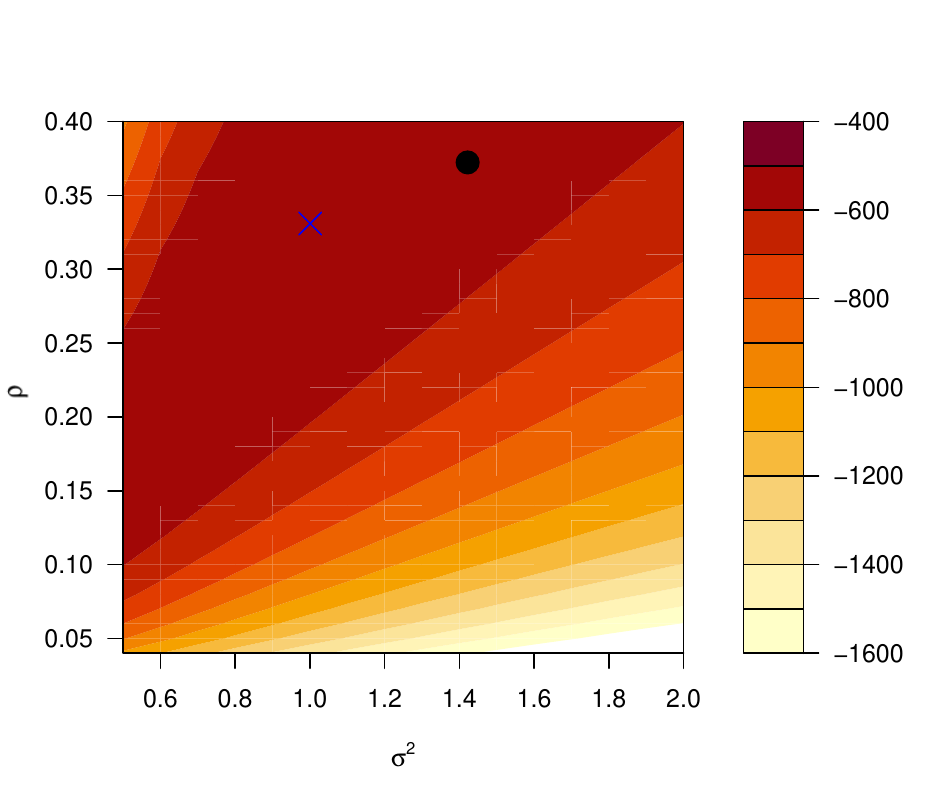}
  \caption*{${\cal M}_3$, strong}
\end{subfigure}
\caption{ Contour plots of the log-likelihood function of 1600 samples simulated from a strong exponential Gaussian field parameterized in ${\cal M}_1$, ${\cal M}_2$, and ${\cal M}_3$ with $\nu=0.5$. Contours of the log-likelihood functions for $\nu = 1$ tell a similar story and, thus, are omitted for concision. Contour plots for the 1600 samples simulated from weak and mediums exponential Gaussian fields are shown in Figure \ref{contour} in the Supplementary Material. The black dot in each figure represents the MLEs and the blue cross represents the true parameter values.}
\label{c}
\end{figure}
First, the MLE is also a statistic that may have different statistical properties, such as asymptotic variance and correlation in various parameterizations for different domains and fields. Second, the differing parameter space of the three variants results in different shapes of the log-likelihood function (shown in Figure \ref{c} and in Figure \ref{contour} in the Supplementary Material) and, therefore, in differences in optimization performance. In particular, although the parameterizations ${\cal M}_1$ and ${\cal M}_3$ are mathematically similar, shapes of their log-likelihood functions are quite different near the optimal values (log-likelihood functions of ${\cal M}_3$ tend to be much flatter). In addition, the log-likelihood functions parameterized in ${\cal M}_2$ are quite different from those of ${\cal M}_1$ and ${\cal M}_3$. Therefore, a possible difference exists in the modeling and prediction efficiency for the three variants.

 In this work, we examine the parameterizations \({\cal M}_1\), \({\cal M}_2\), and \({\cal M}_3\) from the perspective of increasing-domain asymptotics and numerical optimization performance with the assessment tools implemented in {\em ExaGeoStat}, a high-performance parallel software for large-scale spatial statistics \citep{abdulah2018exageostat}. Among these benchmark functions, we rely on the Fisher information matrix to investigate the increasing-domain asymptotics of all parameter estimates, including the smoothness parameter, mean square prediction error (MSPE) for prediction performance, the mean loss of efficiency (MLOE) and the mean misspecification of the mean square error (MMOM) criteria for modeling accuracy \citep{hong2021efficiency} and the total number of optimization iterations for modeling speed. Finally, we adopt the tile low-rank (TLR) covariance matrix approximation method, implemented by \cite{abdulah2018parallel}, to study the estimation accuracy of the three variants under an approximated covariance. Experiments under TLR settings are only conducted for ${\cal M}_1$ and ${\cal M}_3$ because ${\cal M}_2$ encounters numerical issues when compressing the covariance matrix in low-rank. All of the experiments are also repeated and jointly analyzed with nugget effects.

The paper is organized as follows. Section~2 presents a brief background of the {\em ExaGeoStat} software, and Section~3 introduces the embedded benchmarking functions of {\em ExaGeoStat}. Next, Section 4 analyzes the experimental results obtained from the benchmarking functions. Section~5 summarizes and studies the TLR covariance approximation method for the three variants. Lastly, Section 6 applies the three variants to a Saudi wind speed spatial dataset.

\section{Large-Scale Spatial Modeling and Prediction}
A significant challenge of handling large-scale spatial data is operating on the large covariance matrix of size $n\times n$, where $n$ is the number of spatial locations. Computations involving matrices of large dimensions are usually of considerable complexity and become intractable as $n$ becomes large. Specifically, evaluating the Gaussian log-likelihood function requires $O(n^2)$ memory space and $O(n^3)$ operations. \cite{abdulah2018exageostat} developed the geostatistical software \textit{ExaGeoStat} to address this issue. It is a unified multicore high-performance computing software for large-scale geospatial data modeling and predictions on large-scale systems, including graphical processing units (GPUs) \citep{abdulah2018parallel}. \textit{ExaGeoStat} is a powerful tool for researchers working in geostatistics, providing advanced capabilities for large-scale spatial modeling and simulation that are not available in traditional software packages. \textit{ExaGeoStat} adopts a three-layer software structure that includes a dynamic runtime system library, task-based parallel linear algebra solvers, and geospatial operations from bottom to top \citep{abdulah2019exageostatr}. 

Upon its initial release, the developers implemented a stationary GRF data generator on a unit square based on ${\cal M}_1$. The data generator randomly selects a specific number of irregularly distributed locations within the unit square to calculate the distance matrix. Then it generates synthetic GRFs using the covariance matrix computed by applying ${\cal M}_1$ on the distance matrix. The synthetic data generator adopts parallel computing to split the large covariance matrix, which might be cumbersome, into smaller tiles and employs a column-wise filling procedure for each tile. The column-wise filling process operates in parallel on the tiles. In addition, \textit{ExaGeoStat} performs all matrix operations in parallel by relying on state-of-the-art parallel linear algebra libraries to enhance computation efficiency significantly. The {\em ExaGeoStat} software also supports several approximation techniques in the covariance matrix, including independent block approximation (IND), TLR \citep{abdulah2018parallel}, and mixed-precision \citep{abdulah2021accelerating,cao2022reshaping}.  The IND algorithm is based on the covariance approximation method discussed by \cite{stein2014limitations}, which attempts to decompose the large covariance matrix into blocks and annihilate off-diagonal blocks up to a prespecified bandwidth of the diagonal blocks. With TLR approximations, the authors aim to exploit the sparsity of the covariance matrix to reduce the memory footprint and decrease the execution time. With mixed-precision approximations, they aim to have more refined reduction techniques for the complexity of handling the covariance matrix by supporting lower-precision computation on the off-diagonal tiles.


\section{Benchmarking Functions}
The objective of this study is to evaluate various variants of the Matérn covariance function using several metrics. In the subsequent subsections, we outline the specific metrics used in this study.

\subsection{Maximum Likelihood Estimation}
Using the Matérn form ${\cal M}_1$, $\boldsymbol{\theta}_{1}$ can be estimated  by maximizing the log-likelihood function:
\begin{equation}
l(\boldsymbol{\theta}_{1})=-\frac{1}{2}\log|\bSigma(\boldsymbol{\theta}_{1})|-\frac{1}{2}\bZ^\top{\bSigma(\boldsymbol{\theta}_{1})}^{-1}\bZ-\frac{n}{2}\log(2\pi),
\label{e4}
\end{equation}
where $\bZ$ is a realization of a zero-mean GRF with covariance matrix ${\bSigma}(\boldsymbol{\theta}_{1})$. As presented by \cite{abdulah2018parallel}, \textit{ExaGeoStat}, the main software platform in this study, applies ${\cal M}_1$ on the distance matrix to generate $\bSigma(\btheta_1)$ in a column-wise block parallel fashion. Analogously, in this work, we implement the generation function of $\bSigma(\btheta_{2})$ and $\bSigma(\btheta_{3})$ by applying the Matérn covariance functions ${\cal M}_2$ and ${\cal M}_3$ on the distance matrix. The synthetic data generator subsequently receives $\bSigma(\btheta_{2})$ or $\bSigma(\btheta_{3})$ to simulate realizations from the GRFs based on ${\cal M}_2$ or ${\cal M}_3$. Finally, the optimization function in {\em ExaGeoStat} maximizes $l(\btheta_{2})$ and $l(\btheta_{3})$ with the embedded model-based optimization algorithm \textbf{BOBYQA} presented by \cite{powell2009bobyqa} to obtain MLEs of $\btheta_{2}$ and $\btheta_{3}$, respectively. Using \textbf{BOBYQA} in {\em ExaGeoStat} avoids explicitly computing the first derivatives of the log-likelihood function, which can be cumbersome, especially when the sample size is large. To demonstrate this statement, we compute the first derivative of $l(\btheta)$ based on the properties of matrix derivatives derived by \cite{petersen2008matrix} as follows:
\begin{align*}
    \frac{\partial l(\btheta)}{\partial \theta_k}=-\frac{1}{2}\Tr\left\{\bSigma(\btheta)^{-1}\frac{\partial \bSigma(\btheta)}{\partial \theta_k}\right\}+\frac{1}{2}\Tr\biggl\{\bSigma(\btheta)^{-1}\bZ\bZ^\top\bSigma(\btheta)^{-1}\frac{\partial \bSigma(\btheta)}{\partial \theta_k}\biggr\}.
\end{align*}
The expression above involves one call of matrix inversion and five calls of matrix multiplication. Matrix inversion and multiplication have high computational time complexity, which can take up to $O(n^2\log n)$ and $O(n^3)$ operations, respectively,  rendering gradient-based algorithms non-scalable on large spatial datasets. Moreover, the generation of $\frac{\partial \bSigma(\btheta)}{\partial \theta_k}$ relies on equations \ref{e6}-\ref{e15} in the Supplementary Material, which can be slow when involving numerous calls of the modified Bessel function of the second kind especially in the computation of the derivative with respect to $\nu$. {\redcolor Moreover, $\frac{\partial l(\btheta)}{\partial \theta_k}$ is frequently unstable during the optimization, leading to problematic MLEs. Although gradient-based methods can take fewer iterations to converge given a reasonable starting point, they are overshadowed by the amount of time consumed for the gradient computation.}     

In addition, \textbf{BOBYQA} is an iterative local optimization approach subject to bounded constraints. The main rationale behind the mechanism of \textbf{BOBYQA} is to approximate a function $F(\bx),\bx \in \R^n,$ through a quadratic form $Q$ that satisfies $Q(\by_i)=F(\by_i),i=1,\dots,m$, where $\by_i$ denotes an interpolation point. Then, each model update is realized by minimizing the  Frobenius norm of the difference between the second-derivative matrices of $Q$, in symbols $\|\nabla^2 Q_{k+1} - \nabla^2 Q_k\|_{\text{F}}$.  Updating from $Q_k$ to $Q_{k+1}$ only requires $O(n^2)$ operations if the number of interpolation points is $m=2n+1$ and calculating the change $Q_{k+1}-Q_k$ is limited to $O(m^2)$ operations. Therefore, \textbf{BOBYQA} is more computationally efficient than gradient-based algorithms consuming at least $O(n^3)$ operations, and more scalable on large datasets. 

\subsection{Fisher Information Matrix}
The Fisher information matrix measures the amount of information on unknown parameters estimated from an observed sample and provides practical statistical insights into the uncertainty analysis of MLEs. \cite{mardia1984maximum} demonstrated that the Fisher information matrix could accurately approximate the covariance matrix of the MLEs even if the sample sizes are small. \cite{abt1998fisher} illustrated that the covariance matrix of the limiting distribution of the MLEs for Gaussian processes is equal to the limit of the inverse Fisher information matrix, and it accurately approximates the variability and correlation between MLEs even in singularity conditions. To calculate the Fisher information matrix, we define a random variable $X$ characterized by the $m$-dimensional parameter vector $\btheta$ with density function \(f(x;\boldsymbol{\theta})\). The associated Fisher information matrix,
\({\bI}(\boldsymbol{\theta})\)=\(\{\text{I}_{ij}(\boldsymbol{\theta})\}_{i,j=1,\dots,m}\), of $X$ can be computed as follows:
\begin{equation*}
\text{I}_{ij}(\boldsymbol{\theta})=-\text{E}\left[\left\{\frac{\partial}{\partial\theta_i} \log f(X;\boldsymbol{\theta})\right\}\left\{\log\frac{\partial}{\partial\theta_j} f(X;\boldsymbol{\theta})\right\}\right].
\end{equation*}

\cite{geoga2020scalable} recalled the asymptotic theory of MLEs, which states that if the smallest eigenvalue of \(\bI(\boldsymbol{\theta})\) tends to infinity as the sample size increases, then \(\bI(\hat{\boldsymbol{{\theta}}})^{1/2}(\hat{\boldsymbol{{\theta}}}-\boldsymbol{\theta}){\rightarrow} {\cal N} _m(\boldsymbol{0},\bI_m)\), where \(\hat{\boldsymbol{{\theta}}}\) is the estimated parameter vector of the MLE. The asymptotic distribution of MLEs enables calculations of confidence intervals, which provide statistical tools to assess the estimation accuracy and quantify uncertainty. The Fisher information matrix of a GRF depends entirely on the covariance matrix and its first-order derivatives with respect to the parameters. Moreover, in a GRF, the mean $\mu$ is either assumed to be zero in common practice, as we typically model the residuals, or is empirically removed from the observations by subtracting the sample mean. Therefore, a sensible option is to exclude $\mu$ from the Fisher information matrix. Consequently, we can express the Fisher information matrix of a Matérn GRF, \(\bI(\boldsymbol{\theta})=\{\text{I}_{ij}(\boldsymbol{\theta})\}_{i,j=1,2,3,4}\) , in the form of taking the trace of a sequence of matrix multiplications \citep{mardia1984maximum}:
\begin{equation}
\bI_{ij}(\boldsymbol{\theta})=\frac{1}{2}\mbox{tr}\left\{\boldsymbol{\Sigma(\btheta)}^{-1}\boldsymbol{\Sigma}_i\boldsymbol{\Sigma(\btheta)}^{-1}\boldsymbol{\Sigma}_j\right\},
\label{e5}
\end{equation}
where \(\bSigma_i=\frac{\partial\boldsymbol{\Sigma(\theta)}}{\partial\theta_i} \) denotes the derivative of  \(\mathbf{\Sigma}(\boldsymbol{\theta})\) with respect to \(\theta_i\) for $i=1,2,3,4$.

The computation of the Fisher information matrix, involving one call of matrix inversion and three calls of matrix multiplication, is usually costly when the sample size is large. To reduce complexity, \cite{geoga2020scalable} approximated the Fisher information matrix using hierarchical matrices,  allowing for quasi-linear time complexity and memory space. The results from \cite{geoga2020scalable} revealed that the proposed hierarchical matrices accelerate the computing process while maintaining decent accuracy. However, this technique excessively relies on matrix approximations. Therefore, this work presents a parallel computing technique that computes the Fisher information precisely and controls the computational complexity. 

\begin{figure}
 \begin{subfigure}{.5\textwidth}
  \centering
  \includegraphics[width=.8\linewidth]{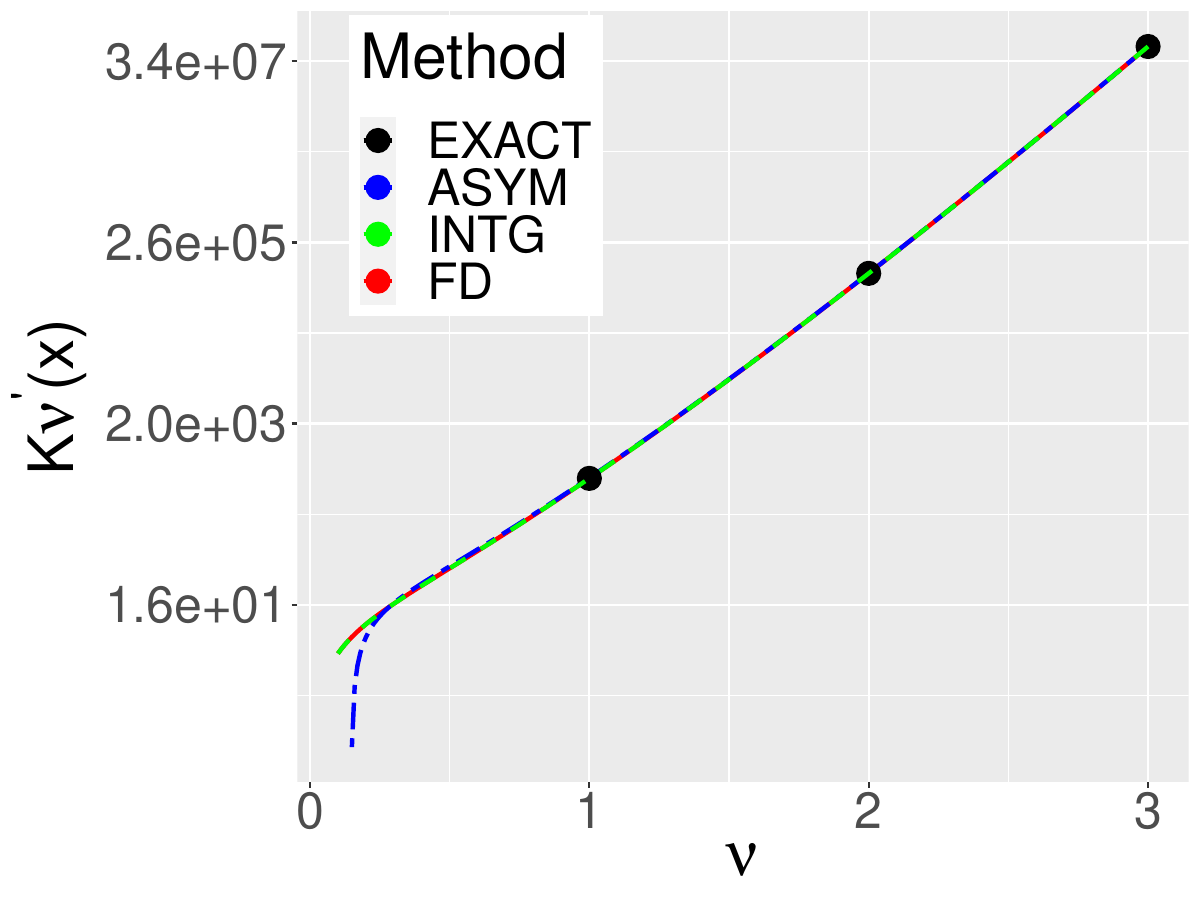}
  \caption{\hspace{15mm} $x=0.01$}
  \label{fig1(a)}
  \hspace{5mm}
\end{subfigure}
 \begin{subfigure}{.5\textwidth}
  \centering
  \includegraphics[width=.8\linewidth]{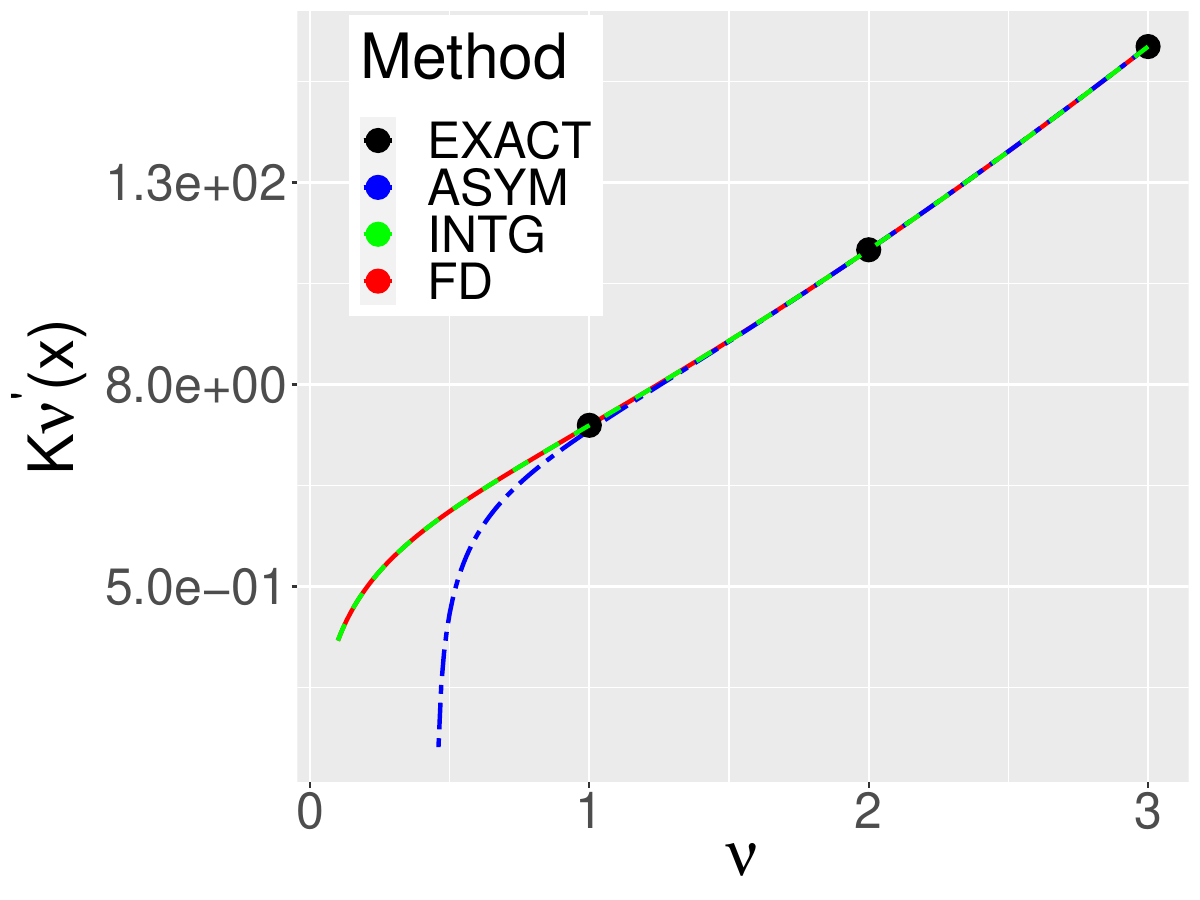}
  \caption{\hspace{15mm}$x=0.3$}
  \label{fig1(b)}
  \hspace{5mm}
\end{subfigure}
 \begin{subfigure}{.5\textwidth}
  \centering
  \includegraphics[width=.8\linewidth]{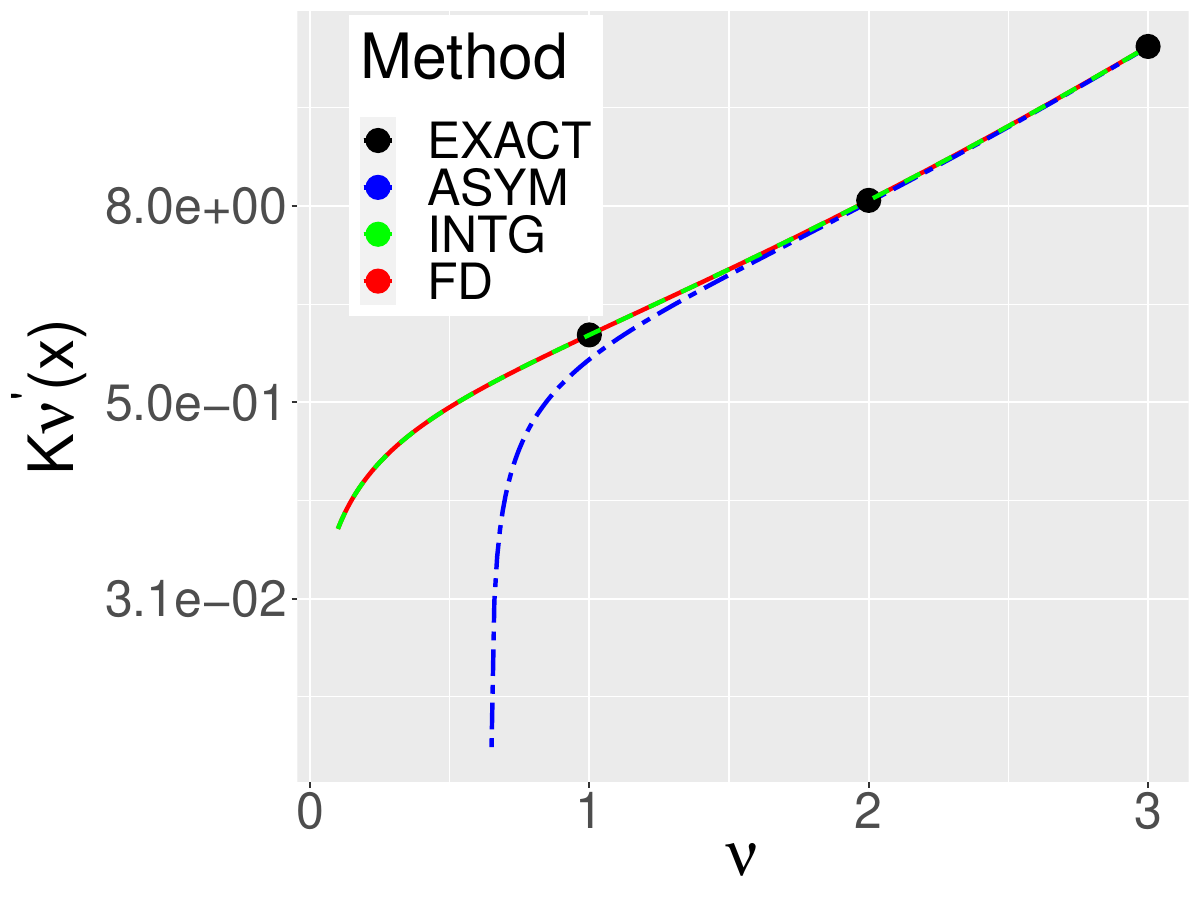}
  \caption{\hspace{15mm}$x=0.6$}
\end{subfigure}
 \begin{subfigure}{.5\textwidth}
  \centering
  \includegraphics[width=.8\linewidth]{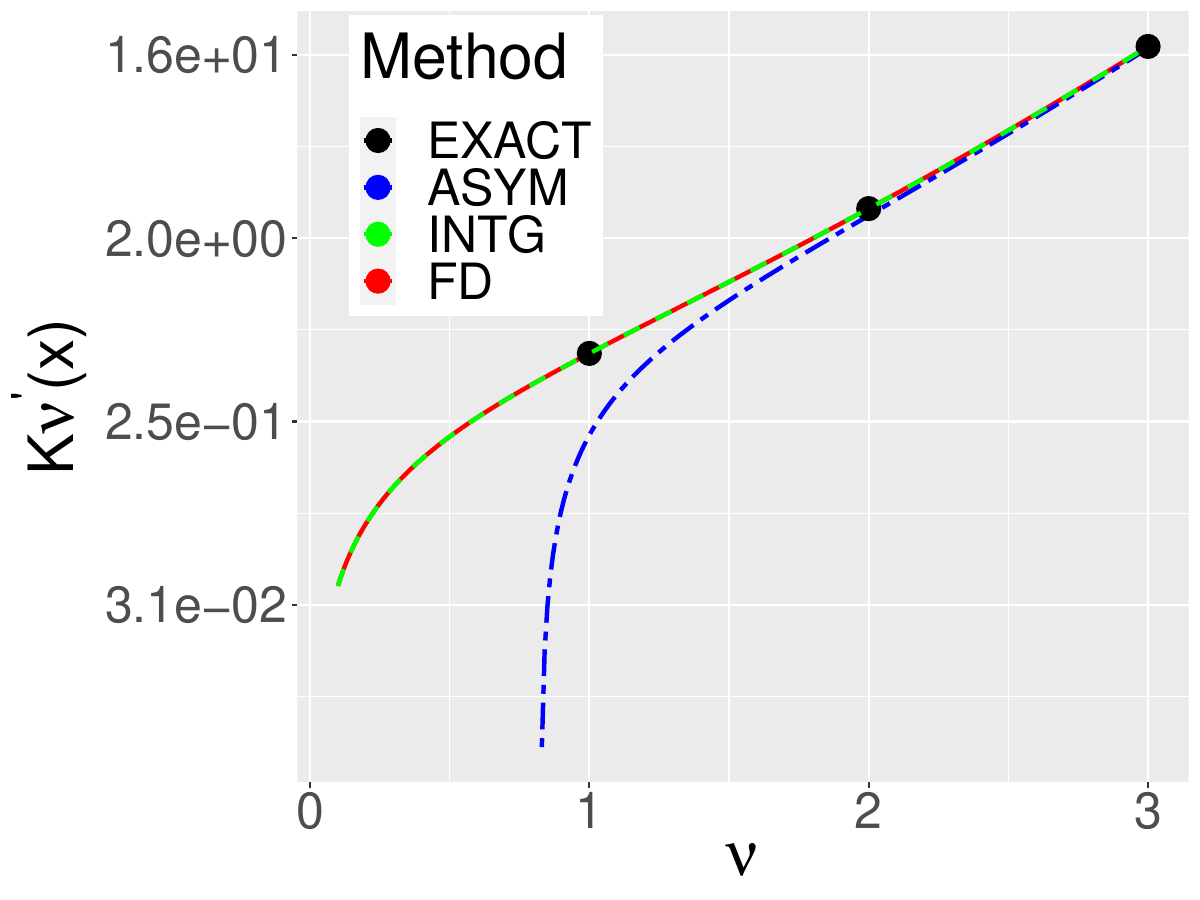}
  \caption{\hspace{15mm}$x=0.9$}
  \label{fig1(d)}
\end{subfigure}
\caption{Transformed $\log_2$ values of \(\frac{\partial {\cal K}_\nu(x)}{\partial \nu}\) returned from the finite difference approach (FD, red), exact integer form (EXACT, black), asymptotic approximation (ASYM, blue), and integral form (INTG, green).}
\label{fig1}
\end{figure}

To compute the Fisher information matrix of Matérn GRFs in parallel, we have to compute the first-order derivatives of ${\cal M}_1$, ${\cal M}_2$, and ${\cal M}_3$ so that we can generate \(\boldsymbol{\Sigma}_i\) in the same tile-based parallel fashion as we generate $\bSigma(\btheta)$ as presented by \cite{abdulah2018exageostat}. Section S1 in the Supplementary Material contains all the required first-order derivatives associated with ${\cal M}_1$, ${\cal M}_2$, and ${\cal M}_3$.  
In these equations, \(\Psi(\cdot)=\Gamma^\prime(\cdot)/\Gamma(\cdot)\) represents the digamma function. In addition, \({\cal K}_\nu^\prime(x)\) denotes the derivative of the modified Bessel function of the second kind with respect to its argument $x$, and \({\cal K}_{\nu^\prime}(x)\) denotes the first-order derivative taken with respect to its order $\nu$. {\em ExaGeoStat} calculates ${\cal K}_\nu(x)$ using the GNU scientific library \citep{gough2009gnu} and first-order derivatives of the modified Bessel function of the second kind with self-implemented functions based on the following identities:
\begin{align*}
{\cal K}^\prime_\nu(x)&=-\frac{1}{2}\{{\cal K}_{\nu-1}(x)+{\cal K}_{\nu+1}(x)\},\\
{\cal K}_{\nu^\prime}(x)&=\frac{\partial {\cal K}_\nu(x)}{\partial\nu}=\lim_{\Delta\rightarrow 0} \frac{{\cal K}_{\nu+\Delta}(x)-{\cal K}_\nu(x)}{\Delta}.
\end{align*}
The first-order (and even higher-order) derivative of the modified Bessel function of the second kind with respect to $\nu$ does have closed-form expressions \citep{gonzalez2018closed}, and there are series approximations that attempt to ease their numerical computation \citep{olver2010nist}. However, we have found no existing well-compatible algorithms that perform this task. In \cite{geoga2022fitting}, the authors provided an automatic differentiation method to compute \(\frac{\partial K_\nu(x)}{\partial \nu}\) by splitting the smoothness parameter $\nu$ into different categories and using the most efficient differentiation methodology for each. Nonetheless, the proposed software is currently only compatible with Julia. The lack of computational tools led us to implement a simple finite difference (FD) approach with an increment of $\Delta=10^{-9}$, and we choose such a small tolerance because ${\cal K}_\nu(x)$, in most cases, is a smooth function of its order $\nu$:
\begin{align*}
\frac{\partial {\cal K}_\nu(x)}{\partial \nu}|_{\nu=n}&=  \frac{n!}{2}\sum_{i=0}^{n-1}\frac{(\frac{x}{2})^{i-n}}{i!(n-i)}{\cal K}_i(x), \quad n\in \mathbb{Z}^{+}, &&\quad \text{Exact Integer Form (EXACT)}
\\
 \frac{\partial {\cal K}_\nu(x)}{\partial \nu} &=  \int_{0}^{\infty} t e^{-x \text{cosh}(t)}\text{sinh}(\nu t)\text{d}t, &&\quad \text{Integral Form (INTG)}
 \\
 {\redcolor \frac{\partial {\cal K}_\nu(x)}{\partial \nu}}  &{\redcolor \approx  {\cal K}_\nu(x)\biggl\{1-\frac{1}{2\nu}-\log \left(\frac{e\cdot x}{2\nu}\right)\biggr\}, \quad{\nu \rightarrow \infty}.} &&\quad \text{Asymptotic Form (ASYM)}
\end{align*}
We aim to analyze GRFs on the unit square $[0,1]^2$; thus, distances between observations are continuous on the interval $[0,1]$. Then, we attempt to draw comparisons by discretizing the argument $x$ as $(0.01,0.3,0.6,0.9)$ and obtaining results across a range of values of the smoothness parameter $\nu$ based on the four approaches. Figure~\ref{fig1} illustrates the comparison results. 

Figure~\ref{fig1} precisely indicates that the finite difference approach is an accurate estimation of $\frac{\partial {\cal K}_\nu(x)}{\partial \nu}$ because the integral form and the exact integer form precisely overlap with the finite difference approach for all $x$. The asymptotic approximation converges to the finite difference approach reasonably rapidly. An additional observation is that asymptotic approximations converge slower when the argument $x$ increases to the actual values. As presented in Figure~\ref{fig1}, the asymptotic approximation curves overlap with the finite difference and integral form curves at increasingly large values of $\nu$. 

\begin{algorithm}

\caption{Fisher Information Matrix Algorithm }\label{alg:sov}
\begin{algorithmic}
\Require Set of locations (\(\bs\)) and Matérn parameter vector (\(\boldsymbol{\theta}_m\))
\Ensure Fisher information matrix \(\bI(\btheta_m)\)
\State Generate the covariance matrix \(\boldsymbol{\Sigma}(\btheta_m)\)    \Comment{using ${\cal M}_1$, ${\cal M}_2$, or ${\cal M}_3$} 
\State POTRF ($\bSigma(\btheta_m)$)     \Comment{Cholesky factorization --- $\bL\bL^\top=\bSigma(\btheta_m)$}

\For{\texttt{$i = 1 : len(\btheta_m)$}}
\State Generate the derivative covariance matrix \(\boldsymbol{\Sigma}_i\)   \Comment{using ${\cal M}_1$, ${\cal M}_2$, or ${\cal M}_3$} 
\State \(\boldsymbol{\Sigma}_i\) $\gets$ TRSM (\(\textbf{L}^{-1}\), \(\boldsymbol{\Sigma}_i\))  \Comment{Triangular solve} 
\State \(\boldsymbol{\Sigma}_i\) $\gets$ TRSM (\(\textbf{L}^{-\top}\), \(\boldsymbol{\Sigma}_i\))  \Comment{Triangular solve} 
\For{\texttt{$j = i : len(\btheta_m)$}}
\State Generate the derivative covariance matrix \(\boldsymbol{\Sigma}_j\)   \Comment{using ${\cal M}_1$, ${\cal M}_2$, or ${\cal M}_3$} 
\State \(\boldsymbol{\Sigma}_j\) $\gets$ TRSM (\(\textbf{L}^{-1}\), \(\boldsymbol{\Sigma}_j\))   \Comment{Triangular solve} 
\State \(\boldsymbol{\Sigma}_j\) $\gets$ TRSM (\(\textbf{L}^{-\top}\), \(\boldsymbol{\Sigma}_j\))  \Comment{Triangular solve} 
\State \(\boldsymbol{\Sigma}_{temp}\) $\gets$ GEMM (\(\boldsymbol{\Sigma}_i\), \(\boldsymbol{\Sigma}_j\))  \Comment{Matrix multiplication}

\State \(\bI_{ij}(\btheta_m)\) and \(\bI_{ji}(\btheta_m)\) $\gets$ $\frac{1}{2}$ TRACE (\(\boldsymbol{\Sigma}_{temp}\))

\EndFor
\EndFor
\end{algorithmic}
\label{alg:fisher}
\end{algorithm}

Algorithm~\ref{alg:fisher} illustrates in detail the task-based parallel implementation of the Fisher information matrix. We rely on ${\cal M}_1$ to establish a basic algorithm architecture for the other parameterizations. The algorithm has two inputs, the set of locations (\(\bs\)) and the Matérn parameter vector (\(\boldsymbol{\theta}_m\)). The algorithm begins with generating the covariance matrix $\bSigma(\btheta_{1})$ using ${\cal M}_1$ in parallel and performs a Cholesky factorization based on the tile-based matrices to obtain $\bSigma(\btheta_{1})^{-1}$. To estimate each Fisher matrix element, the parallel generation function fills in the entries for $\bSigma_i$ or $\bSigma_j$  with Equations~(\ref{e6}), (\ref{e7}), and (\ref{e8}) in the Supplementary Material. Next, the algorithm computes the full Fisher information matrix using a set of parallel linear algebra operations as shown by equation~(\ref{e5}). We implemented two versions of Algorithm~\ref{alg:fisher} for the other two parameterizations ${\cal M}_2$ and ${\cal M}_3$, together with their first-order derivative functions (\ref{e9})--(\ref{e14}) to the existing implementation to compute their associated Fisher information matrices.
The MLEs are obtained using the {\redcolor invariance} property and the link functions provided in Table \ref{table1}. \cite{myung2005information} illustrated that transformations of the Fisher information matrix are linear operations based on the Jacobian matrix; thus,
\begin{equation}
\bI(\boldsymbol{\theta}_m)=\textbf{J}_{k \rightarrow m}^\top \bI(\boldsymbol{\theta}_k)\textbf{J}_{k\rightarrow m},
\label{e17}
\end{equation}
where $k,m=1,2,3$ and \((\textbf{J}_{k \rightarrow m})_{i,j}=\frac{\partial (\boldsymbol{\theta}_m)_i}{\partial (\boldsymbol{\theta}_k)_j}\) represents the Jacobian matrix for $i,j=1,2,3,4$. Then, we compute all the Jacobian matrices that enable transformations between ${\cal M}_1$, ${\cal M}_2$, and ${\cal M}_3$ (see the Supplementary Material). The reverse transformations can be established by inverting the relationship presented in (\ref{e17}) and the computed Jacobian matrices.

The tile-based Fisher information matrix generation algorithm is adaptive to the parameter space. Thus, $\bI(\btheta)$ will be a $3\times 3$ or $4\times 4$ matrix, depending on the inclusion or exclusion of the nugget effect.

\subsection{Mean Squared Prediction Error} \label{MSPE}  
The given statistical model can predict missing measurements at new spatial locations through kriging using the posterior distribution of the missing observations based on the following:
$$\begin{pmatrix}
\bZ_1\\
\bZ_2
\end{pmatrix}\sim{\cal N}_{m+n}(\0,\bSigma(\btheta)), \text{where } \bSigma(\btheta)=\begin{pmatrix}
\bSigma_{11} & \bSigma_{12} \\
\bSigma_{21} & \bSigma_{22}
\end{pmatrix}.$$


The posterior mean, $\text{E}(\bZ_1|\bZ_2=\bz_2)=\bSigma_{12}\bSigma_{22}^{-1}\bz_2$, can be used for the predictions of $\bZ_1$, where $\bZ_1$ and $\bZ_2$ represent the missing and observed values, respectively. The MSPE is:\\
\begin{equation}
    \text{MSPE}=\frac{1}{m}\sum_{i=1}^{m}\{\hat{Z}(\bs_{0i})-Z( \bs_{0i})\}^2,
    \label{e18}
\end{equation}
where \(\hat{Z}(\bs_{0i})\) is the predicted value at location \( \bs_{0i}\) and \(Z(\bs_{0i})\) is the observed value at location \( \bs_{0i}\). The MSPE directly accounts for the degree of deviation between the predicted and true values, a frequently used and standard cross-validation-based criterion to measure prediction errors and, therefore, quantify the prediction performance.

\subsection{Data Modeling Accuracy}
The concepts of loss of efficiency (LOE) and misspecification of the mean square error (MOM) were introduced by \cite{stein1999interpolation} to evaluate the impact of approximating the covariance on prediction error. LOE measures spatial prediction efficiency, while MOM quantifies the difference in MSPEs between a single prediction value calculated using the exact and approximated covariance values. \cite{hong2021efficiency} extended these concepts to mean LOE (MLOE) and mean MOM (MMOM), which average LOE and MOM, respectively, across all locations with missing values. MLOE and MMOM provide a comprehensive assessment of the overall prediction accuracy and modeling error associated with the approximated covariance, and they are sensitive to inaccurate estimates and misspecified models. MLOE and MMOM are explicitly defined as $\text{MLOE}=\frac{1}{m}\sum_{i=1}^m \text{LOE}(\bs_{0i})$ and $\text{MMOM}=\frac{1}{m}\sum_{i=1}^m \text{MOM}(\bs_{0i})$, where $\text{LOE}(\bs_{0i})=\text{E}_t\{e_a^2(\bs_{0i})\}/\text{E}_t\{e_t^2(\bs_{0i})\}-1$ and $\text{MOM}(\bs_{0i})=\text{E}_a\{e_a^2(\bs_{0i})\}/{\text{E}_t\{e_a^2(\bs_{0i})}\}-1$. Here, $\bs_{0i}$ represents the locations of the missing values, $e_a(\bs_{0})$ denotes the residuals computed using an approximated covariance, and $e_t(\bs_{0})$ represents the residuals computed with the true covariance. MLOE is non-negative since $\text{E}_t\{e_t^2(\bs_{0i})\} \leq \text{E}_t\{e_a^2(\bs_{0i})\}$ \citep{stein1999interpolation}. MMOM is zero if parameter estimation is perfect; otherwise, it takes positive or negative values \citep{hong2021efficiency}. Finally, $\text{E}_a\{e_a^2(\bs_0)\}$ is easy to compute, but evaluating $\text{E}_t\{e_a^2(\bs_0)\}$ is challenging.

Two methods were proposed by \cite{stein1999interpolation} for evaluating $\text{E}_t\{e_a^2(\bs_{0})\}$. The first method involves estimating it by the conditional distribution of the estimate given the observations, which can be approximated as $\text{E}_t\{e_a^2(\bs_0)\} \approx \text{E}_{\btheta} \left\{e_{\hat{\btheta}(\bZ)}^2(\bs_0)|\bZ=\bz \right\}= \text{E}_{\btheta}\{e_{\btheta}^2 (\bs_0)\}+\left\{e_{\hat{\btheta}(\bZ)}(\bs_0) -e_{\btheta}(\bs_0)\right\}^2$. Here, $\hat{\btheta}(\bZ)$ refers to the estimate, and $\btheta$ represents the true parameter value. The second method involves a resampling mechanism that repeatedly uses subsamples of $\bZ$ to estimate $\btheta$, computes the prediction error for each subsample, averages these prediction errors, and adds them to $\text{E}_t\{e_t^2(\bs_0)\}$ to estimate $\text{E}_t\{e_a^2(\bs_0)\}$.

We utilized the covariance-based plug-in method proposed by \cite{hong2021efficiency}, which is based on Stein's approaches, to calculate both the true mean square prediction error (MSPE), $\text{E}_t\{e_a^2(\bs_{0})\}$, and the approximated MSPE, $\text{E}_a\{e_a^2(\bs_{0})\}$. To compute these, we employed the covariance matrix $\textbf{K}_{\btheta}=\text{Cov}_{\btheta}\{\bZ,\bZ^\top\}$, the covariance vector $\textbf{k}_{\btheta}=\text{Cov}_{\btheta}\{\bZ,\text{Z}(\bs_0)\}$, and the scalar variance $\text{k}_{0,\btheta}=\text{Var}_{\btheta}\{\text{Z}(\bs_0)\}$. Specifically, we estimated $\text{E}_t\{e_a^2(\bs_0)\}$ as $\text{E}_{\btheta}\left\{e_{\hat{\btheta}}^2(\bs_0)\right\}\approx\text{k}_{0,\btheta}-2\textbf{k}_{\btheta}^\top\textbf{K}_{\hat{\btheta}}^{-1}\textbf{k}_{\hat{\btheta}}+\textbf{k}_{\hat{\btheta}}^\top\textbf{K}_{\hat{\btheta}}^{-1}\textbf{K}_{\btheta}\textbf{K}_{\hat{\btheta}}^{-1}\textbf{k}_{\hat{\btheta}}$, and estimated $\text{E}_a\{e_a^2(\bs_0)\}$ as $\text{k}_{0,\hat{\btheta}}-\textbf{k}_{\hat{\btheta}}^\top\textbf{K}_{\hat{\btheta}}^{-1}\textbf{k}_{\hat{\btheta}}$, where $\hat{\btheta}$ is the estimated parameter value. This matrix-based method allows for parallel computation, making it well-suited for large-scale datasets. We implemented this method in \textit{ExaGeoStat}, which automatically considers the implemented covariance functions (${\cal M}_1$, ${\cal M}_2$, and ${\cal M}_3$) in the synthetic data generator and maximum likelihood estimators (MLEs) to construct true and approximated covariance matrices. The resulting MLOE and MMOM values are computed using the \textit{mloe-mmom} command.

\section{Experimental Results}
In this section, we conduct a thorough evaluation of the Matérn covariance function's three parameterizations: \({\cal M}_1\), \({\cal M}_2 \), \({\cal M}_3\). Our objective is to assess the accuracy of parameter estimation for each variant, both under exact and TLR approximate representations of the covariance matrix. Additionally, we evaluate the predictive performance of each covariance function by employing the MSPE, MLOE, and MMOM criteria. We also investigate the uncertainty quantification capabilities of each parameterization using the Fisher information matrix. Finally, we present our observations based on several experiments.

\subsection{Experimental Testbed}
For the experimentation, we exploited the concept of weak, medium, and strong dependencies in GRF to discretize the field correlation strength to evaluate the three parametrizations of the Matérn covariance function. First, we specified the weak, medium, and strong dependence, using the effective range, which is the spatial lag for the correlation to drop to 5\% \citep{stein1999interpolation}.
The effective range accounts for varying decay rates in the Matérn covariance function. The covariance function is expected to {\redcolor decrease} at an increasingly slower rate as the correlation strengthens, addressing various scopes of dependence. Simulations in \cite{abdulah2018parallel} revealed that the dependence range of GRFs affects the parameter estimation even in large-scale problems.

We applied the concept of effective range to determine the parameters in ${\cal M}_1$, ${\cal M}_2$, and ${\cal M}_3$. To specify the parameters,  \(\sigma^2\) is set to 1, and $\nu$ is set as 0.5 and 1, respectively, to address rough and smooth fields. Then, we computed \(\beta\) by numerically solving the equations, ${\cal M}_1(h;\btheta_1)=0.05$ for $h=0.1,0.3,0.7$. With all of the parameters specified in ${\cal M}_1$, we applied the link functions provided in Table~\ref{table1} to specify parameters in ${\cal M}_2$ and ${\cal M}_3$. The nugget effect $\tau^2$ is set to 0.1. Table~\ref{table5} illustrates the parameters that describe identical Matérn GRFs parameterized in ${\cal M}_1$, ${\cal M}_2$, and ${\cal M}_3$. 
\begin{table}[b!]
    \centering
    \caption{Parameters specification in ${\cal M}_1$, ${\cal M}_2$, and ${\cal M}_3$ for different correlation strengths for $\nu=0.5$ and $\nu=1$, and all with $\sigma^2=1$.}
    \begin{tabular}{|l|c|c|c|c||c|c|c|c|}
    \hline
Dependence & \multicolumn{4}{c||}{$\nu=0.5$}                     & \multicolumn{4}{c|}{$\nu=1$}\\

             & \(\beta\) & \(\rho\)  &\(\phi\) &\(\alpha\) & \(\beta\) & \(\rho\)  &\(\phi\) &\(\alpha\) \\ \hline
       Weak  & 0.0330 & 0.0467     & 9.6458 &  30.3030 & 0.0250 & 0.0500     & 800.0000 &  40.0000\\  \hline
       Medium & 0.1000 & 0.1414&  3.1831     & 10.0000 & 0.0750 & 0.1500     & 88.8889 &  13.3333\\  \hline
       Strong & 0.2340  & 0.3309 &  1.3603   & 4.2735 & 0.1750 & 0.3500     & 16.3265 &  5.7143\\
       \hline
    \end{tabular}
    \label{table5}
\end{table}
Table \ref{table5} reveals that the range and scale parameters \((\beta,\rho)\) and \(\alpha\) pertain to the correlation strength and, therefore, the decay rate of the covariance function. Moreover, \(\beta\) and \(\rho\) are direct measures of the dependence range. Increases in $(\beta, \rho)$ can independently strengthen field correlation. In contrast, \(\alpha\), on the other hand, decreases as correlation strengthens. In addition, $\alpha$ cannot autonomously govern correlation strength. Decreasing values of $\phi$ are also imperative to increase the correlation strength. 

\subsection{Efficiency Assessment}
This section assesses the efficiency of the ${\cal M}_1$, ${\cal M}_2$, and ${\cal M}_3$ covariance function variants concerning MLEs, the number of iterations to convergence, MLOE and MMOM criteria, and lastly MSPEs. We employed the synthetic data generator to simulate 300 replicates of GRFs of size 1600 in weak, medium, and strong fields with ${\cal M}_1$, ${\cal M}_2$, and ${\cal M}_3$ to explore these measures. The synthetic data generator uses $[\{i-0.5\times \text{unif}(-0.4,0.4)_{{\redcolor i}}\}/\sqrt{n},\{j-0.5\times \text{unif}(-0.4,0.4)_{\redcolor i}\}/\sqrt{n}]$ with $i,j=1,\dots,\sqrt{n}$ for the random locations, where $\text{unif}$ denotes the uniform distribution and $n$ is the sample size.  Efficiency assessments are conducted on irregular grids because the behaviors of the metrics, such as MSPE, MLOE, MMOM, and MLEs, are similar on regular grids with dense observations. We used a 40-core Intel Cascade Lake machine with four V100 GPUs.
\subsubsection{Parameter Estimation Accuracy \label{acc}}
\begin{figure}[t!]
\begin{subfigure}{0.33\linewidth}
  \centering
  \includegraphics[width=1\textwidth,]{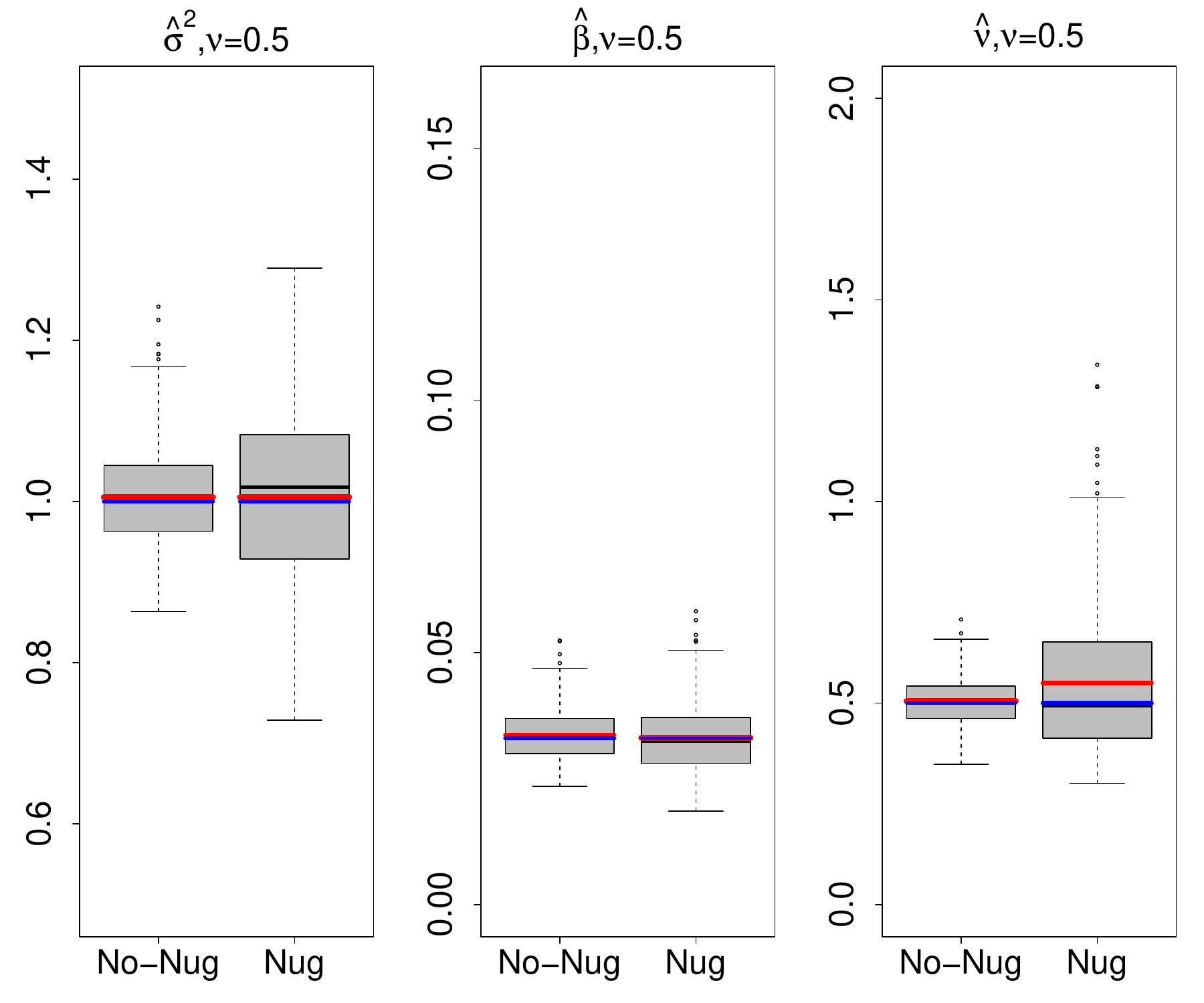}
  \caption{$\hat{\btheta}_1,{\cal M}_1$, weak correlation }
  \label{fig2(a)}
  \vspace{5mm}
\end{subfigure}
\begin{subfigure}{0.33\linewidth}
  \centering
  \includegraphics[width=1\textwidth,]{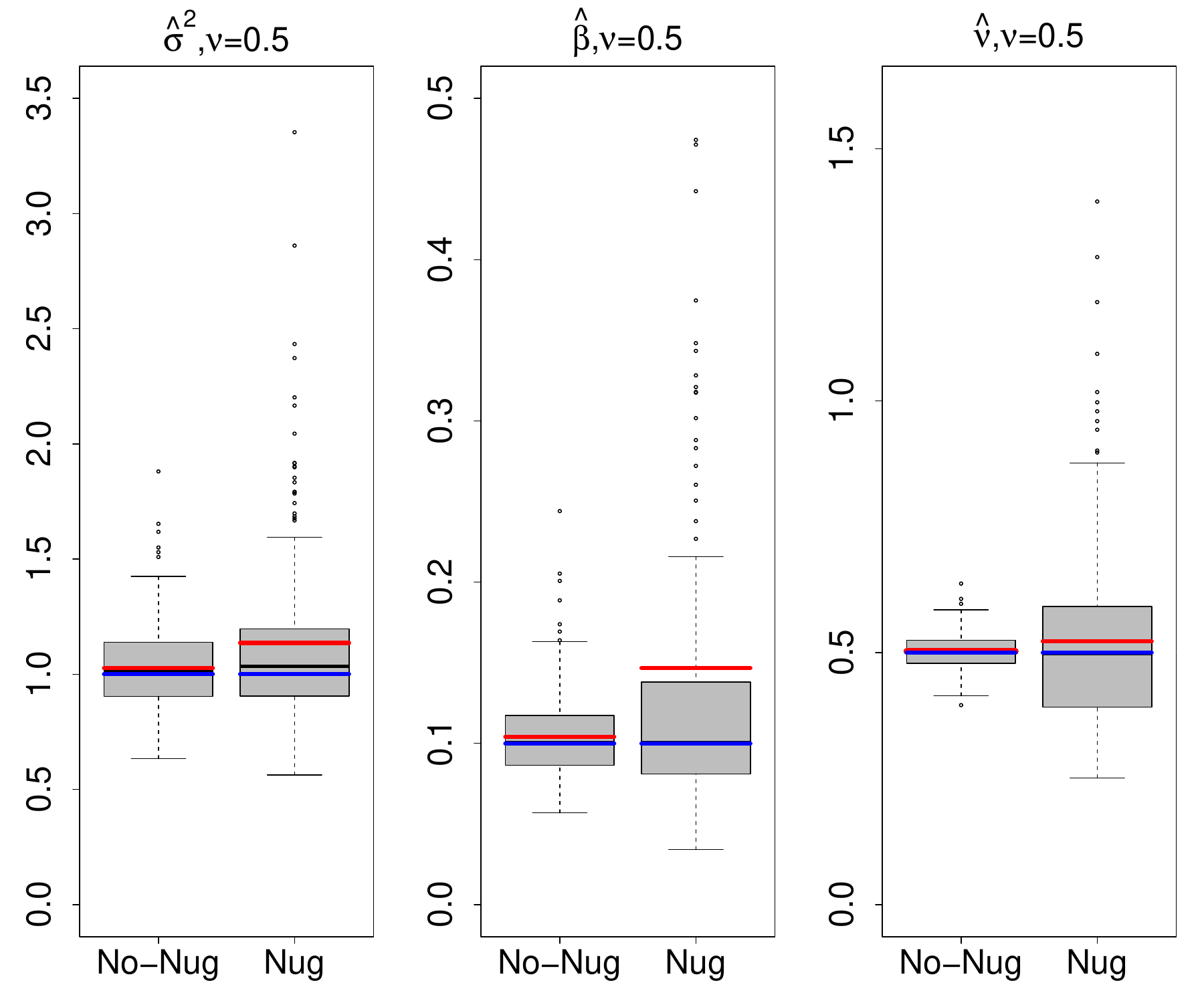}
  \caption{$\hat{\btheta}_1,{\cal M}_1$, medium correlation }
  \label{fig2(b)}
  \vspace{5mm}
\end{subfigure}
\begin{subfigure}{0.33\linewidth}
  \centering
  \includegraphics[width=1\textwidth,]{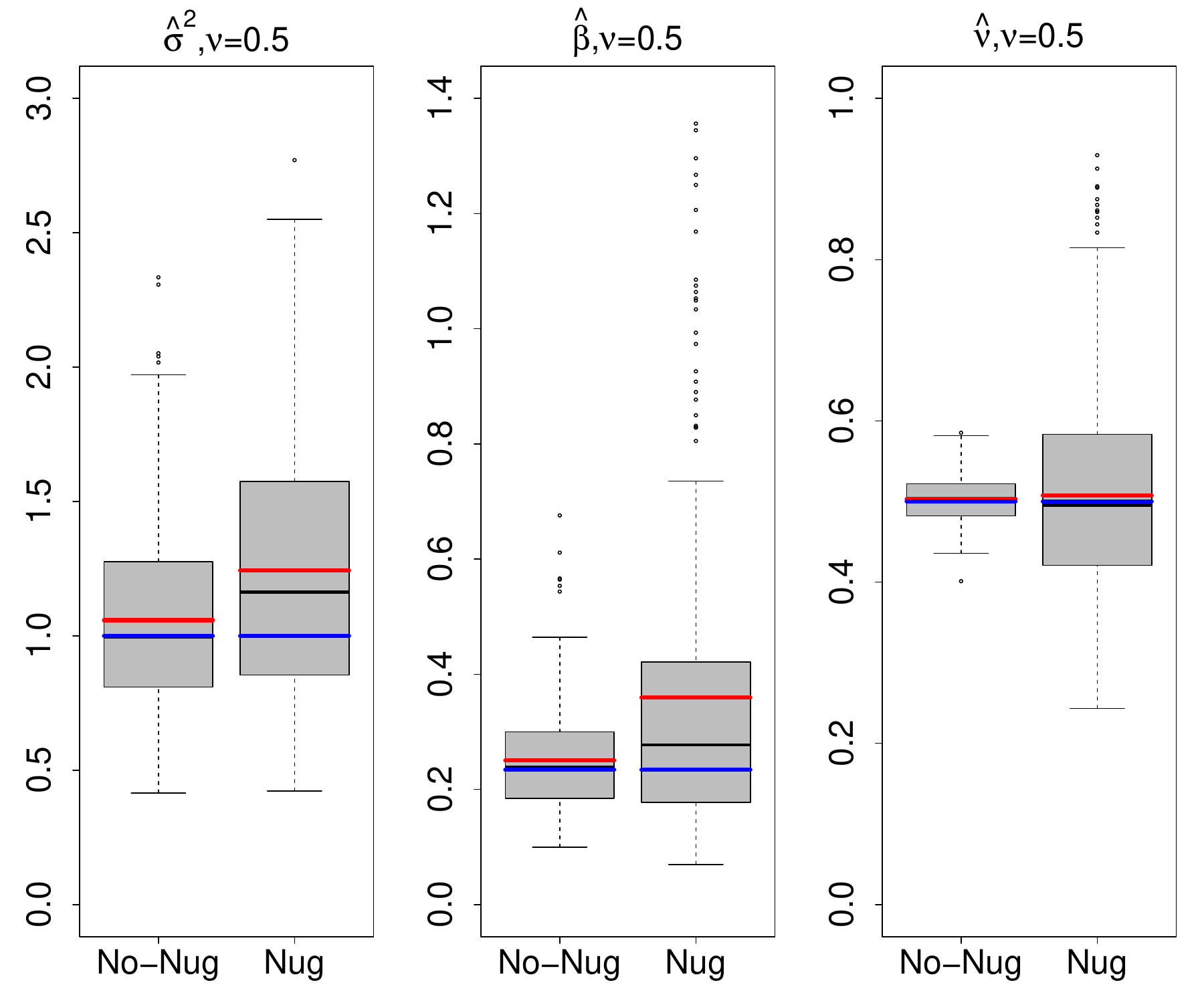}
  \caption{$\hat{\btheta}_1,{\cal M}_1$, strong correlation }
  \label{fig2(c)}
  \vspace{5mm}
\end{subfigure}
\begin{subfigure}{0.33\linewidth}
  \centering
  \includegraphics[width=1\textwidth,]{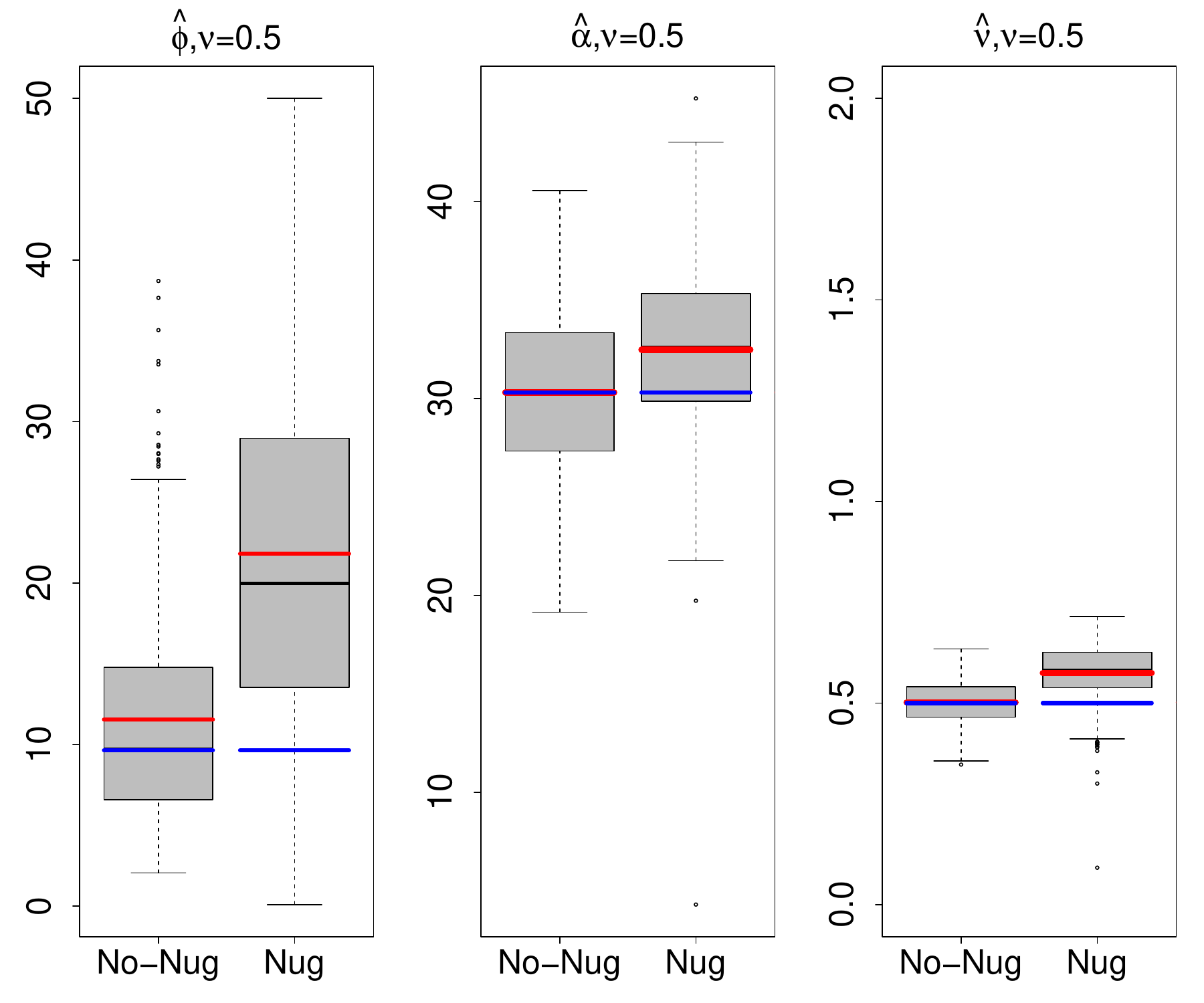}
  \caption{$\hat{\btheta}_2,{\cal M}_2$, weak correlation }
  \label{fig2(d)}
  \vspace{5mm}
\end{subfigure}
\begin{subfigure}{0.33\linewidth}
  \centering
  \includegraphics[width=1\textwidth,]{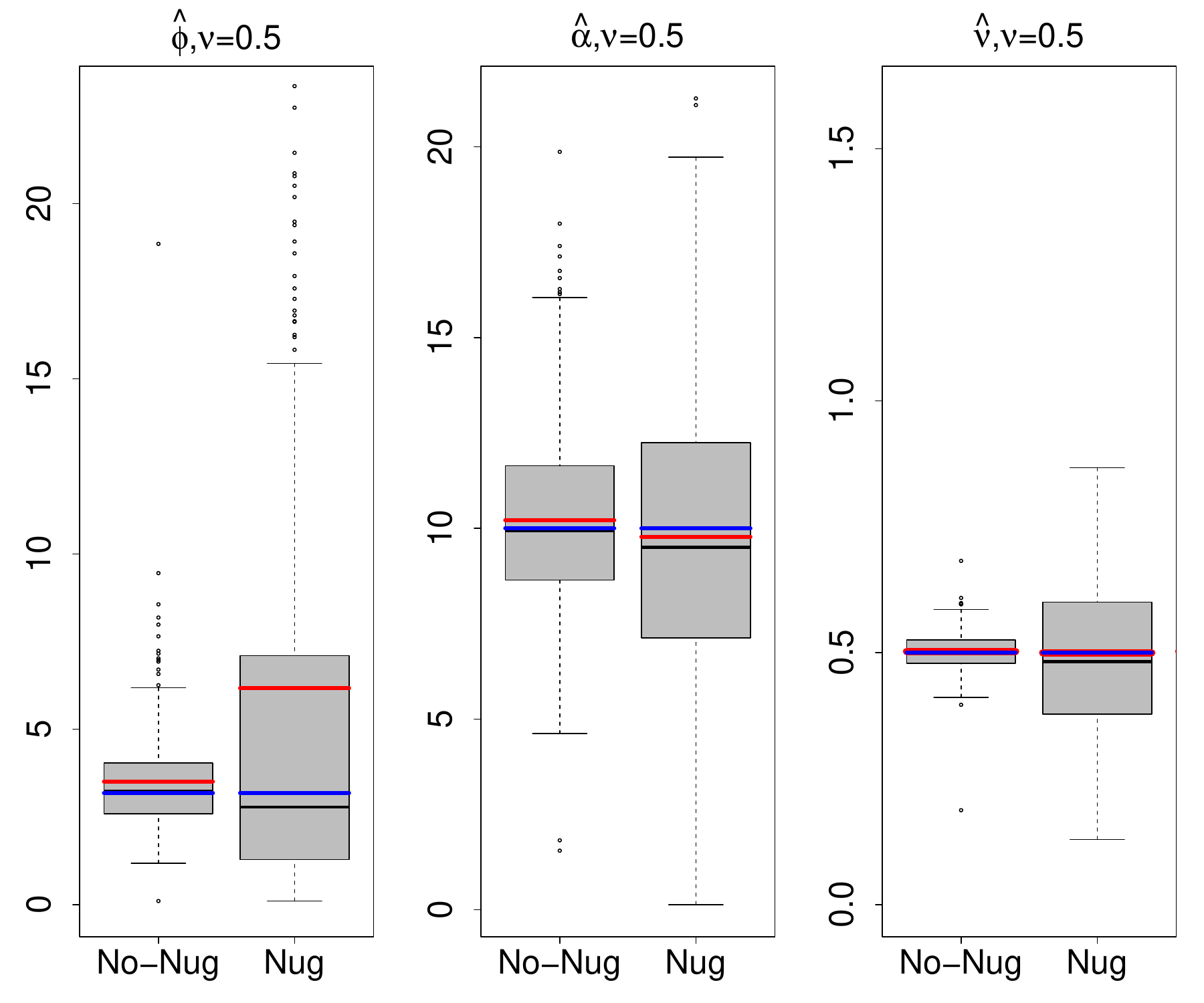}
  \caption{$\hat{\btheta}_2,{\cal M}_2$, medium correlation }
  \label{fig2(e)}
  \vspace{5mm}
\end{subfigure}
\begin{subfigure}{0.33\linewidth}
  \centering
  \includegraphics[width=1\textwidth,]{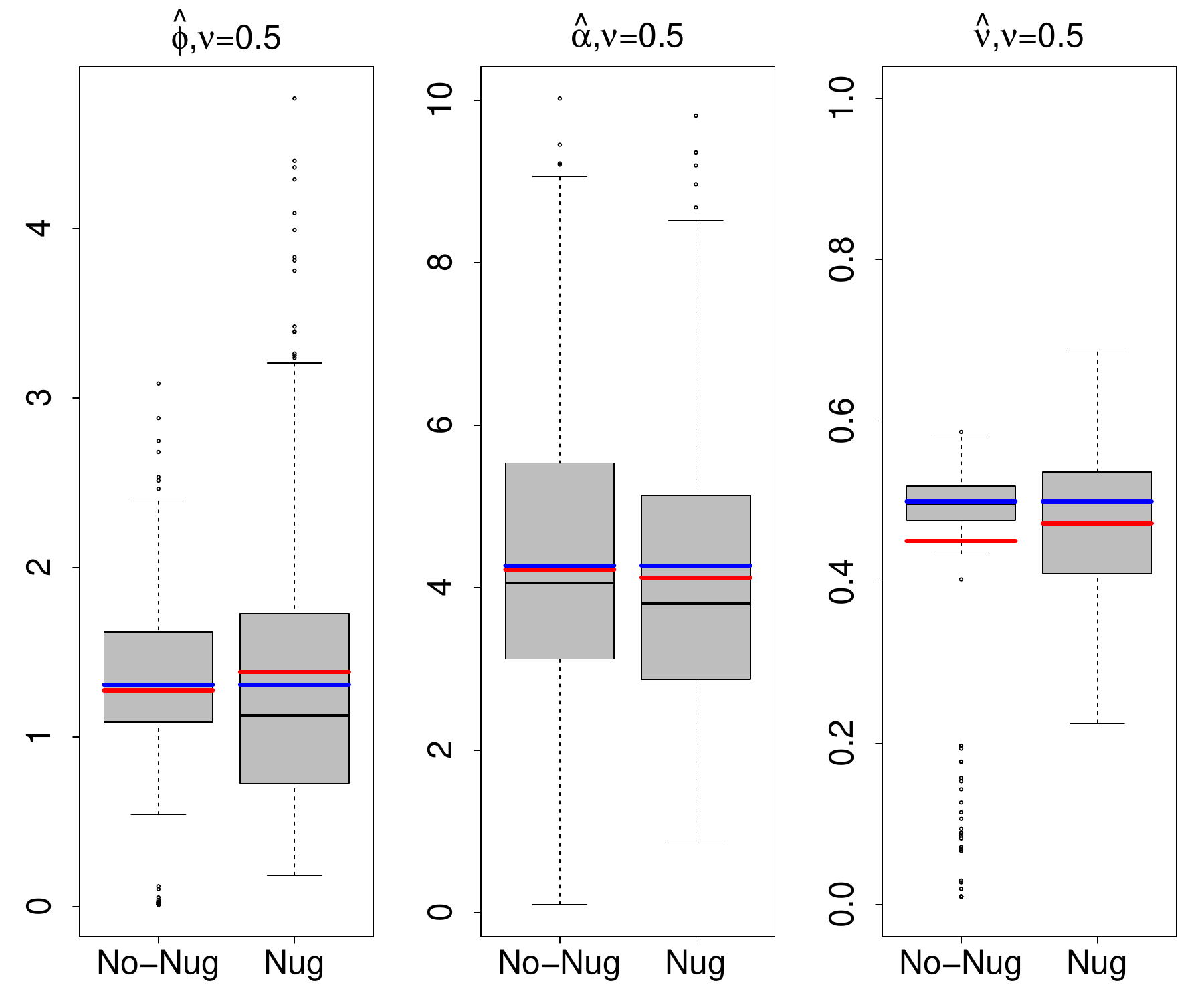}
  \caption{$\hat{\btheta}_2,{\cal M}_2$, strong correlation }
  \label{fig2(f)}
  \vspace{5mm}
\end{subfigure}
\begin{subfigure}{0.33\linewidth}
  \centering
  \includegraphics[width=1\textwidth,]{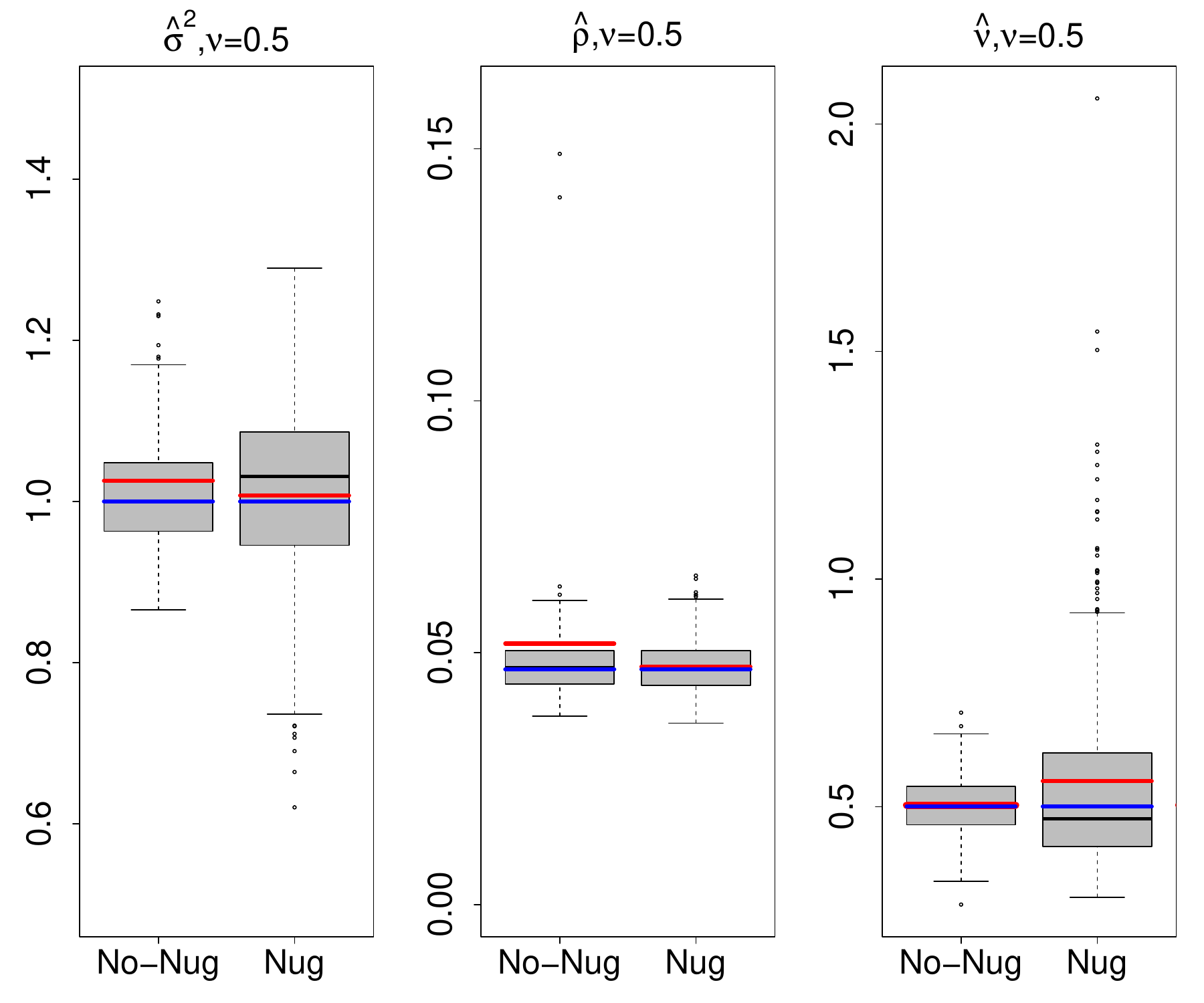}
  \caption{$\hat{\btheta}_3,{\cal M}_3$, weak correlation }
  \label{fig2(g)}
\end{subfigure}
\begin{subfigure}{0.33\linewidth}
  \centering
  \includegraphics[width=1\textwidth,]{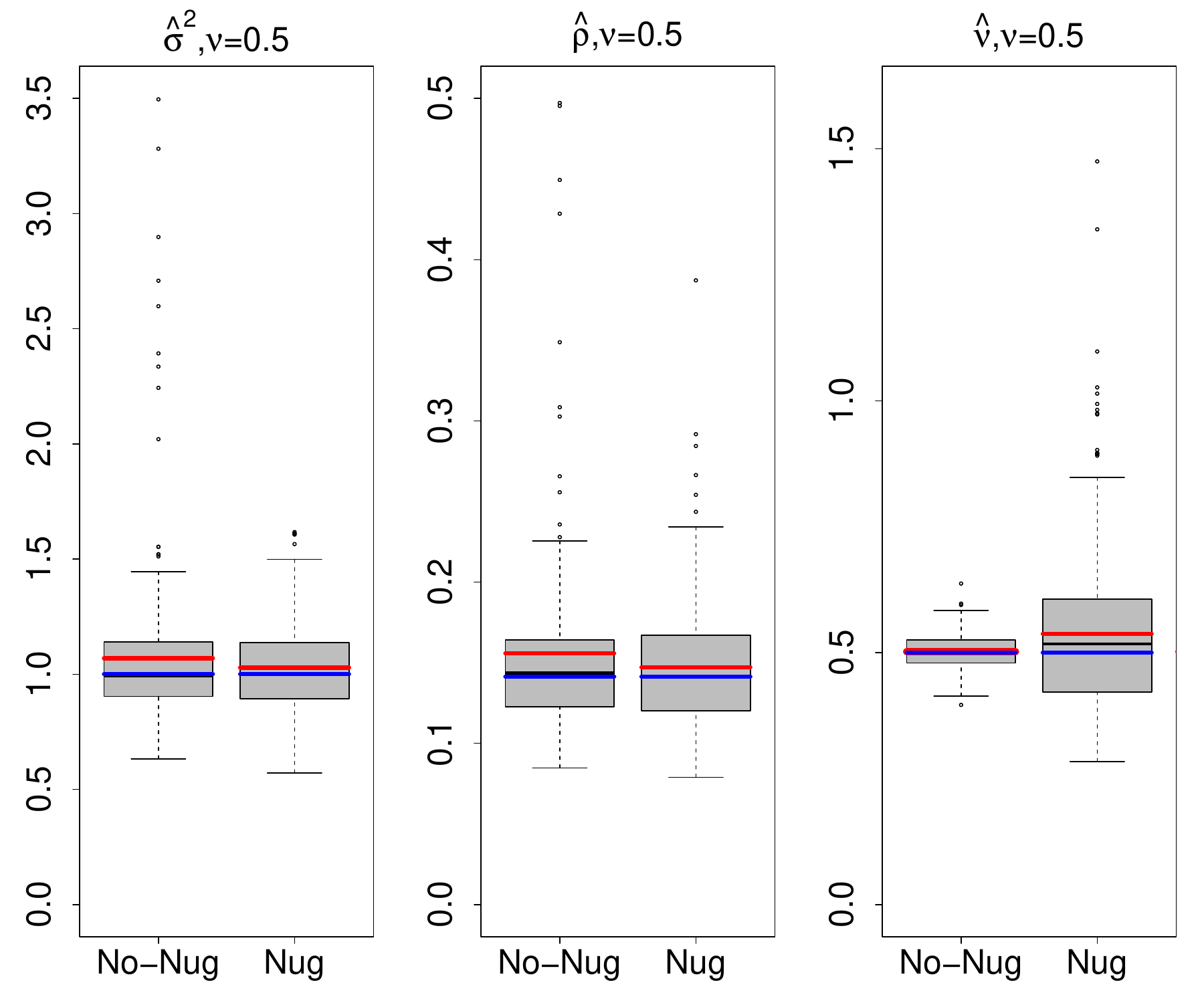}
  \caption{{\redcolor $\hat{\btheta}_3,{\cal M}_3$, medium correlation }}
  \label{fig2(h)}
\end{subfigure}
\begin{subfigure}{0.33\linewidth}
  \centering
  \includegraphics[width=1\textwidth,]{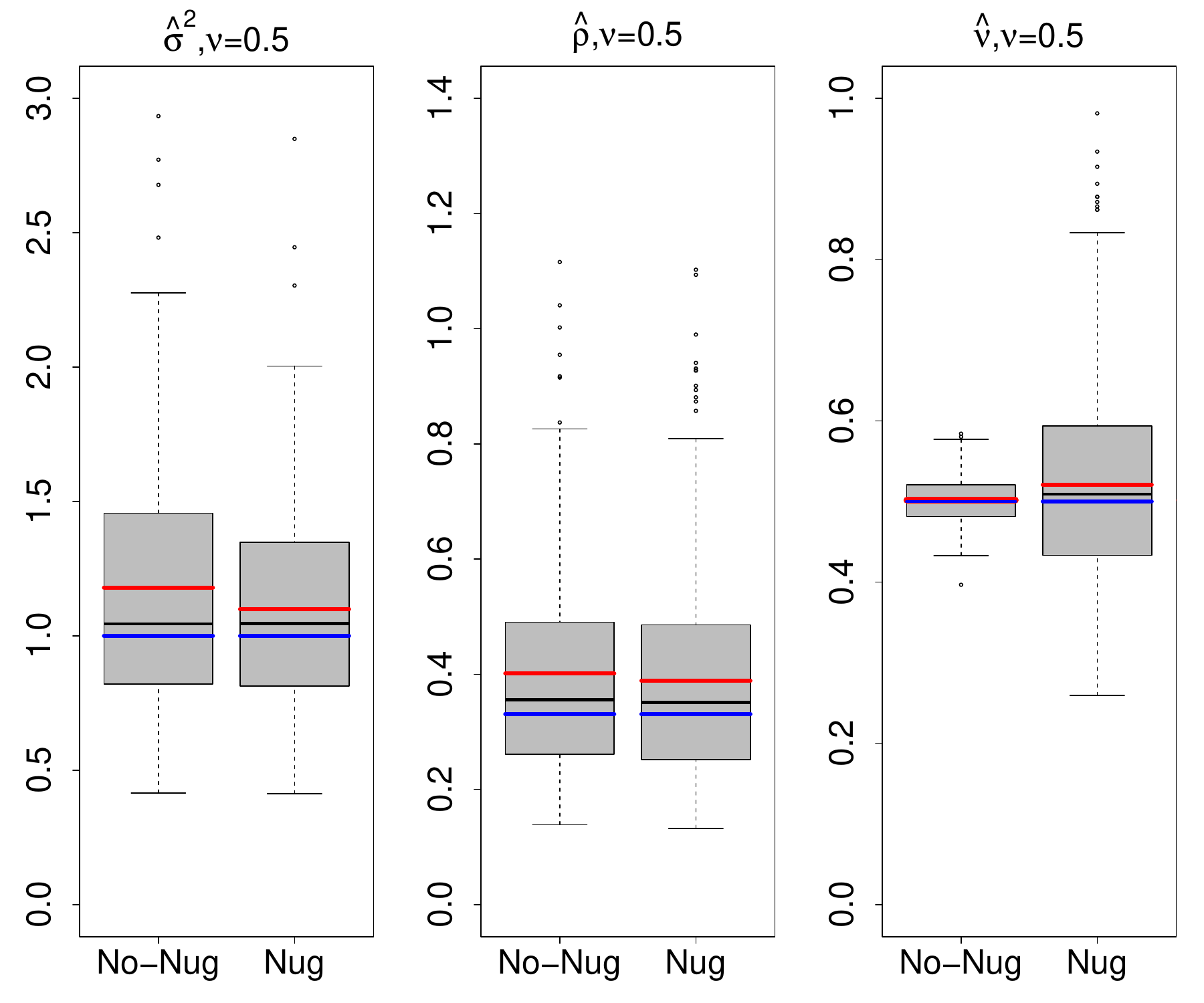}
  \caption{$\hat{\btheta}_3,{\cal M}_3$, strong correlation }
  \label{fig2(i)}
\end{subfigure}
\caption{ Boxplots of MLEs on the exponential Gaussian fields ($\nu=0.5$) using 300 replicates. Red lines represent the sample means; blue lines denote the true values.  No-Nug denotes non-nugget effect models and Nug denotes nugget effect models.}
\label{fig2}
\end{figure}

 Figure \ref{fig2} presents the empirical description of the probabilistic distributions for the MLEs. We set the optimization tolerance to $10^{-5}$ and the tile size ($NB$) to be $50$ for parallel execution. Tile size (NB) is tuned for the execution time performance. We adjusted the parameter search space by scaling the default optimization boundary in {\em ExaGeoStat}, $(0.01, 5)$, by a constant factor to cover the MLEs adequately. For instance, we used the default boundaries if the true parameter was less than 2. For true parameter values between 2 and 20, we scaled the boundaries by 10. For the values within $(20,50)$ and $(50,500)$, we scaled the boundaries by 20 and 200, respectively. Lastly, for values greater than 500, we scaled the boundaries by 1,000.  The optimization boundaries are selected to properly contain the true parameter values and their respective MLEs. Because \textbf{BOBYQA} is a local optimization algorithm, each step will seek to find a more desirable position in the search space. As a result, \textbf{BOBYQA} will not end up on the boundaries if the MLEs are well-contained in the search space. {\redcolor Notice that in practice, we have no prior information on the true parameter values, but \textbf{BOBYQA} will always get stuck on the upper bounds (lower bounds can always be set to small values, like 0.01) if the MLEs are jointly or marginally larger than the upper bounds. Therefore, we recommend redoing the optimization with increased upper bounds under such circumstances until the MLEs are well-contained in the search space.}
 In addition, we computed the difference ratios (DRs), 
 $
{|\text{Mean}-\text{True Value}|}/{\text{True Value}},
 $
 and the approximated effective range (AER), ${\cal M}_i(h;\hat{\btheta}_i)$ for $h=0.1,0.3,0.7$ and $i=1,2,3$.
 In the Supplementary Material, parameter estimates on the Whittle Gaussian fields are plotted in Figure~\ref{fig9}. The DRs calibrate differing parameter scales for fair comparisons and refer to the performance of point estimations. The AER measures the overall accuracy of the mean estimates.

 For the non-nugget effect models, Figure~\ref{fig2} (No-Nug) and Table~\ref{table7} (first and third horizontal blocks separated by double horizontal lines in the Supplementary Material) demonstrate that estimates of ${\cal M}_3$ are the least accurate compared with their counterparts in ${\cal M}_1$ and ${\cal M}_2$. In particular, $\hat{\sigma}^2$ in ${\cal M}_3$ appears more biased than $\hat{\sigma}^2$ in ${\cal M}_1$. Also $\hat{\rho}$ is more variable and biased than $\hat{\beta}$. Although parameters in ${\cal M}_2$ are on different scales, Table~\ref{table7} illustrates that $(\hat{\sigma}^2,\hat{\rho})$ are more biased than $(\hat{\phi},\hat{\alpha})$ . The biases of $(\hat{\sigma}^2,\hat{\rho})$ are due to flatter shapes of the associated log-likelihood functions near the maxima (leading the optimization algorithm to stop at non-optimal values) as shown in Figures \ref{c} and \ref{contour} in the Supplementary Material, which a more stringent convergence condition can resolve (i.e., $10^{-9}$) at the expense of modeling speed. Table~\ref{AER} (vertical block on the left separated by double vertical lines) also reveals that ${\cal M}_3$ provides undesirable AERs compared with ${\cal M}_1$ and ${\cal M}_2$, reflecting its less ideal overall estimation efficiency. ${\cal M}_2$ provides more refined AERs than ${\cal M}_1$ but performs poorly on point estimations of $(\hat{\phi},\hat{\nu})$ in many cases. Furthermore, $(\hat{\phi},\hat{\alpha})$ exhibits considerably higher volatility than $(\hat{\sigma}^2,\hat{\beta},\hat{\rho})$, which can be observed through their large interquartile ranges (IQRs) shown in Figures~\ref{fig2(d)}-\ref{fig2(f)} and \ref{fig9(d)}-\ref{fig9(f)}.

 For the nugget effect models, the nugget effect significantly influences the estimates, but the effect varies in terms of ${\cal M}_1$, ${\cal M}_2$, and ${\cal M}_3$. From the experiments, the influence of the nugget effect is more crucial on ${\cal M}_1$ and ${\cal M}_2$. Figure~\ref{fig2} (Nug) and Tables~\ref{table7} (second and fourth horizontal blocks separated by double horizontal lines in the Supplementary Material), \ref{AER} (vertical block on the right separated by double vertical lines in the Supplementary Material) demonstrate that the nugget effect models of ${\cal M}_1$ and ${\cal M}_2$ both suffer from biased point estimates for $(\hat{\sigma}^2,\hat{\beta},\hat{\phi},\hat{\alpha})$, and undesirable AERs. In addition, $(\hat{\sigma}^2,\hat{\beta},\hat{\phi})$ also becomes more variable in the presence of the nugget effect for ${\cal M}_1$ and ${\cal M}_2$. In contrast, parameter estimates in ${\cal M}_3$ maintain more robustness. In the nugget effect models, $\hat{\sigma}^2$ is less biased and less variable. $\hat{\rho}$ is less biased and about equally variable under the nugget effect. Impact of the nugget effect on $\hat{\nu}$ is similar across all parameterizations; $\hat{\nu}$ in all parameterizations tends to be more variable and biased under the nugget effect.

\begin{figure}[t!]
\begin{subfigure}{0.5\textwidth}
    \centering
    \includegraphics[width=0.8\textwidth,]{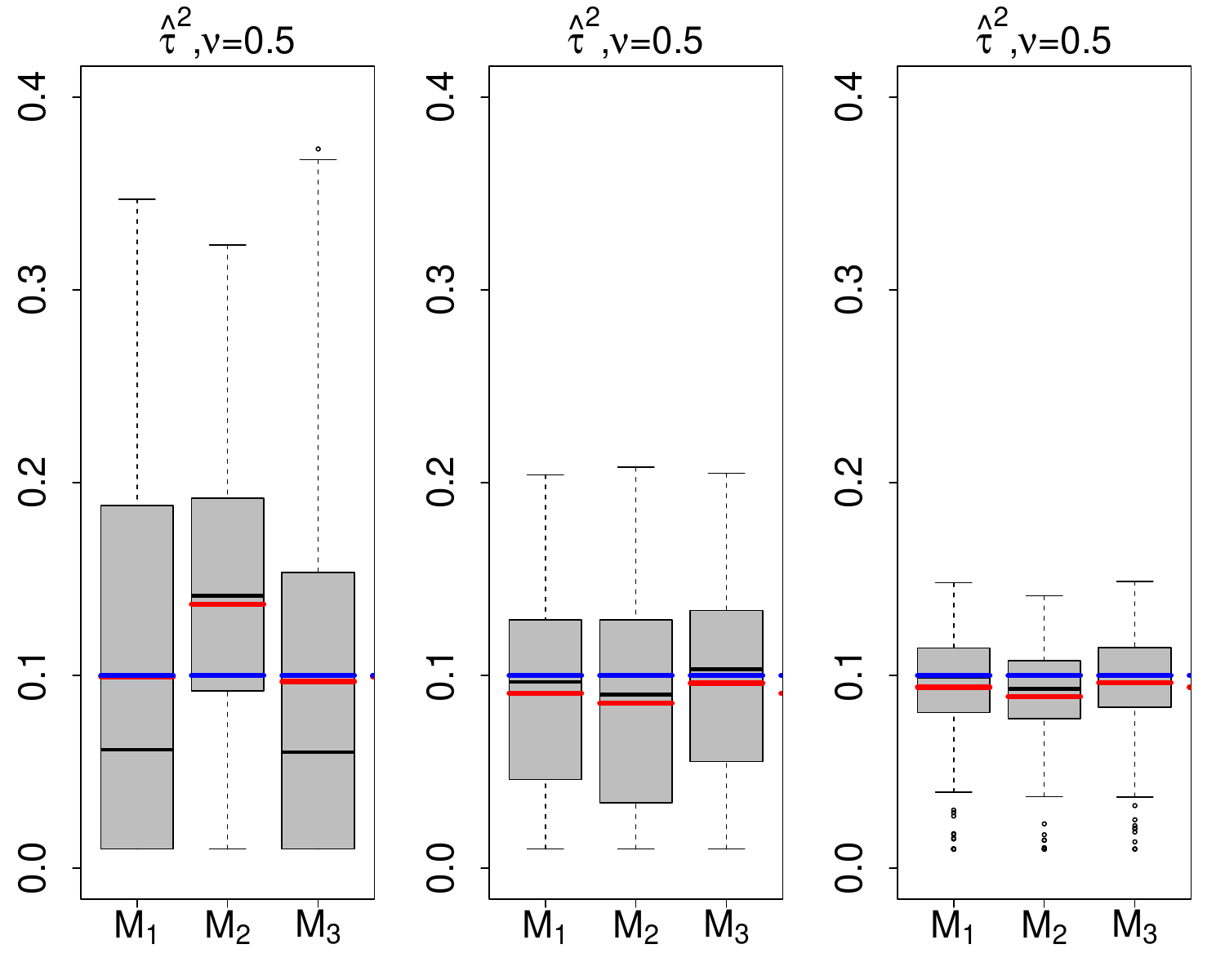}
    \caption*{\hspace{3mm}Weak\hspace{13mm}Medium\hspace{12mm}Strong}
    \label{tau_n1}
\end{subfigure}
\begin{subfigure}{0.5\textwidth}
    \centering
    \includegraphics[width=0.8\textwidth,]{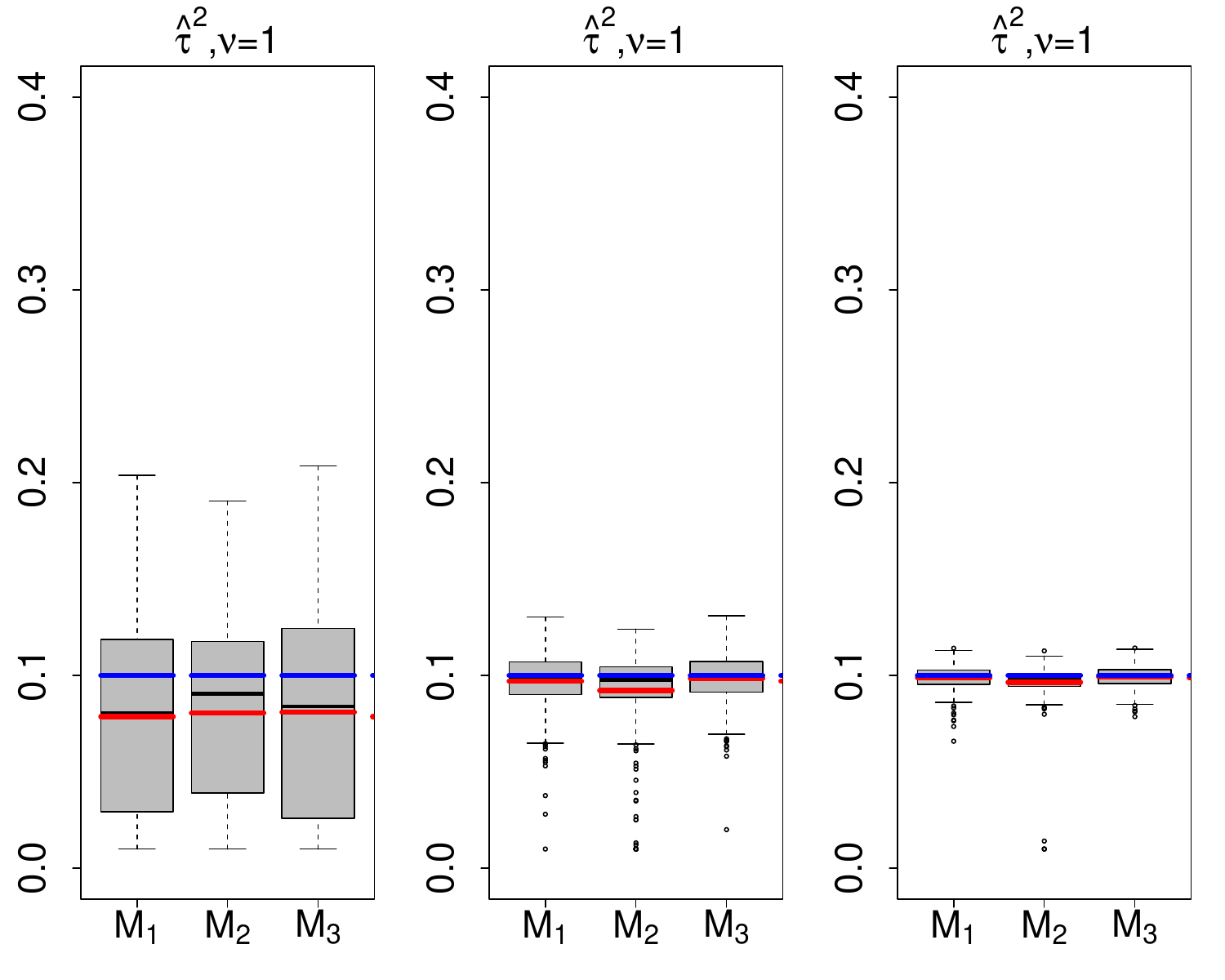}
    \caption*{\hspace{3mm}Weak\hspace{13mm}Medium\hspace{12mm}Strong}
    \label{tau_n2}
\end{subfigure}
\caption{Boxplots of MLEs for the nugget effect $\tau^2$ on the exponential Gaussian fields ($\nu=0.5$) (left three subfigures) and Whittle Gaussian fields ($\nu=1$) (right three subfigures) using 300 replicates. Red lines represent the sample means; blue lines denote the true parameter values.}
\label{taukk}
\end{figure}

Figure~\ref{taukk} confirms the conclusion of the nugget effect proposed by \cite{stein1999interpolation}, stating that $\tau^2$ should be easier to estimate in a smooth process.
Figure~\ref{taukk} also indicates that the IQRs of $\tau^2$ decrease from an exponential Gaussian field to a Whittle Gaussian field. In addition, within the same field, the IQRs decline as the field correlation strengthens. The decrease in IQRs implies that $\tau^2$ has a lower uncertainty and, therefore, more information in a smooth process, agreeing with the decreasing $L^{\tau}/\tau^2$ along with increasing $\nu$ calculated on p. 186--187 in work by \cite{stein1999interpolation}. Moreover, the estimations of $\tau^2$ are less accurate in ${\cal M}_2$ compared with ${\cal M}_1$ and ${\cal M}_3$, which tends to have more significant deviations between the sample means and the true value.
Lastly, IQRs decrease for $(\hat{\phi},\hat{\alpha},\hat{\nu})$ but increase for $(\hat{\sigma}^2,\hat{\beta},\hat{\rho})$ as the field strengthens, indicating opposite behaviors of their uncertainties. Therefore, the estimation is more challenging for $(\hat{\sigma}^2,\hat{\beta},\hat{\rho})$ but easier for $(\hat{\phi},\hat{\alpha},\hat{\nu})$ when the observations are correlated. Such differing behaviors of the estimates are due to variations in the data efficiency of their estimations, which we detail in Section~\ref{4.3.1} through their Fisher information.

\begin{figure}[t!]
\begin{subfigure}{0.33\textwidth}
  \centering
  \includegraphics[width=1\textwidth,]{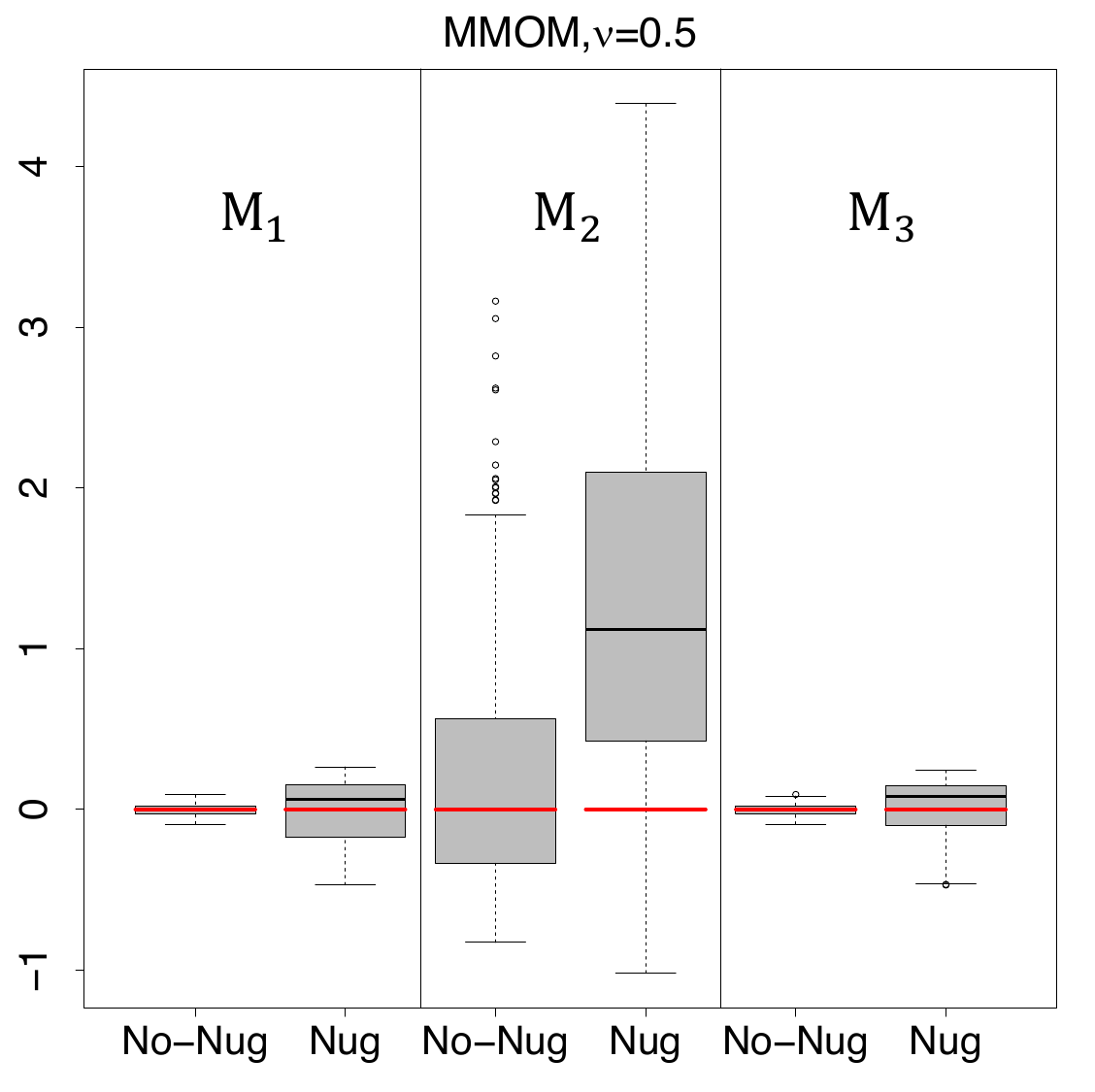}
  \caption*{Weak}
  \label{figmmom(a)}
\end{subfigure}
\begin{subfigure}{0.33\textwidth}
  \centering
  \includegraphics[width=1\textwidth,]{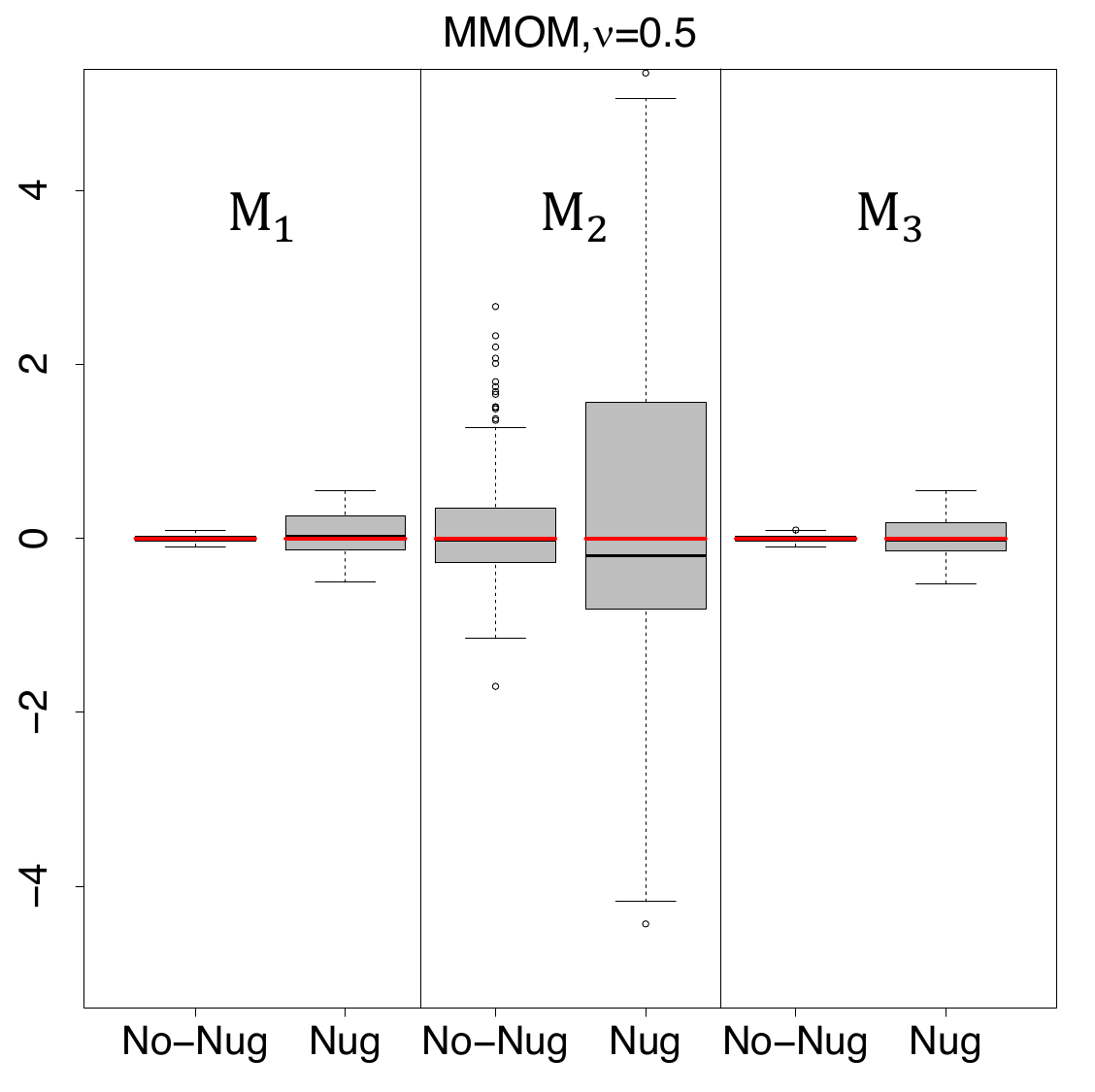}
  \caption*{ Medium  }
  \label{figmmom(b)}
\end{subfigure}
\begin{subfigure}{0.33\textwidth}
  \centering
  \includegraphics[width=1\textwidth,]{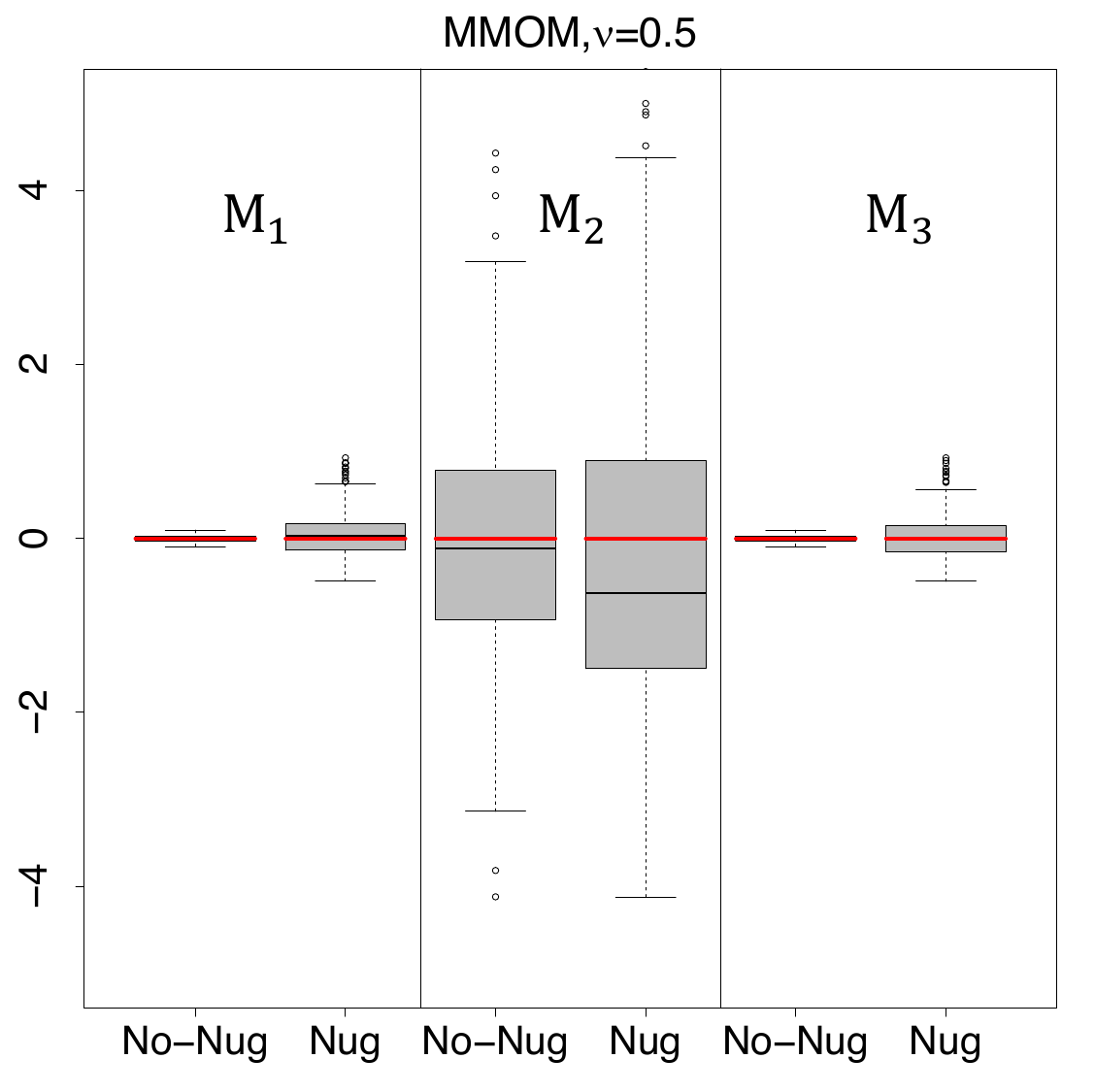}
  \caption*{Strong}
  \label{figmmom(c)}
\end{subfigure}
\caption{Boxplots of MMOM criteria using 300 replicates. For all subfigures above,  ${\cal M}_1$ is in the first block, and ${\cal M}_2$ and ${\cal M}_3$ fill in the second and third blocks, respectively. Vertical lines separate the blocks. Red lines represent the expected MMOM value when true parameter values are used.}
\label{figmmom}
\end{figure}

The MLOE, which is small (i.e., $<10^{-3}$), does not differ among the three parameterizations for all the non-nugget effect models, indicating qualified prediction efficiency for all three parameterizations. Nugget effects increase the MLOE for all three parameterizations, incurring prediction efficiency loss, and they are comparable around $5\times 10^{-2}$. We also calculated the MMOM criteria to scrutinize the modeling accuracy of ${\cal M}_1$, ${\cal M}_2$, and ${\cal M}_3$. Figure~\ref{figmmom} presents boxplots of MMOM obtained from the 300 replicates. For each replicate, we randomly single out 320 locations (20\%) as missing values to calculate $\text{E}_t\{e_a^2(\bs_{0i})\}$ and $\text{E}_a\{e_a^2(\bs_{0i})\}, i=1,\dots,320$. The MMOM on the Whittle Gaussian fields displays similar patterns. Hence, we only provide the MMOM on the exponential Gaussian fields. 

Figures~\ref{figmmom} indicate that the MMOM of ${\cal M}_2$ has numerous outliers and the largest IQRs across all field strengths, signifying frequent appearances of biased estimates. This result agrees with the high volatility issue of $\hat{\btheta}_2$ presented in Figures~\ref{fig2(d)}--\ref{fig2(f)}. Large variability of the MMOM can potentially diminish the reliability of the MSPE as a measure of prediction performance because the approximated MSPE is likely to differ from the true MSPE. Therefore,  cross-validation approaches based on mean squared errors may fail to capture the actual performance of $\hat{\btheta}_2$ in practice. 
Nugget effects influence MMOMs, and the effects are more critical on ${\cal M}_2$. First, nugget effects generally increase the IQRs of MMOMs, increasing the possibility for the approximated MSPE to deviate from the true MSPE. Wider IQRs of MMOMs can be ascribed to the higher volatility of $\hat{\btheta}$ in the presence of nugget effects, as presented in Figure ~\ref{fig2}. Second, the MMOMs of ${\cal M}_2$ with nugget effects are more inclined to significant IQR increases and are more likely to be nonzero, implying that the approximated MSPE can frequently overestimate or underestimate the true prediction error. In contrast, MMOMs of ${\cal M}_1$ and ${\cal M}_3$ are zero-centered with milder IQRs' increases for all field strengths, indicating more stable estimates and a reliable approximated MSPE. Consequently, modeling ${\cal M}_2$ with nugget effects is more likely to result in misspecified models, especially for weak and strong fields. 

\subsubsection{Prediction}\label{4.2.2}
In this part of the experiment, we calculate the MSPEs to quantify the prediction performance of ${\cal M}_1$, ${\cal M}_2$, and ${\cal M}_3$ under various scenarios. We visualize empirical distributions of the MSPEs of the 300 replicates in Figures~\ref{fig4} and \ref{figmspe} in the Supplementary Material.

Figures~\ref{fig4} and \ref{figmspe} demonstrate that the prediction performance is stable and comparable for ${\cal M}_1$, ${\cal M}_2$, and ${\cal M}_3$ for all settings. The MSPEs have approximately identical distributions with relatively narrow IQRs and few outliers. Complicated parameterizations, such as ${\cal M}_2$ and ${\cal M}_3$, do not produce improved prediction errors. However, as mentioned above, the MSPE of ${\cal M}_2$ is not as representative of the prediction performance as the MSPEs of ${\cal M}_1$ and ${\cal M}_3$. Although the three parameterizations have varying behaviors of the MMOM, identical MSPE values are possible because the MSPE is only a superficial measure. \cite{hong2021efficiency} demonstrated similar results. 
\begin{figure}[t!]
\begin{subfigure}{0.5\textwidth}
  \centering
  \includegraphics[width=0.8\textwidth]{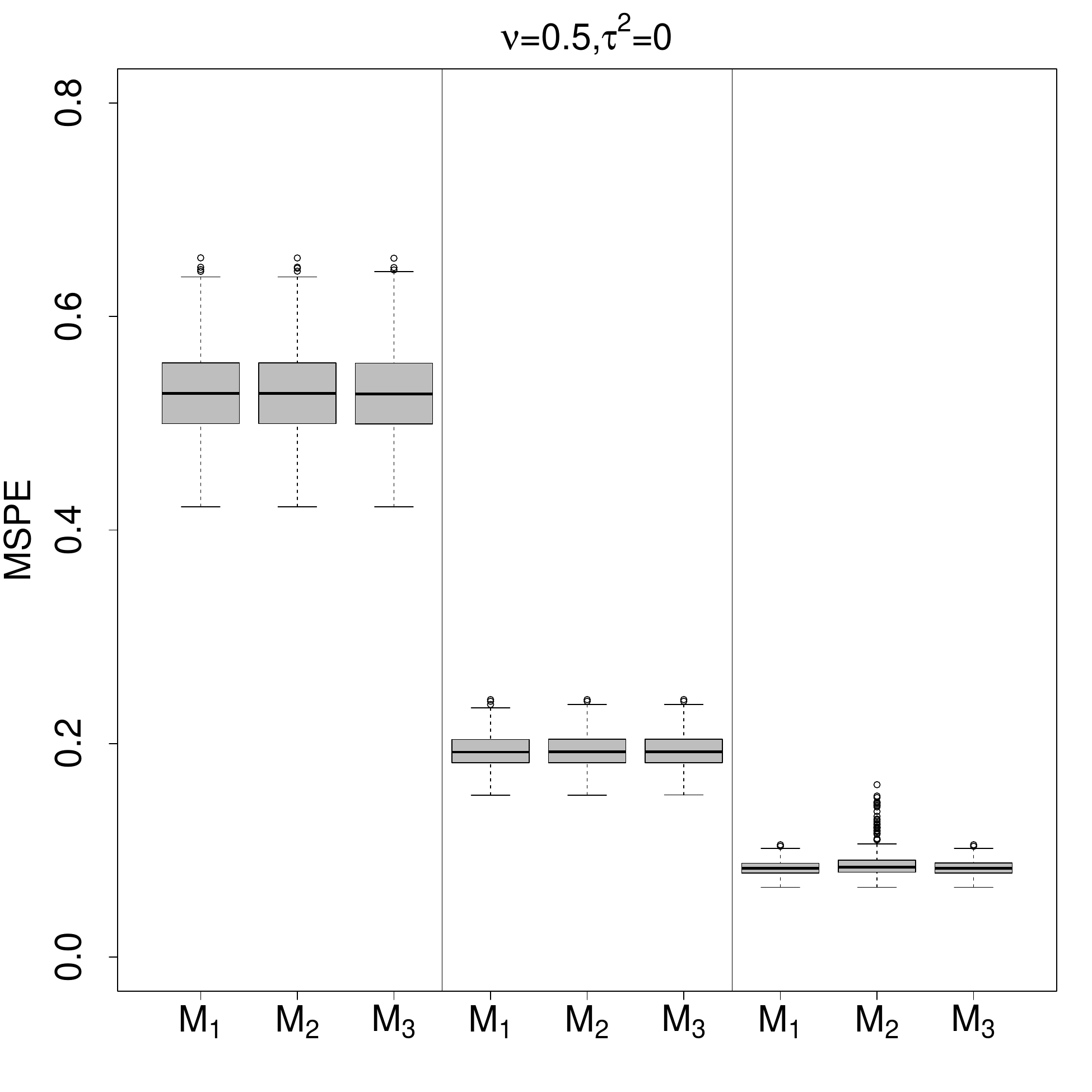}
\end{subfigure}
\begin{subfigure}{0.5\textwidth}
  \centering
  \includegraphics[width=0.8\textwidth]{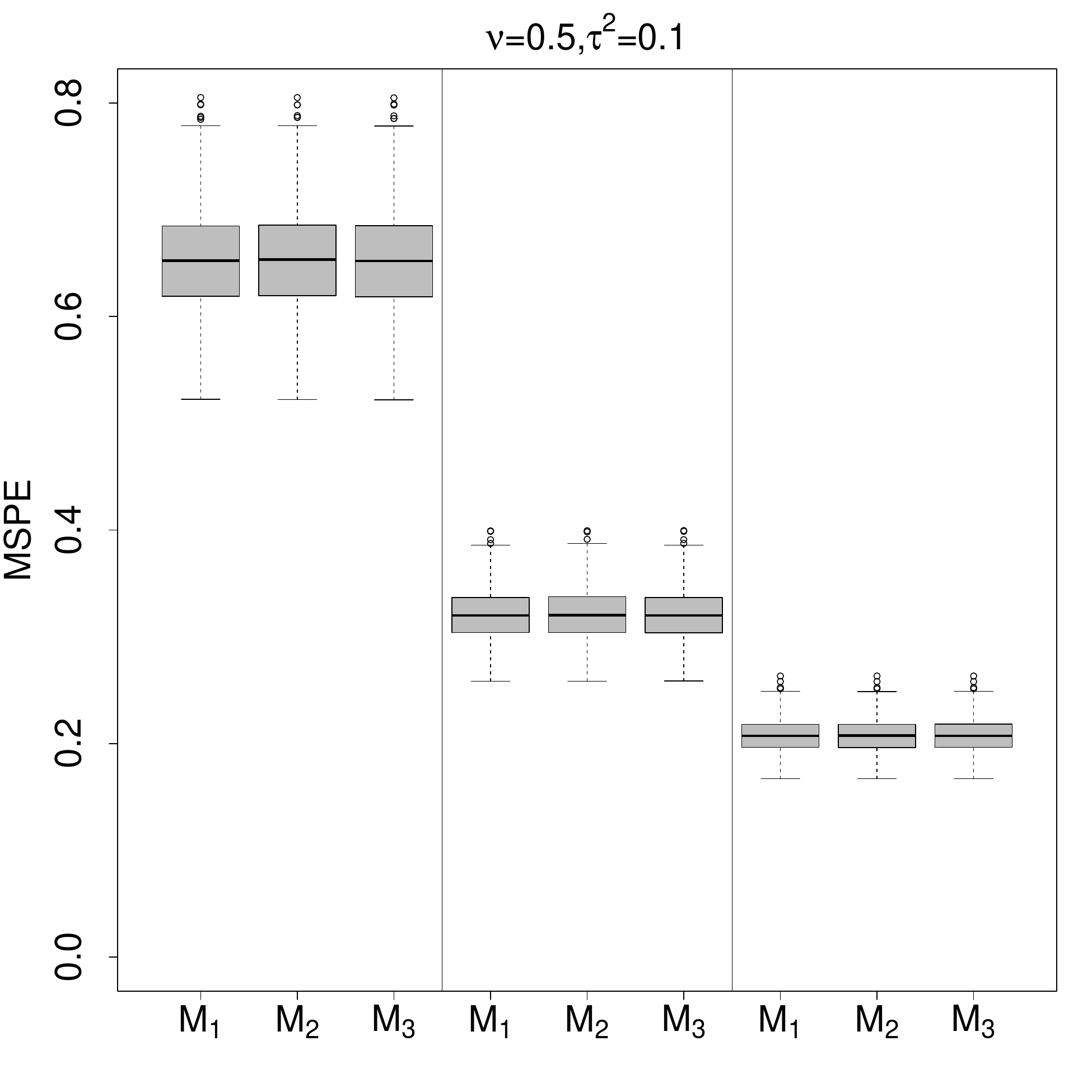}
\end{subfigure}
\caption{Boxplots of the MSPEs using 300 replicates for non-nugget effect models ($\tau^2=0$) and nugget effect models ($\tau^2=0.1$) under weak (First block), medium (second block), and strong (third block) correlations, separated by vertical lines in each sub-figure, on the exponential Gaussian fields ($\nu=0.5$).}
\label{fig4}
\end{figure}
In addition, within the same field, there is a downward trend of the MSPE regarding the field strength. In different fields, the MSPE decreases as the spatial process becomes smoother (exponential to Whittle Gaussian field). These results are consistent with the property of the MSPE of best linear unbiased predictor (BLUP) reported in \cite{zimmerman1992mean}, which stated that the BLUP is the most efficient (lowest MSPE) when the spatial correlation is strong. Nugget effects increase prediction errors due to the addition of noise in the spatial process.

The estimation accuracy of $(\sigma^2,\phi)$ does not affect the MSPE. In Section \ref{MSPE}, spatial predictions are calculated as $\bSigma_{12}\bSigma_{22}^{-1}\bz_2$, which can be rewritten as $\Tilde{\bSigma}_{12}\Tilde{\bSigma}_{22}^{-1}\bz_2 $, where $\bSigma=\sigma^2\Tilde{\bSigma}$ in ${\cal M}_1$ and ${\cal M}_3$ or $\bSigma=\phi\Tilde{\bSigma}$ in ${\cal M}_2$. Nonetheless, variations of their values do crucially influence the MSPE. For instance, if the rest of the parameters are fixed, increases in $\sigma^2$ or $\phi$ yield a larger MSPE. In contrast, neither variations of $\alpha$ nor its estimation accuracy can have significant effects on the MSPE, which has been accounted for by \cite{stein1999interpolation}, stating that high-frequency parameters primarily affect spatial interpolations. As for the range parameters, the MSPE is more sensitive to $\beta$ than $\rho$. It takes a sharper increase in $\rho$ for the field parameterized in ${\cal M}_3$ to attain the same strength as in ${\cal M}_1$. In other words, the same amount of MSPE variation requires a more dramatic change in $\rho$ than $\beta$. A higher sensitivity of $\beta$ should always hold except for extremely rough fields where $\nu < 0.25$. The 0.25 bound comes from the link function between $\beta$ and $\rho$ provided in Table \ref{table1}, $\rho=2\nu^{1/2}\beta$. 
\subsubsection{Evaluating Convergence Rate}
We visualized the number of iterations the optimization algorithm requires to converge and found the optimal values of the underlying parameters with a given tolerance. We  henceforth refer to the number of iterations to converge as the convergent iterations for concision. Moreover, we only display the convergent iterations on the exponential Gaussian field because they exhibit the same patterns on the Whittle Gaussian field. 

Figures~\ref{figiter} and \ref{iterw}, in the Supplementary Material, indicate that ${\cal M}_2$ has the most significant convergent iterations. This low convergence rate can be anticipated from its ample parameter search space. Nevertheless, the large parameter space is inevitable for decent {\redcolor convergence} of $(\hat{\phi},\hat{\alpha})$. According to the theory of asymptotic normality, we have for some settings that $\hat{\phi}\sim{\cal N}(800,596200)$ and $\hat{\alpha}\sim{\cal N}(40,27.5)$ as calculated in Table~\ref{table13}. It is possible to accelerate the modeling process by reducing the search space at the risk of losing optimal values because the variances of $\hat{\btheta}_2$ are quite extraordinary. 
\begin{figure}[t!]
\begin{subfigure}{0.5\textwidth}
  \centering
  \includegraphics[width=0.8\textwidth]{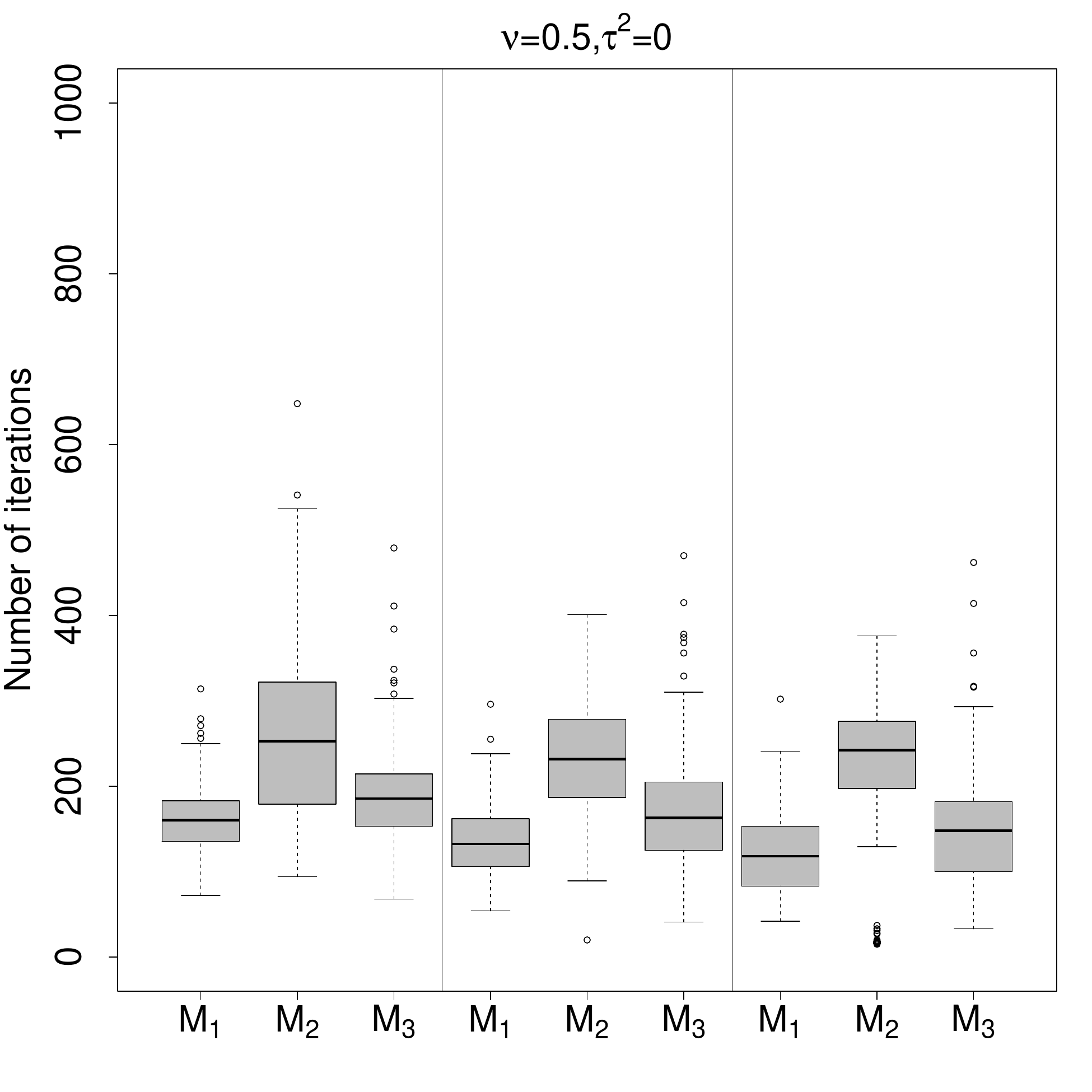}
\end{subfigure}
\begin{subfigure}{0.5\textwidth}
  \centering
  \includegraphics[width=0.8\textwidth]{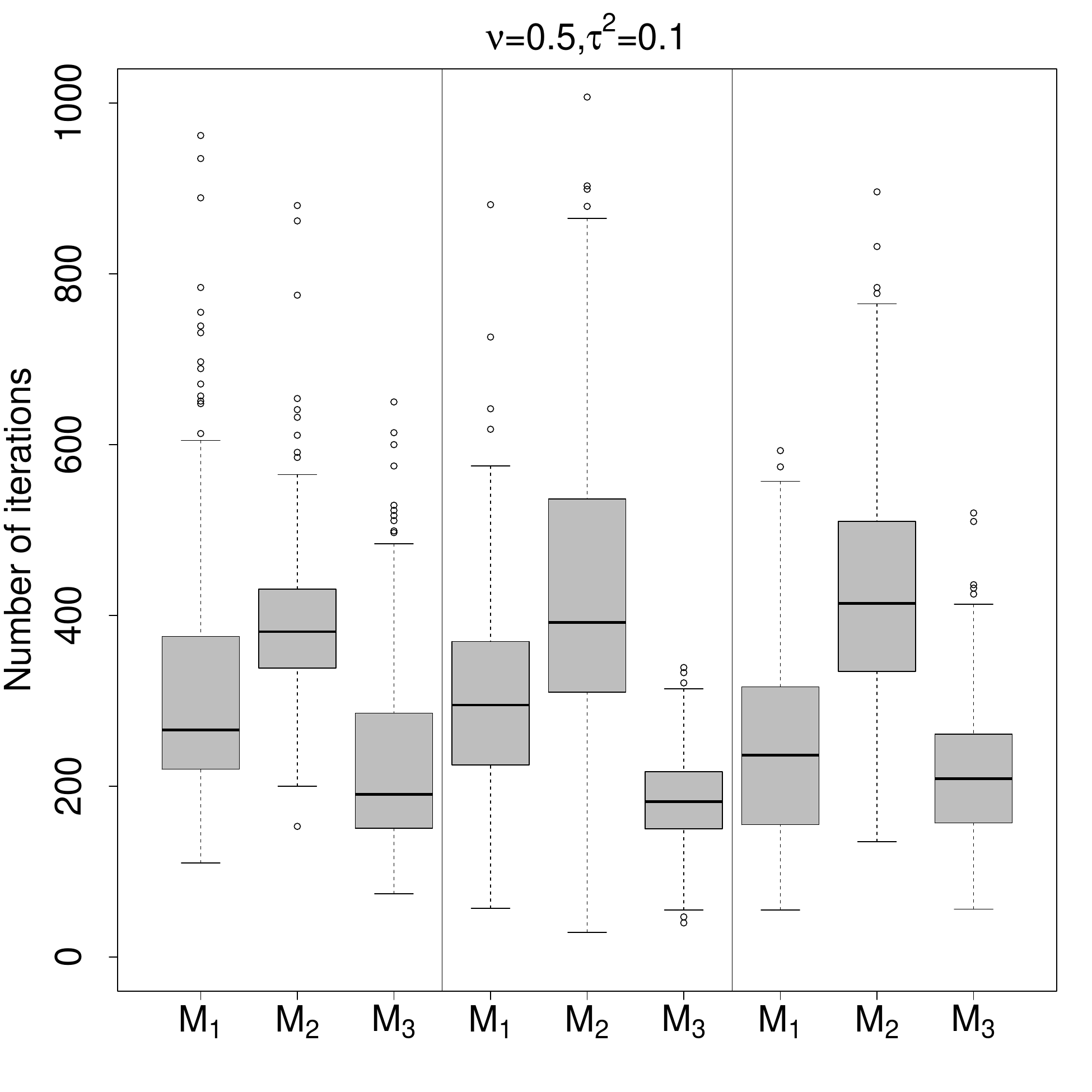}
\end{subfigure}
\caption{Boxplots of the convergent iterations obtained using the 300 replicates for non-nugget effect models ($\tau^2=0$) and nugget effect models
($\tau^2 = 0.1$) under weak (first block), medium (second block), and strong (third block)
correlations, separated by vertical lines in each sub-figure, on the exponential Gaussian fields ($\nu=0.5$). }
\label{figiter}
\end{figure}
The other crucial factor that contributes to the slow convergence of ${\cal M}_2$ is its high parameter correlation \citep{cox1987parameter}, indicated by the parameter asymptotics in Section~\ref{4.3.1}. The evidence of parameter correlation-induced slow convergence is that sharp decreases in the search space for $(\hat{\phi},\hat{\alpha})$ do not lead to improved convergence rates. The convergence rate of ${\cal M}_1$ is superior in non-nugget effect models on the exponential Gaussian fields and identical to ${\cal M}_3$ on the Whittle Gaussian fields. Nonetheless, the convergence rate for ${\cal M}_1$ becomes more variable and exceeds that of ${\cal M}_3$ with nugget effects, which is further evidence of the vulnerability of ${\cal M}_1$ to nugget effects. Nugget effect models significantly decrease the convergence rate because there is one additional parameter $\tau^2$ to estimate.

\subsection{Increasing-Domain Asymptotics}
We aim to investigate the asymptotic distributions of $\hat{\btheta}$. According to the asymptotic normality theory, we anticipate these estimates to follow a Gaussian distribution, with the covariance matrix being closely approximated by the inverse of the Fisher information matrix once the sample size $n$ surpasses a certain threshold. Moreover, the off-diagonal elements of the inverse Fisher information matrix indicate the explicit correlation between estimates. In their research, \cite{stein1999interpolation} examined the Fisher information of $\hat{\btheta}_2$ with sample sizes of 40 and 80 on both regular and irregular grids with various spacing factors.

The current experiment follows a similar procedure, but the contrast between regular and irregular grids is less noticeable due to the high density of observations  ($n=1600$). Irregular grids tend to create clusters of observations that provide more informative insights into the roles of $(\nu,\beta,\rho)$ in governing the spatial process' local behaviors \citep{stein1999interpolation}. On the other hand, regular grids provide more information on $\alpha$, consistent with the results in \cite{stein1999interpolation}. Following \cite{stein1999interpolation}, ${\bI}_\phi$ is fixed as $n/(2\phi^2)$, where $\bI_\phi$ denotes the corresponding diagonal element of the Fisher information matrix. Similarly, ${\bI}_{\sigma^2}$ is also fixed across grid types and field smoothness at $n/(2\sigma^4)$. We observe that the Fisher information patterns do not vary with smoothness and nugget effects. Therefore, we provide Fisher information visualizations for the exponential Gaussian field on an irregular grid without nugget effects using 1600 samples. We use a spacing factor $\delta$ that varies from $(0.02, 0.05, 0.1, 0.2, 0.5, 1)$, multiplied by the two coordinates to obtain grids with different spacings.  Moreover, we evaluate the uncertainty quantification's quality using the true asymptotic variance (TAV) and calculate the difference ratio ({\redcolor DRV}) with respect to the sample variance (SV):
$
    \text{DR}={|\text{SV}-\text{TAV}|}/{\text{TAV}}. \nonumber 
    \label{e22}
$
\subsubsection{Asymptotic Variances and Correlations \label{4.3.1}}
This part of the experiment extends the asymptotic studies by \cite{stein1999interpolation} to a larger sample size in a two-dimensional field. Moreover, we also included ${\cal M}_1$ and ${\cal M}_3$ for the same analysis. We only used the first replicate for the analysis because the simulations reveal that the variations of $\bI(\btheta)$ are minimal for various replicates. 
    \begin{figure}[t!]
\begin{subfigure}{.33\textwidth}
  \centering
  \begin{tikzpicture}
  \node (img){\includegraphics[width=0.9\linewidth]{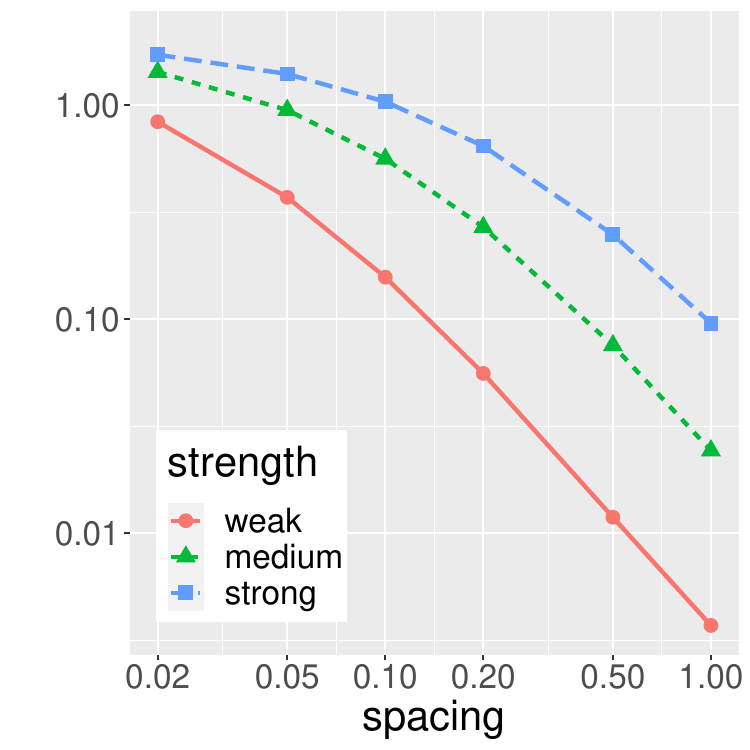}};
  \node[left=of img, node distance=0cm, rotate=90, anchor=center, yshift=-1cm,font= \color{black}]{${\cal M}_1$};
 \end{tikzpicture}
  \caption{\hspace{5mm}$\hat{\sigma}^2$}
  \label{fig6(a)}
  \vspace{5mm}
\end{subfigure}
\begin{subfigure}{.33\textwidth}
  \centering
  \includegraphics[width=0.9\linewidth]{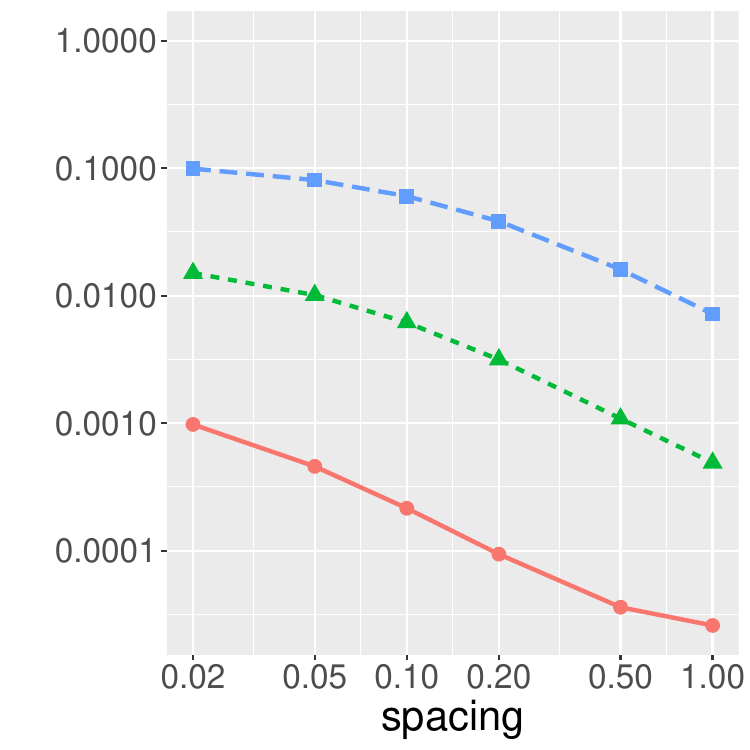}
  \caption{\hspace{5mm}$\hat{\beta}$}
  \label{fig6(b)}
  \vspace{5mm}
\end{subfigure}
\begin{subfigure}{.33\textwidth}
  \centering
  \includegraphics[width=0.9\linewidth]{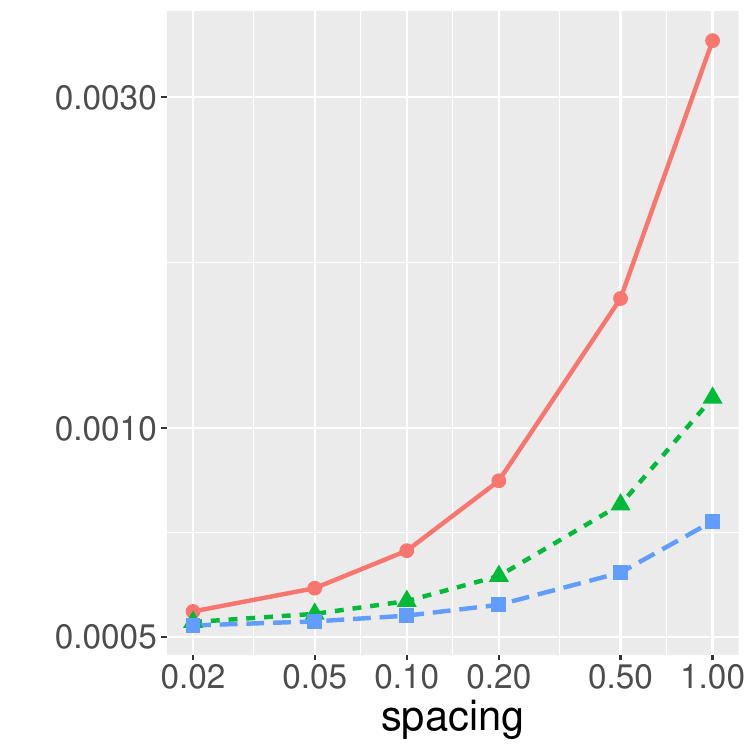}
  \caption{\hspace{5mm}$\hat{\nu}$}
  \label{fig6(c)}
  \vspace{5mm}
\end{subfigure}
\begin{subfigure}{.33\textwidth}
  \centering
  \begin{tikzpicture}
  \node (img){\includegraphics[width=0.9\linewidth]{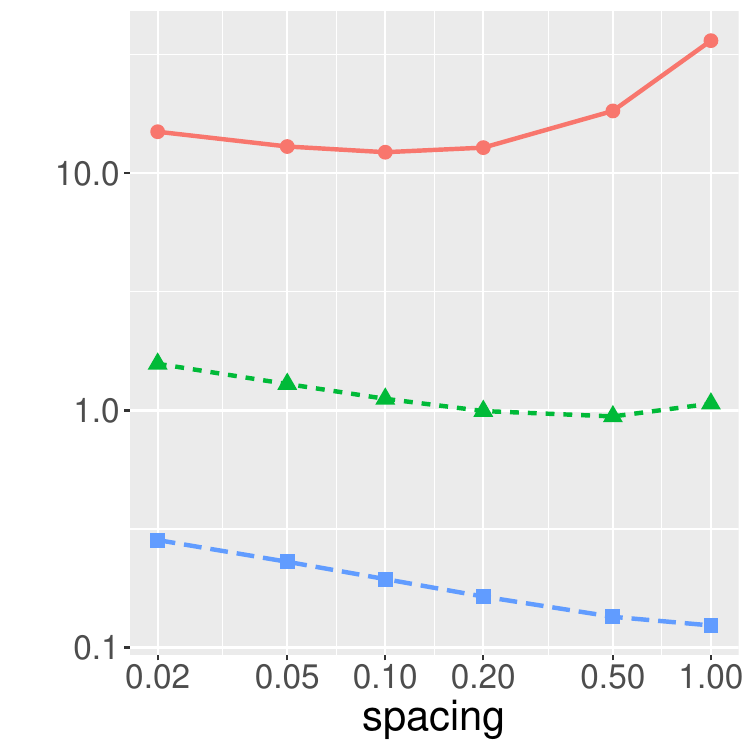}  };
  \node[left=of img, node distance=0cm, rotate=90, anchor=center, yshift=-1cm,font= \color{black}]{${\cal M}_2$};
 \end{tikzpicture}
  \caption{\hspace{5mm}$\hat{\phi}$}
  \label{fig6(g)}
  \vspace{5mm}
\end{subfigure}
\begin{subfigure}{.33\textwidth}
  \centering
  \includegraphics[width=0.9\linewidth]{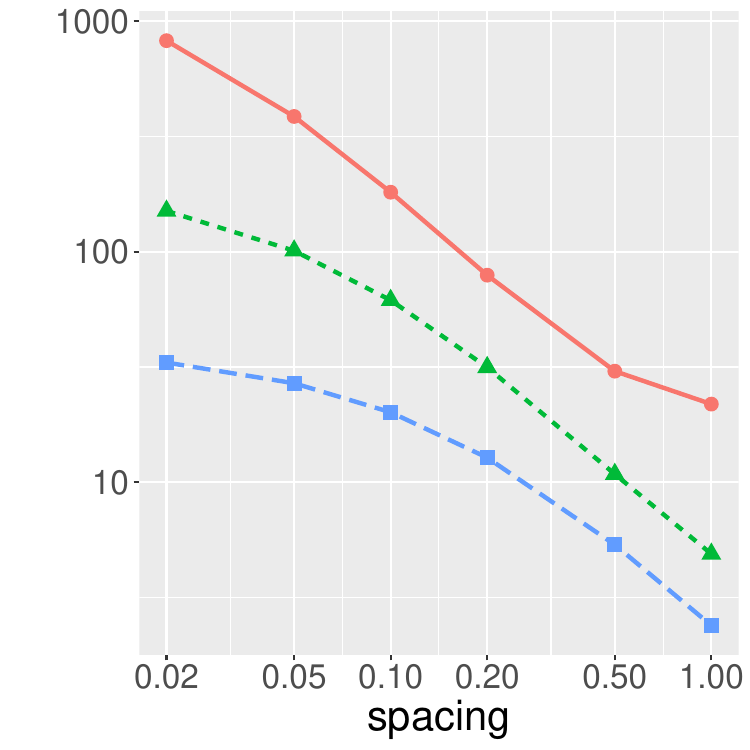} 
  \caption{\hspace{5mm}$\hat{\alpha}$}
  \label{fig6(h)}
  \vspace{5mm}
\end{subfigure}
\begin{subfigure}{.33\textwidth}
  \centering
  \includegraphics[width=0.9\linewidth]{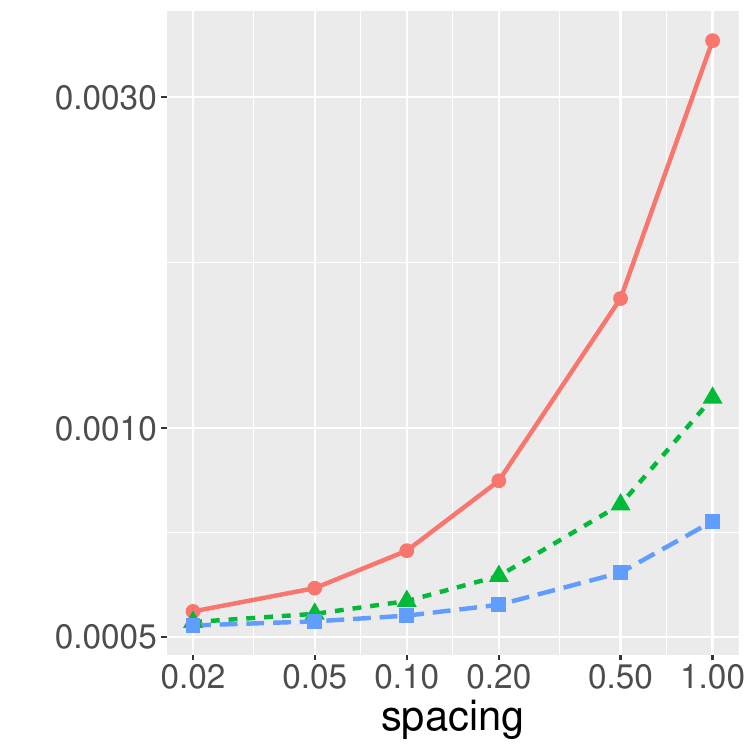}  
  \caption{\hspace{5mm}$\hat{\nu}$}
  \label{fig6(i)}
\end{subfigure}
\begin{subfigure}{.33\textwidth}
  \centering
      \begin{tikzpicture}
  \node (img){\includegraphics[width=0.9\linewidth]{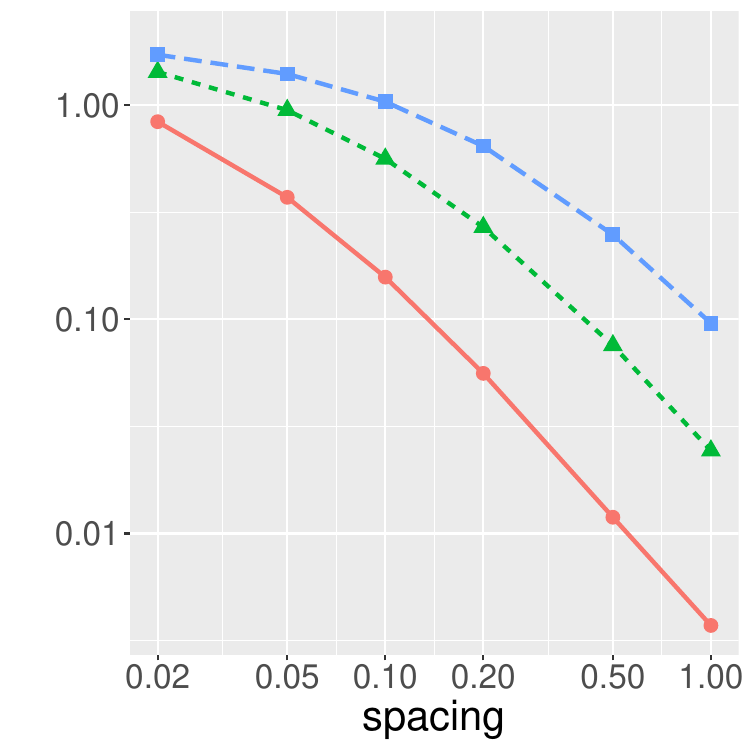}  };
  \node[left=of img, node distance=0cm, rotate=90, anchor=center, yshift=-1cm,font= \color{black}]{${\cal M}_3$};
 \end{tikzpicture}
  \caption{\hspace{5mm}$\hat{\sigma}^2$}
  \label{fig6(m)}
\end{subfigure}
\begin{subfigure}{.33\textwidth}
  \centering
  \includegraphics[width=0.9\linewidth]{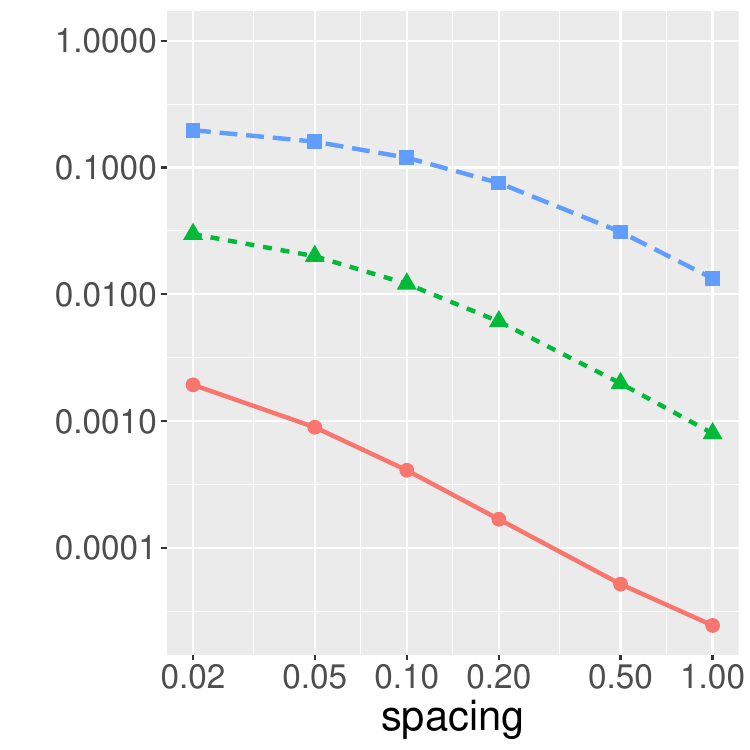}  
  \caption{\hspace{5mm}$\hat{\rho}$}
  \label{fig6(n)}
\end{subfigure}
\begin{subfigure}{.33\textwidth}
  \centering
  \includegraphics[width=0.9\linewidth]{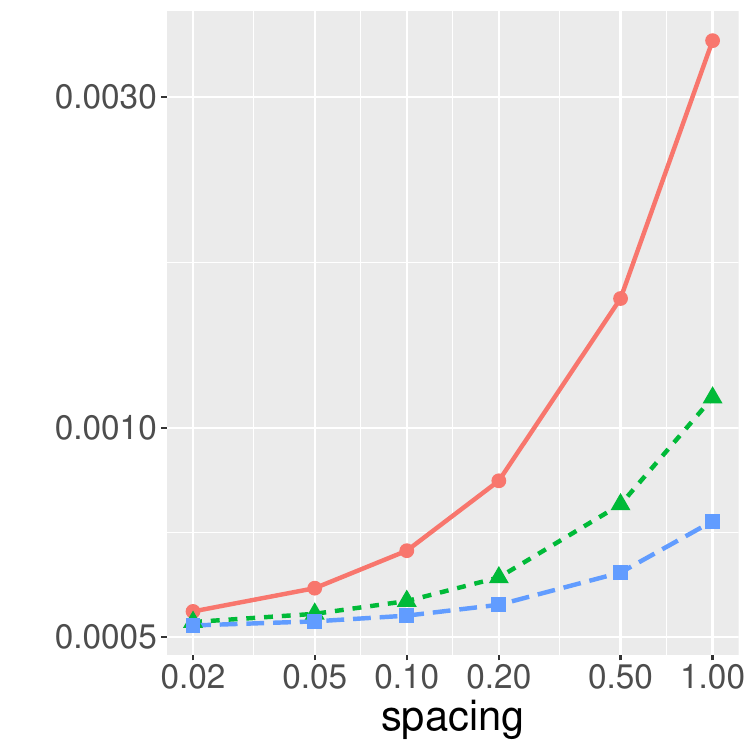} 
  \caption{\hspace{5mm}$\hat{\nu}$}
  \label{fig6(o)}
\end{subfigure}
\caption{True Asymptotic Variances (TAVs) for the MLEs of ${\cal M}_1$, ${\cal M}_2$, and ${\cal M}_3$ under weak (red, circle), medium (green, triangle), and strong (blue, square) exponential Gaussian fields. {\redcolor The x-axis (spacing) and y-axis (TAV) except for $(\hat{\beta},\hat{\rho})$ are on a logarithmic scale to better demonstrate the trend behaviors of the TAVs. The y-axis of $(\hat{\beta},\hat{\rho})$ are on a linear scale because the decreasing trends are already obvious.}}
\label{7}
\end{figure}
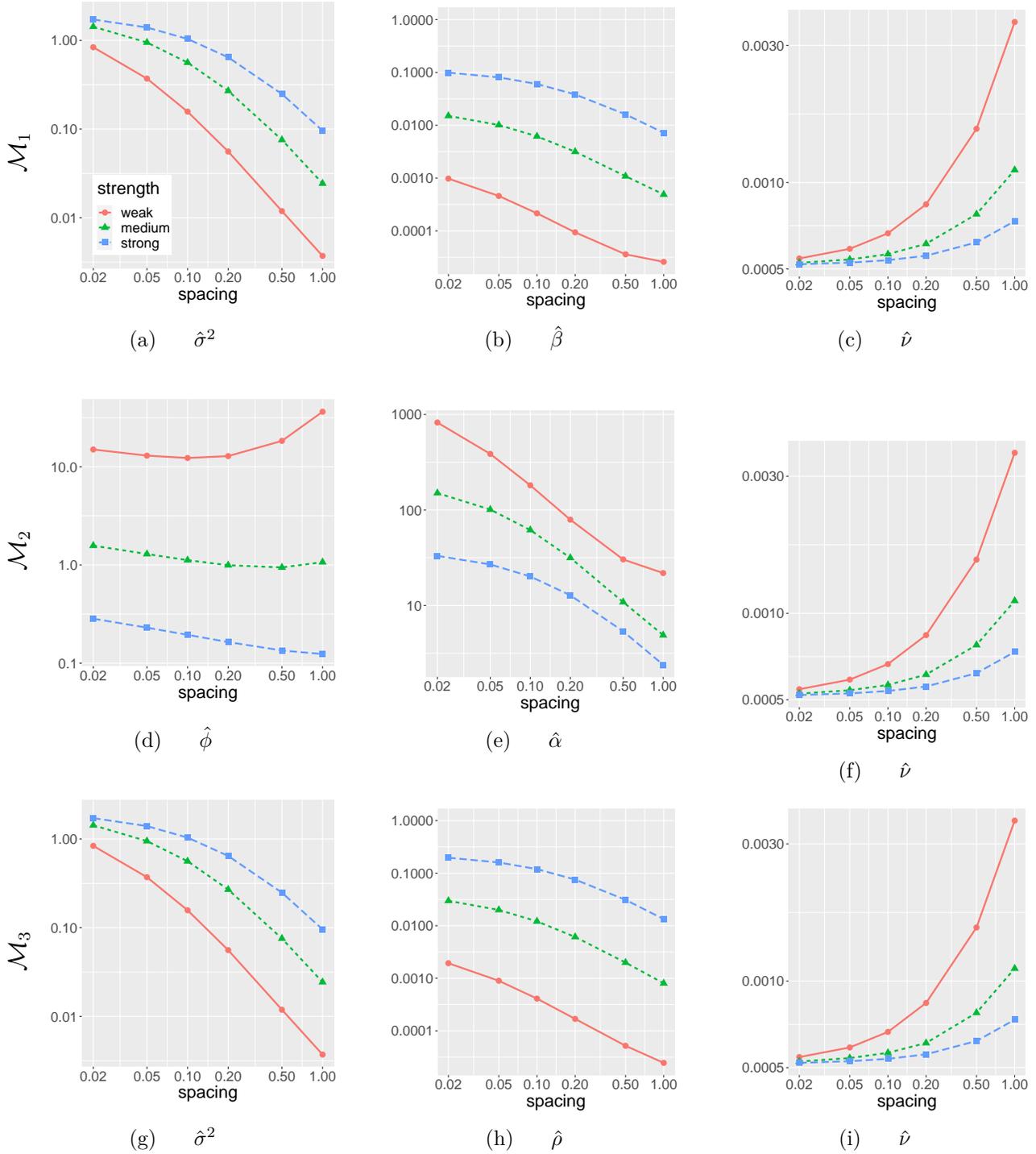

\begin{figure}[t!]
\begin{subfigure}{.33\textwidth}
  \centering
    \begin{tikzpicture}
  \node (img){\includegraphics[width=0.9\linewidth]{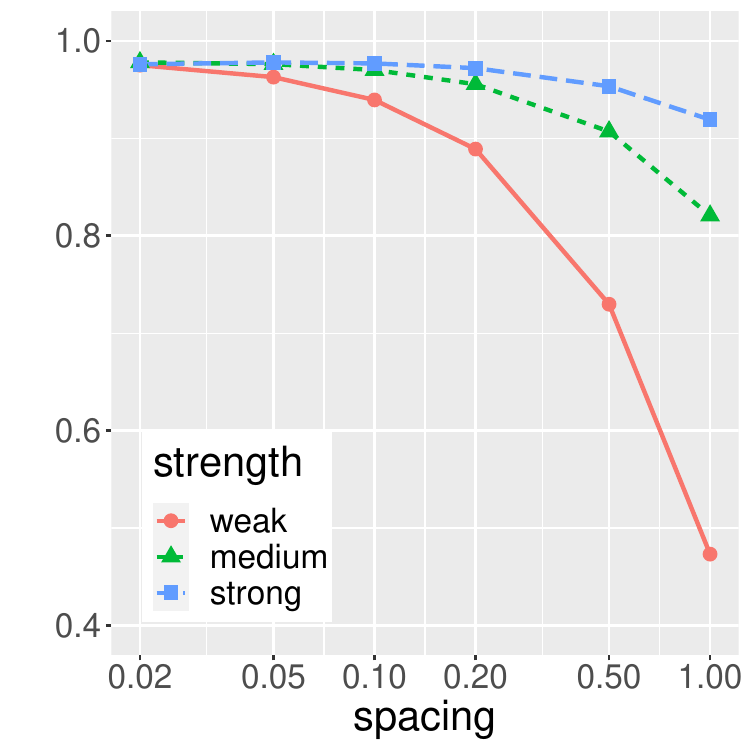} };
  \node[left=of img, node distance=0cm, rotate=90, anchor=center, yshift=-1cm,font= \color{black}]{${\cal M}_1$};
 \end{tikzpicture}
  \caption{\hspace{5mm}$\hat{\sigma}^2,\hat{\beta}$}
  \label{fig6(d)}
  \vspace{5mm}
\end{subfigure}
\begin{subfigure}{.33\textwidth}
  \centering
  \includegraphics[width=0.9\linewidth]{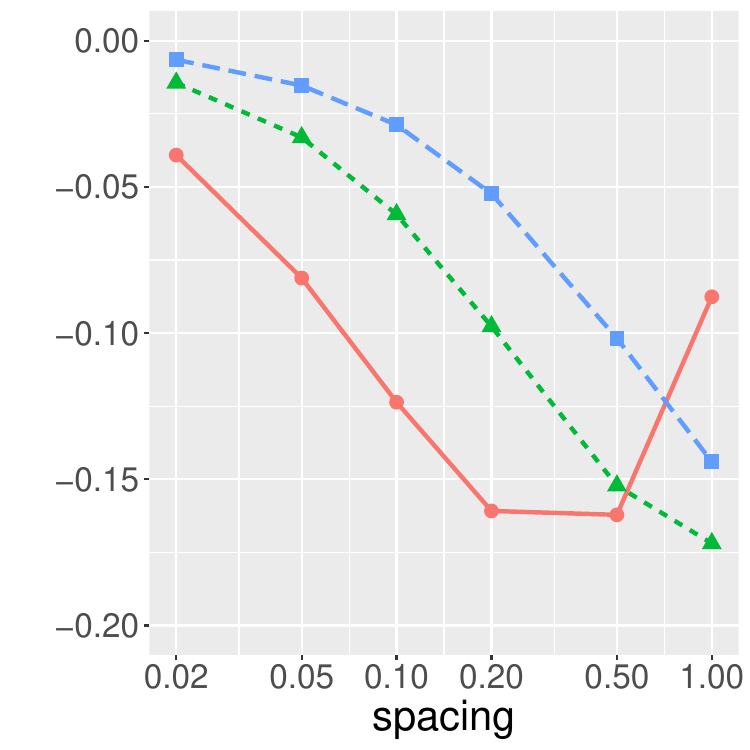}
  \caption{\hspace{5mm}$\hat{\sigma}^2,\hat{\nu}$}
  \label{fig6(e)}
  \vspace{5mm}
\end{subfigure}
\begin{subfigure}{.33\textwidth}
  \centering
  \includegraphics[width=0.9\linewidth]{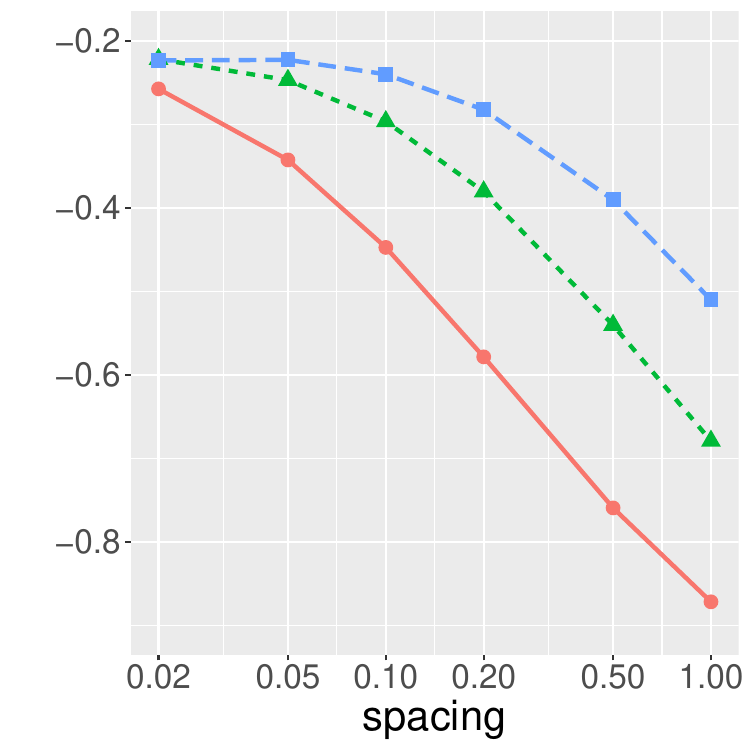} 
  \caption{\hspace{5mm}$\hat{\beta},\hat{\nu}$}
  \label{fig6(f)}
  \vspace{5mm}
\end{subfigure}
\begin{subfigure}{.33\textwidth}
  \centering
    \begin{tikzpicture}
  \node (img){\includegraphics[width=0.9\linewidth]{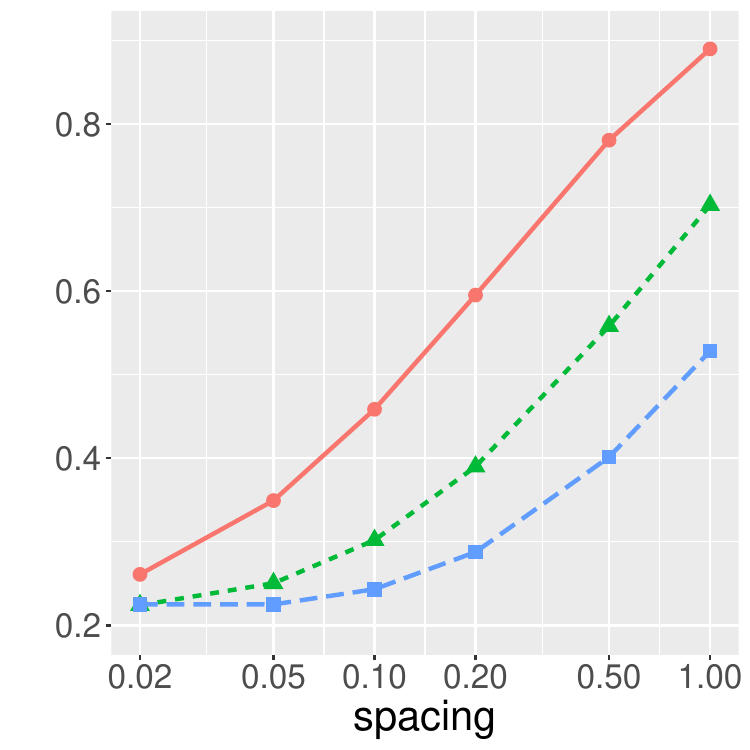}};
  \node[left=of img, node distance=0cm, rotate=90, anchor=center, yshift=-1cm,font= \color{black}]{${\cal M}_2$};
 \end{tikzpicture}
  \caption{\hspace{5mm}$\hat{\phi},\hat{\alpha}$}
  \label{fig6(j)}
  \vspace{5mm}
\end{subfigure}
\begin{subfigure}{.33\textwidth}
  \centering
  \includegraphics[width=0.9\linewidth]{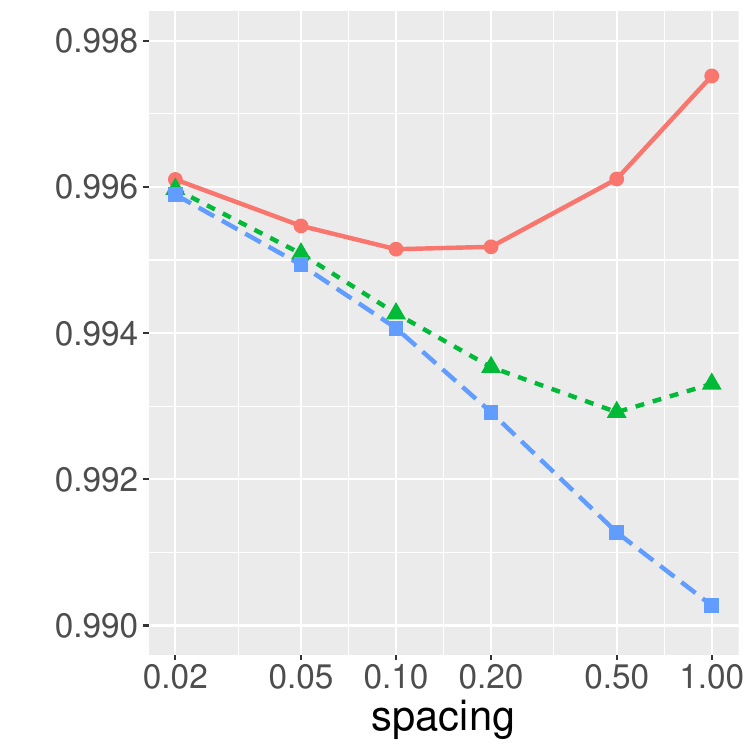}  
  \caption{\hspace{5mm}$\hat{\phi},\hat{\nu}$}
  \label{fig6(k)}
  \vspace{5mm}
\end{subfigure}
\begin{subfigure}{.33\textwidth}
  \centering
  \includegraphics[width=0.9\linewidth]{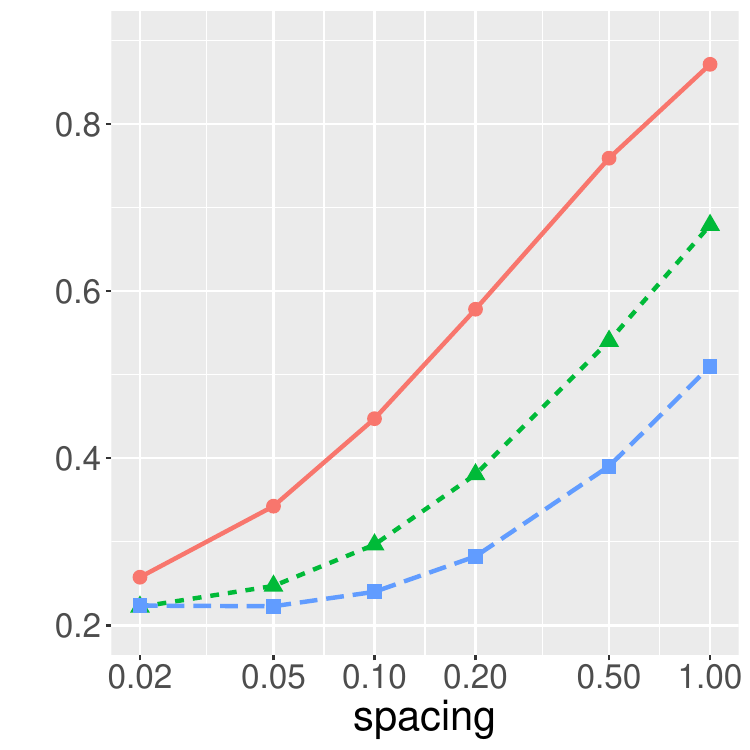}  
  \caption{\hspace{5mm}$\hat{\alpha},\hat{\nu}$}
  \label{fig6(l)}
  \vspace{5mm}
\end{subfigure}
\begin{subfigure}{.33\textwidth}
  \centering
 \begin{tikzpicture}
  \node (img){\includegraphics[width=0.9\linewidth]{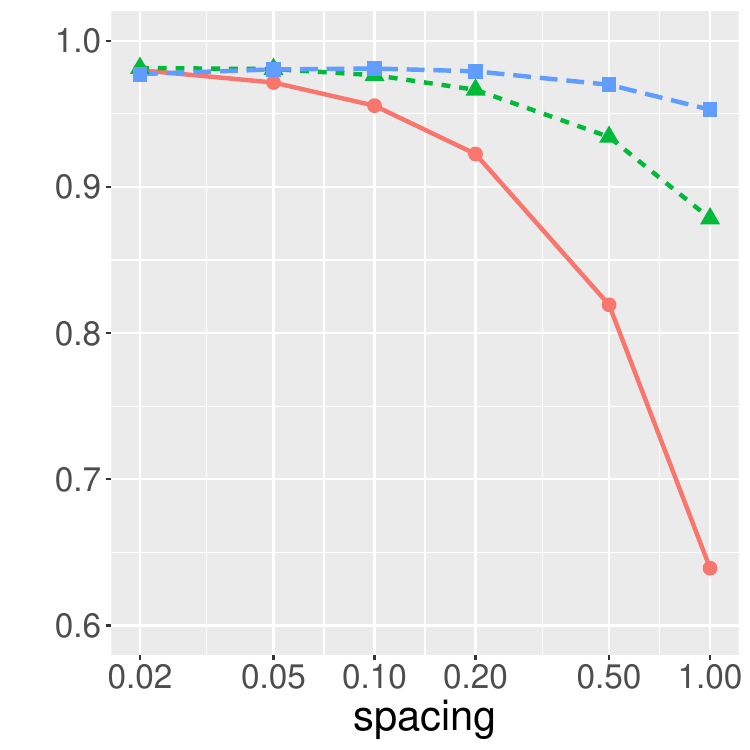}   };
  \node[left=of img, node distance=0cm, rotate=90, anchor=center, yshift=-1cm,font= \color{black}]{${\cal M}_3$};
 \end{tikzpicture}
  \caption{\hspace{5mm}$\hat{\sigma}^2,\hat{\rho}$}
  \label{fig6(p)}
\end{subfigure}
\begin{subfigure}{.33\textwidth}
  \centering
  \includegraphics[width=0.9\linewidth]{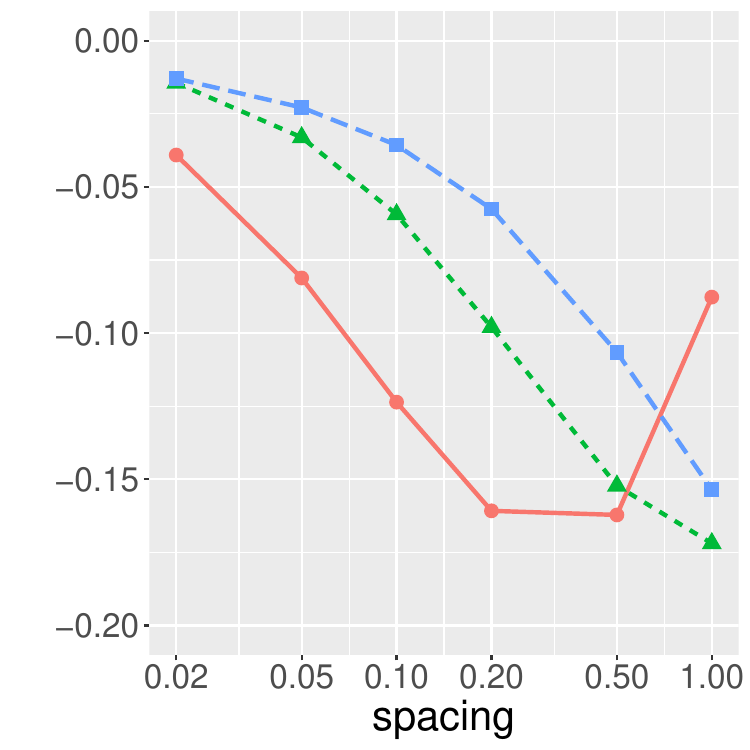}  
  \caption{\hspace{5mm}$\hat{\sigma}^2,\hat{\nu}$}
  \label{fig6(q)}
\end{subfigure}
\begin{subfigure}{.33\textwidth}
  \centering
  \includegraphics[width=0.9\linewidth]{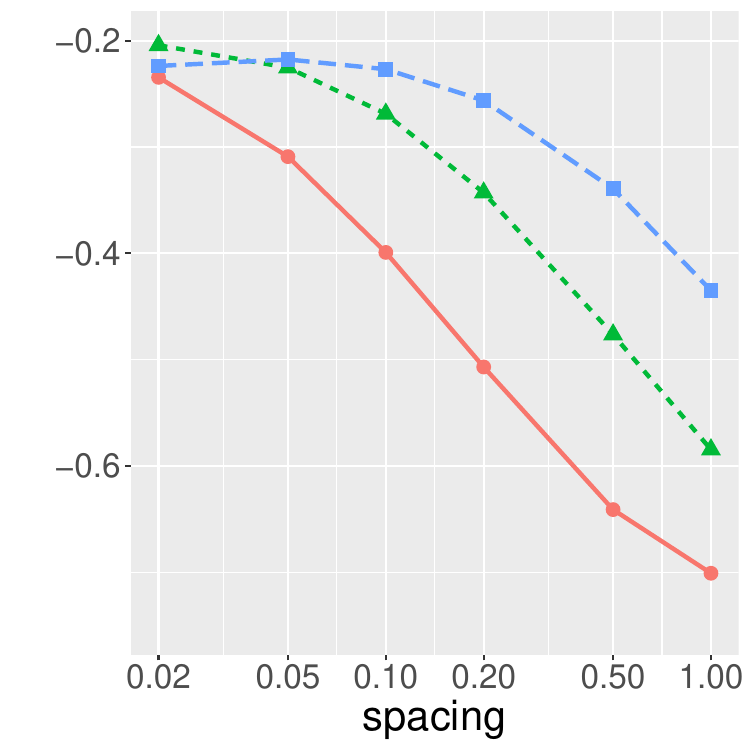}  
  \caption{\hspace{5mm}$\hat{\rho},\hat{\nu}$}
  \label{fig6(r)}
\end{subfigure}
\caption{True Asymptotic Correlations between MLEs of ${\cal M}_1$, ${\cal M}_2$, and ${\cal M}_3$ under weak (red, circle), medium (green, triangle), and strong (blue, square) exponential Gaussian fields. }
\label{fig6}
\end{figure}

 Figures~\ref{7} and \ref{fig6} illustrate that the MLEs of ${\cal M}_2$ distinguish themselves from those of ${\cal M}_1$ and ${\cal M}_3$, which have a high resemblance.
 Instead of taking up a continuously decreasing trend, such as $(\hat{\sigma}^2,\hat{\beta},\hat{\rho},\hat{\alpha})$, the TAV of $\hat{\phi}$ decreases first and then takes an upward trend on weak and medium fields, as presented in Figure~\ref{fig6(g)}. Such upward-turning behavior results from strengthening the positive correlations of $(\hat{\phi},\hat{\alpha})$ in $\delta$ (Figure ~\ref{fig6(j)}), which also appeared in the work by \cite{stein1999interpolation}. 
 In addition, such behavior of $\hat{\phi}$ is not explicit on the strong field due to a weaker correlation of $(\hat{\phi},\hat{\alpha})$, as depicted in Figures ~\ref{fig6(j)}. Moreover, $\hat{\alpha}$ exhibits opposite behavior to $(\hat{\beta},\hat{\rho})$ in terms of the correlation with $\hat{\nu}$ displayed in Figures~\ref{fig6(f)}, \ref{fig6(l)}, and \ref{fig6(r)}. 

 The TAVs of $(\hat{\phi},\hat{\alpha})$ are also opposite to $(\hat{\sigma}^2,\hat{\beta},\hat{\rho})$ concerning field strengths, which agrees with the patterns of their IQRs shown in Figure~\ref{fig2}.  Diagonal elements of the Fisher information matrix for all these parameters at $\delta=1$ are presented in Table~\ref{info} to address the differences together with the asymptotic correlations of the MLEs shown in Figure \ref{fig6}. 
 
Second, all parameters in ${\cal M}_2$ are positively correlated. However, in ${\cal M}_1$ and ${\cal M}_3$, the variance and range parameters are positively correlated (Figures~\ref{fig6(d)} and \ref{fig6(p)}), but the range and smoothness parameters are negatively correlated (Figures~\ref{fig6(f)} and \ref{fig6(r)}). 
In addition, their variance and smoothness parameters are close to being orthogonal (Figures \ref{fig6(e)} and \ref{fig6(q)}). Although ${\cal M}_1$ and ${\cal M}_3$ display many similarities,  
some differences exist. A weaker negative correlation of $(\hat{\rho},\hat{\nu})$ and a stronger positive correlation of $(\hat{\sigma}^2,\hat{\rho})$ than their counterparts in ${\cal M}_1$ are displayed in Figures~\ref{fig6(f)}, \ref{fig6(r)}, \ref{fig6(p)}, and \ref{fig6(d)}.
A stronger correlation of $(\hat{\beta},\hat{\nu})$ is somewhat intuitive because $\beta$ can be viewed as an aggregation of $(\rho, \nu)$, which in principle, should be more correlated with $\nu$. In ${\cal M}_3$, such a correlation is diminished by taking $\nu^{1/2}$ out of the range parameter. There is no simple reason for a weaker correlation of $(\hat{\sigma}^2,\hat{\beta})$, but one possibility is that $\hat{\beta}$ contains $\hat{\nu}$, which is independent of~$\hat{\sigma}^2$.

The TAVs of $\hat{\nu}$ rise in $\delta$ for all three parameterizations shown in Figures~\ref{fig6(c)}, \ref{fig6(i)}, and \ref{fig6(o)}, which is consistent with the results from \cite{stein1999interpolation} because tightly spaced clusters are more informative for $\hat{\nu}$. In addition, the TAVs of $\hat{\nu}$ decrease when field correlation strengthens for all three parameterizations for various reasons.  The decreases for ${\cal M}_1$ and ${\cal M}_3$ result from increase of $\bI_\nu$ shown in Table \ref{info} and also $\hat{\nu}$'s decreasing correlation with $\hat{\beta}$ and $\hat{\rho}$ shown in Figures \ref{fig6(f)} and \ref{fig6(r)}. However, for $\hat{\nu}$ in ${\cal M}_2$, the decrease of its TAV is due to its decreasing correlations with $(\hat{\phi},\hat{\alpha})$ (Figures \ref{fig6(k)} and \ref{fig6(l)}) because $\bI_\nu$ decreases, as listed in Table~\ref{info}. Decreasing TAVs of $\hat{\alpha}$ in $\delta$ agrees with the results from \cite{stein1999interpolation}'s experiment, which was explained by its relevance to low-frequency behaviors. Therefore, a sparse grid is more informative of $\hat{\alpha}$. The upward-turning behavior of the TAVs of $\hat{\alpha}$ in the work by \cite{stein1999interpolation} do not appear in the current studies because the asymptotic correlations of $(\hat{\phi},\hat{\alpha})$ and $(\hat{\alpha},\hat{\nu})$ are lower on irregular grids. Large TAVs of $(\hat{\beta},\hat{\rho})$ at a small $\delta$ can be attributed to the failure to recognize the dependence range with observations squeezed into a small region and therefore, large uncertainties on the parameter estimates. The dependence range becomes easier to pinpoint as the study region expands. The TAV reductions of $\hat{\sigma}^2$ in $\delta$ are not intrinsic but rather due to $\hat{\sigma}^2$'s interactions with $(\hat{\beta},\hat{\rho})$ because $\bI_{\sigma^2}$ is fixed at 800 despite of varying $\delta$ and field correlation strengths. In addition, $(\hat{\sigma}^2,\hat{\nu})$ are nearly orthogonal as shown in Figures \ref{fig6(e)} and \ref{fig6(q)}. In particular, TAV of $\hat{\sigma}^2$ decreases in increasing $\delta$ primarily because of its decreasing correlation with $(\hat{\beta},\hat{\rho})$ (Figures \ref{fig6(d)} and \ref{fig6(p)}). Furthermore, TAVs of $\hat{\sigma}^2$ increase as field correlation strengthens, which can be ascribed to the rising correlations with $(\hat{\beta},\hat{\rho})$.
\begin{table}[t!]
    \centering
    \caption{ Fisher information for $\btheta$ for ${\cal M}_1$, ${\cal M}_2$, and ${\cal M}_3$ on the exponential Gaussian fields ($\nu=0.5$) at $\delta=1$.}
    \begin{tabular}{|l|c|c|c|}
    \hline
        Dependence range/correlation strength (${\cal M}_1$) & $\bI_{\sigma^2}$ & $\bI_{\beta}$ &  $\bI_{\nu}$\\
        \hline
       $\beta=0.033$ (weak) & 800.0 & 474155.5    & 2653.6  \\
       \hline
       $\beta=0.1$ (medium) &  800.0 & 71786.2    & 10694.5 \\
       \hline
       $\beta=0.234$ (strong) & 800.0 & 14146.3   & 21987.3 \\
       \hline
         \hline

        Dependence range/correlation strength (${\cal M}_1$) & $\bI_{\phi}$ & $\bI_{\alpha}$ & $\bI_{\nu}$\\
        \hline
        \hline
        $\phi=9.6458$, $\alpha=30.3030$ (weak) & 8.6 & 0.3 & 74835.8\\
        \hline
        $\phi=3.1831$, $\alpha=10.0000$ (medium) & 79.0 & 0.4 & 71747.9\\
        \hline
        $\phi=1.3603$, $\alpha=4.2735$ (strong) & 432.3 & 0.6 & 71265.5\\
       \hline
       \hline
       Dependence range/correlation strength (${\cal M}_1$) & $\bI_{\sigma^2}$ & $\bI_{\rho}$ & $\bI_{\nu}$\\
       \hline
        $\rho=0.0467$ (weak) & 800.0 & 236782.6    & 956.3  \\
        \hline
       $\rho=0.1414$ (medium) &  800.0 & 35917.6   & 6118.6 \\
       \hline
       $\rho=0.3309$ (strong) & 800.0 & 7080.5   & 14784.9\\
    \hline
    \end{tabular}
    \label{info}
\end{table}

 As previously mentioned, Table~\ref{info} demonstrates that increase in field strength incurs tangible loss on $\bI_\beta$ and $\bI_\rho$. The correlation between range and variance parameters also increases in ${\cal M}_1$ and ${\cal M}_3$ (Figure \ref{fig6(d)} and \ref{fig6(p)}). Consequently, TAVs of $(\hat{\beta},\hat{\rho})$ tend to increase as field correlation strengthens. In contrast, $\bI_\phi$ increases as field correlation strengthens. $\hat{\phi}$ also is less correlated with $(\hat{\alpha},\hat{\nu})$ (Figures \ref{fig6(j)} and \ref{fig6(k)}). Therefore, $\hat{\phi}$ exhibits opposite TAV behaviors compared with $(\hat{\beta},\hat{\rho})$ as indicated in Figure \ref{7}. In addition, $\bI_\alpha$ is only slightly increased and therefore, the TAV reductions originate from $\hat{\alpha}$'s decreasing correlations with $(\hat{\phi},\hat{\nu})$ (Figures \ref{fig6(j)} and \ref{fig6(l)}).
\subsubsection{ Quality of Uncertainty Quantification}\label{4.3.2}
This section calculates the SV, TAV, and {\redcolor DRV} values to measure the quality of uncertainty quantification with TAV at finite samples, $n=1600$. The computed results are summarized in Table \ref{table13} in the Supplementary Material. {\redcolor We have also repeated the experiment with different $n$ and recorded the results in Tables \ref{1}, \ref{2}, and \ref{3} to address the concerns of increasing $n$ because when $n$ is large enough, the variances should converge in theory regardless of parameterizations. We have shown that differences in matches of SV and TAV consistently persist as in $n=1600$ with various values of $n$. Therefore, the comparisons drawn in this subsection are representative.}

 For the non-nugget effect models, Table~\ref{table13} reveals that the DRs of $\hat{\nu}$ in ${\cal M}_2$ are the most significant in all settings, indicating apparent discrepancies between SV and TAV. Such discrepancies can be attributed to the strong parameter correlations (almost linear) of $(\hat{\phi},\hat{\nu})$ as illustrated in Figure~\ref{fig6(k)}. High parameter correlations critically affect MLEs, causing mathematical complexities and errors in their interpretations \citep{willmot1988parameter}. Difficulties will also arise in numerical optimizations, causing slow convergence or even divergence \citep{cox1987parameter} and non-identifiability \citep{li2013identification}. Consequently, the uncertainty for $\hat{\nu}$ in ${\cal M}_2$ is not properly approximated by the TAVs at finite samples.
 
 In contrast, {\redcolor DRVs} for $\hat{\nu}$ in ${\cal M}_1$ and ${\cal M}_3$ are small because $(\hat{\sigma}^2,\hat{\nu})$ are nearly independent, as depicted in Figures~\ref{fig6(e)} and \ref{fig6(q)}. The strong correlation also affects $\hat{\phi}$, where the {\redcolor DRVs} appear large in numerous settings. In addition, the DRs for $(\hat{\sigma}^2,\hat{\beta},\hat{\rho})$ are notably larger when the variance and range estimates have  almost linear dependence at medium and strong correlations. Although this disturbance is veiled by the inappropriate convergence tolerance for ${\cal M}_3$, the same issue arises when a lower tolerance is applied. Large DRs of $\hat{\alpha}$ are more likely to result from low Fisher information and, therefore, are challenging to estimate because $\hat{\alpha}$ does not display linear dependence with other parameters like $(\hat{\beta},\hat{\rho})$.Furthermore, significant {\redcolor DRVs} for $\left(\hat{\sigma}^2,\hat{\rho}\right)$ confirm their high-tolerance induced bias shown in Figure~\ref{fig2}. Overall, ${\cal M}_1$ achieves the best match between SVs and TAVs in the non-nugget effect models.

 In addition, the TAV of $\hat{\sigma}^2$ and $\hat{\nu}$ are comparable in both ${\cal M}_1$ and ${\cal M}_3$ but the TAVs of $\hat{\beta}$ are obviously smaller than those of $\hat{\rho}$ in most cases.  An appropriate explanation of the smaller TAVs of $\hat{\beta}$ can also be drawn from the link function, $\beta=2\nu^{1/2}\rho$, provided in Table ~\ref{table1}. If we assume $\nu$ as known and apply the {\redcolor invariance} property of MLEs, we have $\hat{\beta}=2\nu^{1/2}\hat{\rho}$. Then, $\text{Var}(\hat{\rho})=4\nu\text{Var}(\hat{\beta})$. In principle,  $\text{Var}(\hat{\rho})$ is expected to exceed $\text{Var}(\hat{\beta})$ except on extremely rough fields (i.e., $\nu < 0.25$). The situation appears on the weak exponential Gaussian field but not on the others, as shown in Table~\ref{table13}. 
 
Nugget effects increase the TAVs and result in considerable DRs for $\hat{\btheta}_1$, $\hat{\btheta}_2$, which is consistent with the heavier effects of $\tau^2$ on ${\cal M}_1$ and ${\cal M}_2$ discussed in Section \ref{acc}. In numerous settings, TAVs of $\hat{\beta}$ can exceed those of $\hat{\rho}$.  {\redcolor DRVs} of $\hat{\btheta}_3$ are relatively contained. Moreover, the nugget effects are more critical on strongly high-frequency relevant parameters. For instance, TAVs predominantly scale by ten times or more for $(\hat{\nu},\hat{\phi})$. However, TAV increases are moderate on the rest, primarily scaling by 4 to 5 times at maximum. The TAVs for $\hat{\beta}$ tend to scale larger than those for $\hat{\rho}$, again indicating the susceptibility of ${\cal M}_1$ to the nugget effects compared with ${\cal M}_3$.  Furthermore, Table~ \ref{tau} in the Supplementary Material indicates that ${\cal M}_3$ offers the most stable {\redcolor DRVs} for $\hat{\tau}^2$, whereas ${\cal M}_1$ and ${\cal M}_2$ display critical deviations in some cases. Therefore, TAVs of $\hat{\btheta}_3$ approximate the uncertainties more efficiently at finite samples in the nugget effect models. The TAVs of $\hat{\tau}^2$ follow the same pattern as the IQRs shown in Figure~\ref{taukk}. The TAV of $\hat{\tau}^2$ reduces as the field correlation and smoothness increase and drops to the lowest on a strong-correlated smooth field. It is difficult to draw conclusions on the asymptotic distributions of the MLEs because 1600 samples might not be sufficient for the MLEs to reach normality. The number of samples required for asymptotic normality requires further explorations.

\section{Tile Low-Rank Approximation}
The direct maximization of the log-likelihood function (\ref{e4}) is usually time-consuming and computationally inefficient.
\cite{abdulah2018parallel} implemented a tile-based parallel TLR approximation method for the MLE operation to compress off-diagonal tiles up to a pre-specified accuracy. First, the TLR decomposes the large covariance matrix into tiles of a given tile size, $\bSigma(\btheta)=\{{\textbf{D}(\btheta)}_{i,j}\}$, and maintains the diagonal tiles $\textbf{D}(\btheta)_{i,i}$ in dense forms. Singular value decomposition is performed on the individual off-diagonal tiles so that only a certain number of eigenvalues and eigenvectors can be preserved to reach the specified accuracy measured by the Frobenius norm, $\|\textbf{D}(\btheta)_{i,j}-\Tilde{\textbf{D}}(\btheta)_{i,j}\|_\text{F}< \epsilon$. The low-rank tile $\Tilde{\textbf{D}}(\btheta)_{i,j}$ occupies less memory space during computation, accelerating matrix operations. \cite{abdulah2018parallel} demonstrated that the TLR method could significantly reduce modeling time while maintaining decent estimation accuracy. Furthermore, \cite{hong2021efficiency} argued that TLR outperforms the Gaussian predictive process proposed by \cite{banerjee2008gaussian} and the composite likelihood method developed by \cite{vecchia1988estimation} and \cite{curriero1999composite}.


In this section, we evaluate the performance of the TLR approximation on the previously generated datasets from the three parameterizations under sparse covariance settings. We set the tile size to 400 and the accuracy level to $10^{-7}$. Therefore, the covariance matrix was divided into 16 tiles, each containing $400^2$ entries. We applied the TLR approximation to the 12 off-diagonal tiles while keeping the diagonal tiles dense. As reported in Section~\ref{acc}, the measures MSPE and MLOE are insensitive to the parameterizations. Hence, we only present the DRs between the sample means and true values, the AER, and MMOM on the exponential Gaussian fields (similar results were observed on Whittle Gaussian fields).
 \begin{figure}[t!]
\begin{subfigure}{0.5\textwidth}
\centering
\includegraphics[width=0.8\textwidth]{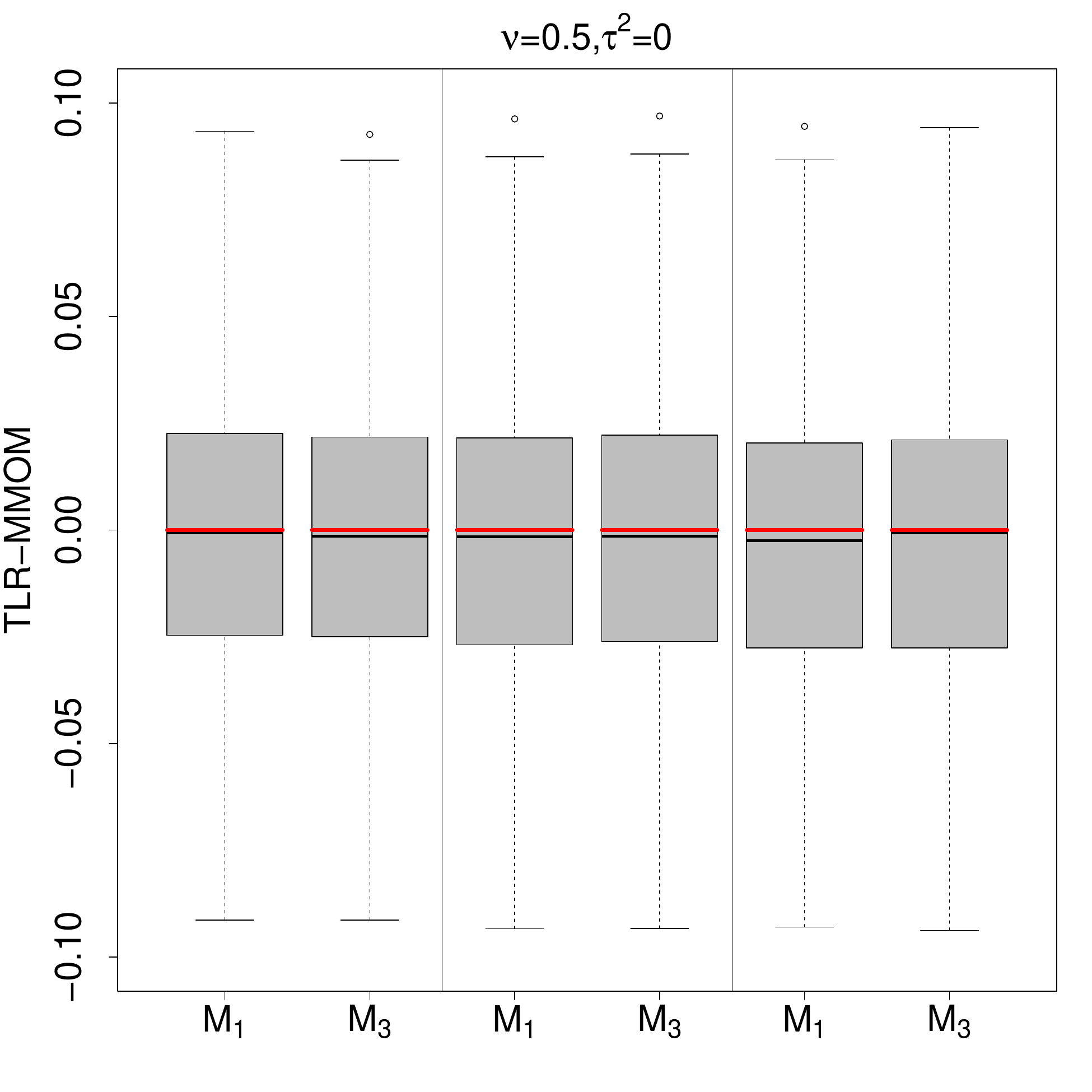}
\label{tlrM(a)}
\end{subfigure}
\begin{subfigure}{0.5\textwidth}
\centering
\includegraphics[width=0.8\textwidth]{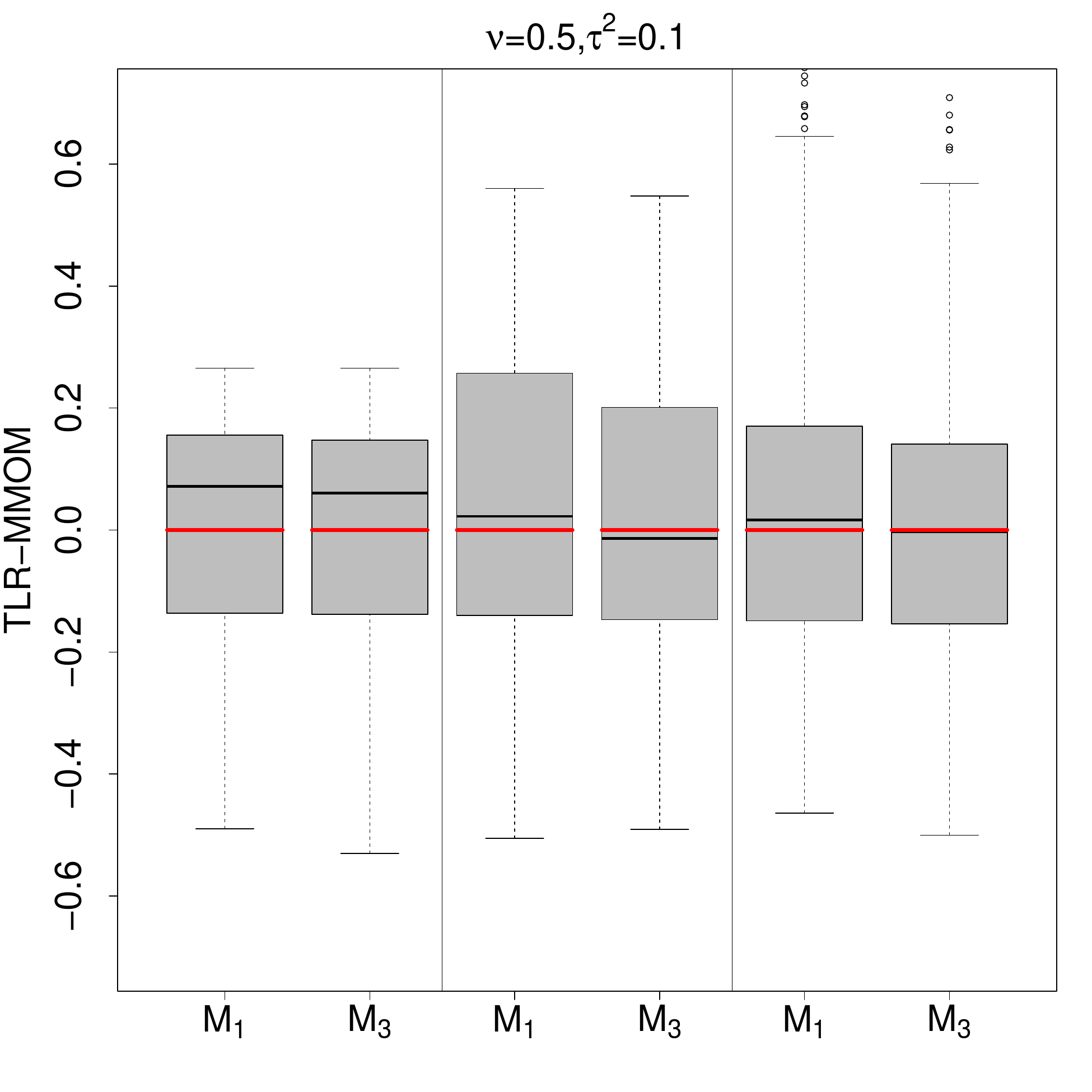}
\label{tlrM(b)}
\end{subfigure}
\caption{Boxplots of the MMOM using TLR approximation with 300 replicates for the non-nugget effect models ($\tau^2=0$) and nugget effect models ($\tau^2=0.1$) under weak (first block), medium (second block), and strong (third block) field correlations, separated by vertical lines in each sub-figure, on the exponential Gaussian fields ($\nu=0.5$). Red lines denote the expected MMOM value when the true covariance is applied.}
\label{tlrM}
\end{figure} 
Overall, the TLR method resembles the results produced by the exact method. In addition, ${\cal M}_1$ is slightly more accurate than ${\cal M}_3$ for the non-nugget effect models in terms of point estimations (Table~\ref{tlrT}, Supplementary Material) and AERs (Table~\ref{AER_tlr}, Supplementary Material) but is more vulnerable to the nugget effects. The MMOMs also reflect the same situation: the MMOMs are mostly comparable for the non-nugget effect models, whereas the MMOMs of ${\cal M}_1$ tend to deviate more from zero compared with MMOMs of ${\cal M}_3$ in the nugget effect models shown in Figure~\ref{tlrM}, indicating that using ${\cal M}_1$ in TLR settings is more prone to model specification errors. 

 Another crucial difference in parameterizations is that applying ${\cal M}_2$ in TLR settings is quite challenging. The problem is that  $\phi$ and $\alpha$ cannot land at small values in the search space because they will then characterize fields of significantly strong correlations, for which the resulting covariance matrix has a high possibility of being numerically singular if a loose accuracy level is applied to the off-diagonal blocks. Indeed, we encountered this problem when the accuracy level was $10^{-5}$, $10^{-7}$, and $10^{-9}$. 
 
 The results from \cite{abdulah2018parallel} demonstrated that a low accuracy level increases the modeling time and diminishes the purpose of the complexity reduction. Numerical singularity is not a critical issue for ${\cal M}_1$ and ${\cal M}_3$ because small range parameter values typically characterize weakly correlated fields. Although numerical singularity is not a significant concern, using the nugget effect models in TLR settings is preferred. Noise added to the diagonals can improve the matrix condition number and, thus, better avoid numerical singularity, enabling even looser accuracy conditions and further accelerating the modeling process.

\section{Application to Saudi Wind Speed Data}

\begin{figure}[b!]
  \centering
  \includegraphics[width=0.4\linewidth]{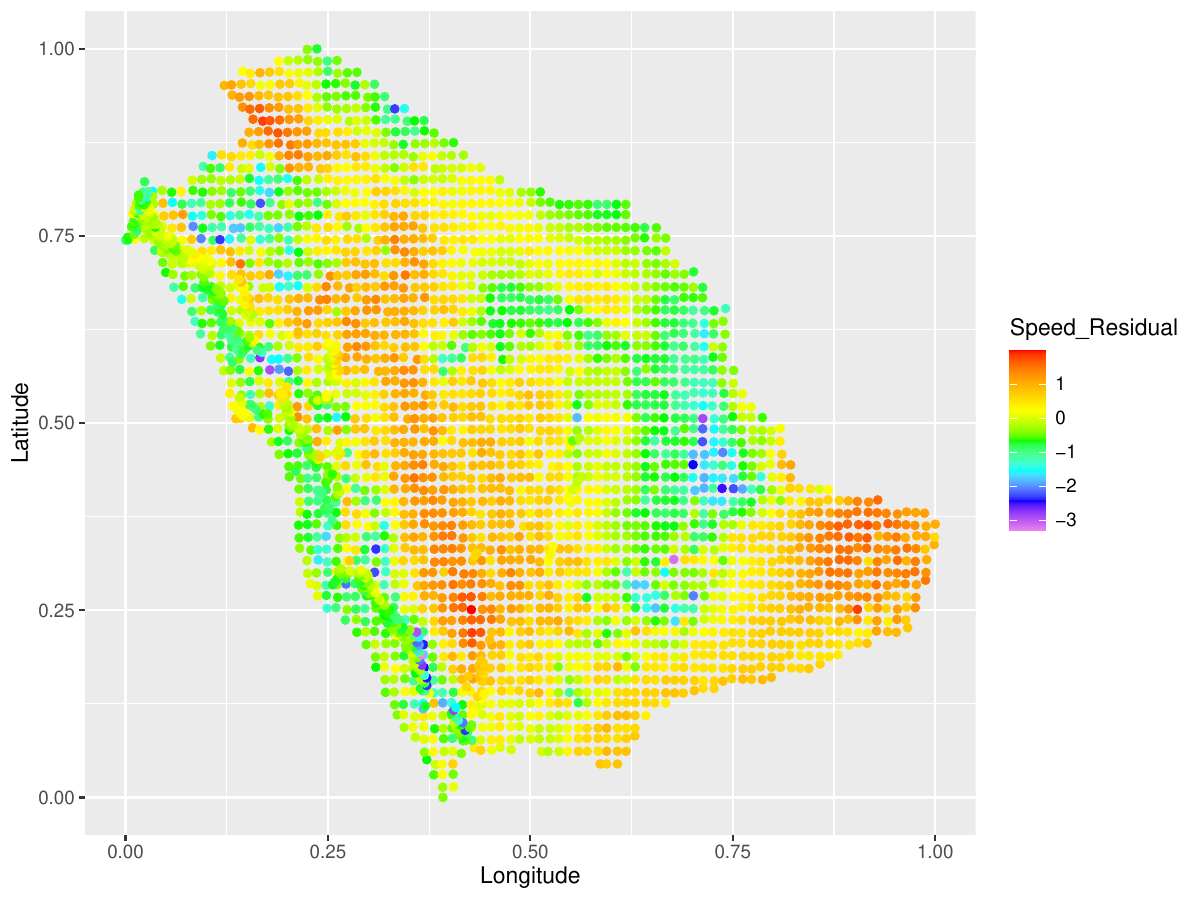}
\caption{Saudi wind speed residual data on January 1st, 2013, at 3,173 locations.}
\label{fig8}
\end{figure}

In this section, we apply ${\cal M}_1$, ${\cal M}_2$, and ${\cal M}_3$ to Saudi Arabia wind speed residual data from \cite{huang} visualized in Figure~\ref{fig8}, containing 3,173 observations across the entire country on January 1st, 2013. 
The locations are rescaled within $[0,1]^2$ to match the previous experiment settings. In addition, we randomly split the data into 10 sets of training and testing data, accounting for 80\% (2,538) and 20\% (635) of the entire dataset. Then, we measured the convergent iterations and asymptotic variance (AV) estimated from the testing data and MSPE on the training data. The tuning parameters are the same as in the simulation studies.
 \begin{table}[t!]
    \centering
    \caption{Comparisons of convergent iterations of the MLEs for ${\cal M}_1$, ${\cal M}_2$, and ${\cal M}_3$ using Saudi wind speed residual data under exact covariance estimation with and without nugget effects.}
    \begin{tabular}{|l|c|c|c|}
    \hline
    Model/parameterization     & ${\cal M}_1$ & ${\cal M}_2$ & ${\cal M}_3$ \\
    \hline
    Non-nugget     & 187.4 (64.5) & 314.7 (85.2) & 189.7 (99.9) \\
    \hline
    Nugget    & 308.9 (168.2) & 545.1 (224.1) & 296.3 (101.0) \\
    \hline
    \end{tabular}
    \label{real1}
\end{table}
\begin{table}[t!]
    \centering
    \caption{Comparisons of the MSPE for ${\cal M}_1$, ${\cal M}_2$, and ${\cal M}_3$ using Saudi wind speed residual data under an exact covariance estimation with and without nugget effects.}
    \begin{tabular}{|l|c|c|c|}
    \hline
    Model/parameterization     & ${\cal M}_1$ & ${\cal M}_2$ & ${\cal M}_3$ \\
    \hline
    Non-nugget     & 0.1115 (0.0166) & 0.1089 (0.0142) & 0.1076 (0.0156) \\
    \hline
    Nugget    & 0.1121 (0.0163) & 0.1118 (0.0166) & 0.1107 (0.0156) \\
    \hline
    \end{tabular}
    \label{real2}
\end{table}
 The numbers in Tables~\ref{real1} to \ref{real0} indicate the means computed across all 10 testing and training sets. The numbers in parentheses indicate their associated standard deviations. Tables~\ref{real1} to \ref{real0} demonstrate that many of the conclusions from numerical simulations remain true in practice. 
 \begin{table}[b!]
    \centering
    \caption{Comparisons of AVs for ${\cal M}_1$, ${\cal M}_2$, and ${\cal M}_3$ using Saudi wind speed residual data under an exact covariance estimation with and without nugget effects.  }
    \begin{tabular}{|c|c|c|c|c|}
    \hline
      
    ${\cal M}_1$ &  ${\sigma}^2$ & $\beta$ & $\nu$ & $\tau^2$ \\ 
      \hline
     &  0.1040 (0.1450) & 0.0546 (0.0873) & 0.000258 (0.0000215) & 0  \\
    \hline
  
   ${\cal M}_2$ &   $\phi$ & $\alpha$ & $\nu$ & $\tau^2$ \\ 
      \hline
     &  0.00649 (0.00221) & 5.7512 (0.9247) & 0.000281 (0.0000222) & 0 \\
    \hline
    
    ${\cal M}_3$ &  ${\sigma}^2$ & $\rho$ & $\nu$ & $\tau^2$\\ 
      \hline
    &    0.5272 (0.7103) & 0.5897 (0.8085) & 0.000249 (0.0000233) & 0  \\
    \hline
      \hline 
    ${\cal M}_1$ &  ${\sigma}^2$ & $\beta$ & $\nu$ & $\tau^2$ \\ 
      \hline
     &   0.1671 (0.3272) & 0.0413 (0.0892) & 0.00423 (0.00157) & 0.000123 (0.0000326)  \\
    \hline
    
    ${\cal M}_2$ &  $\phi$ & $\alpha$ & $\nu$ & $\tau^2$ \\ 
      \hline
     &   3.3101 (5.6246) & 11.1619 (2.2612) &  0.00557 (0.00157)  & 0.000119 (0.0000346) \\
    \hline
      
    ${\cal M}_3$ &  ${\sigma}^2$ & $\rho$ & $\nu$ & $\tau^2$\\ 
      \hline
     &   0.0683 (0.1209) & 0.0304 (0.0647) & 0.00478 (0.00154) &  0.000127 (0.0000440) \\
    \hline
    \end{tabular}
    \label{real0}
\end{table}
 First, we observe that ${\cal M}_2$ requires the most convergent iterations for both non-nugget and nugget effect models. In addition, ${\cal M}_1$ is the most efficient for the non-nugget effect models but becomes volatile and gets outperformed by ${\cal M}_3$ in the presence of nugget effects. Further, $(\hat{\sigma}^2,\hat{\rho})$ in ${\cal M}_3$ encounter inaccuracy issues in the non-nugget model, as the AVs significantly deviate from those of their counterparts in ${\cal M}_1$ with the nugget effect. The nugget effect $\tau^2$ and its AV are consistently estimated to be around 0.03 and 0.00012 across all parameterizations. Therefore, the overall nugget effect is limited. Hence, the AVs of $(\hat{\sigma}^2,\hat{\rho})$ should be similar in both scenarios like those of $(\hat{\sigma}^2,\hat{\beta})$ in ${\cal M}_1$.

Second, $\hat{\alpha}$ in both scenarios and $\hat{\phi}$ in the nugget effect model display much more significant AVs than the other estimates. Third, MSPE barely changes across differing parameterizations and increases under the exposure of nugget effects. In addition, the nugget effect, even minimally, significantly affects $(\hat{\phi},\hat{\nu})$ but has limited influences on the uncertainties of $(\hat{\sigma}^2,\hat{\beta},\hat{\rho},\hat{\alpha})$ in comparison. Last, the AV of $\hat{\beta}$ is slightly more significant but roughly identical to that of $\hat{\rho}$ in the nugget effect model because $\hat{\nu}$ is estimated to be around 0.3, which is a rough field with some minor nugget effects.

\section{Discussion}
In this study, we conducted a comprehensive comparison of the three most widely used parameterizations of the Matérn covariance function. We utilized both simulated and real data to investigate various aspects such as modeling efficiency and accuracy, the behavior of maximum likelihood estimators, and other crucial factors. Our analysis was made possible by utilizing the high-performance computing software, \textit{ExaGeoStat}, which allowed us to perform matrix operations in parallel and generate large-scale spatial data efficiently. We also employed benchmarking functions such as MMOM and the Fisher information matrix, which are sensitive to different parameterizations and can aid in selecting the optimal parameterization.

In summary, ${\cal M}_1$ is preferable over ${\cal M}_2$ and ${\cal M}_3$ in modeling speed. Further, ${\cal M}_1$ also provides decent and stable estimation accuracy in non-nugget effect models, offering lower uncertainty than ${\cal M}_2$ and avoiding the high-tolerance inaccuracy issues occurring in ${\cal M}_3$.  However, in the presence of nugget effects, ${\cal M}_1$ and ${\cal M}_2$ are more disturbed than ${\cal M}_3$ with biased estimates and apparent deviations between TAVs and SVs. Moreover, the resistance of ${\cal M}_3$ to the nugget effects also holds in TLR approximations. The TAV of MLEs for ${\cal M}_1$ is highly likely to be the lowest in non-nugget effect models, allowing for more informative confidence intervals. Prediction efficiency and the MSPE of the three parameterizations are comparable, but the MSPEs of ${\cal M}_1$ and ${\cal M}_3$ are the most trustworthy of the three. Moreover, ${\cal M}_1$ and ${\cal M}_3$ offer better parameter orthogonality than ${\cal M}_2$;  therefore, TAVs of $\hat{\nu}$ are closer to the SVs. Furthermore, parameters for ${\cal M}_1$ and ${\cal M}_3$ have more statistical-intuitive interpretations, in which $\sigma^2+\tau^2$=$\text{Var}\{Z(\bs)\}$ and $\beta$ and $\rho$ are direct measures of the dependence range. In addition, the TLR method can be applied to ${\cal M}_1$ and ${\cal M}_3$, whereas ${\cal M}_2$ often runs into numerical singularity issues. Lastly, per the small variances of MLEs for ${\cal M}_1$ and ${\cal M}_3$, it would be relatively safe to use a smaller search space in optimization, further enhancing modeling speed. However, such action is hazardous for ${\cal M}_2$, whose MLEs might take a large range of values and vary dramatically. Due to the above reasoning, we recommend ${\cal M}_1$ for the non-nugget effect model and ${\cal M}_3$ for the nugget effect model. If there is an absolute preference, the link functions presented in Table \ref{table1} and the Jacobian transformations of $\bI(\btheta)$ provided in the Supplementary Material can enable flexible transitions between any of them. As for the Bayesian context, we believe similar issues as in the frequentist context will likely arise due
to the likelihood part unless very strong priors are imposed. {\redcolor As for the Bayesian context, we believe that similar issues as in the frequentist context will likely arise due to the likelihood part unless very strong priors are imposed. A further analysis of the Mat\'ern parametrizations in the Bayesian context is worth exploring but beyond the scope of the current paper.}   
\section{Acknowledgement}
We would like to thank King Abdullah University
of Science and Technology (KAUST) and SDAIA-KAUST Center of Excellence in Data Science and Artificial Intelligence (SDAIA-KAUST AI) for supporting the research in this work.

\bibliographystyle{apalike}
\bibliography{reference}

\begin{thebibliography}{}

\bibitem[Abdulah et~al., 2021]{abdulah2021accelerating}
Abdulah, S., Cao, Q., Pei, Y., Bosilca, G., Dongarra, J., Genton, M.~G., Keyes,
  D.~E., Ltaief, H., and Sun, Y. (2021).
\newblock Accelerating geostatistical modeling and prediction with
  mixed-precision computations: A high-productivity approach with {PaRSEC}.
\newblock {\em IEEE Transactions on Parallel and Distributed Systems},
  33(4):964--976.

\bibitem[Abdulah et~al., 2023]{abdulah2019exageostatr}
Abdulah, S., Li, Y., Cao, J., Ltaief, H., Keyes, D.~E., Genton, M.~G., and Sun,
  Y. (2023).
\newblock Large-scale environmental data science with {ExaGeoStatR}.
\newblock {\em Environmetrics}, 34(1):e2770.

\bibitem[Abdulah et~al., 2018a]{abdulah2018exageostat}
Abdulah, S., Ltaief, H., Sun, Y., Genton, M.~G., and Keyes, D.~E. (2018a).
\newblock {ExaGeoStat}: A high performance unified software for geostatistics
  on manycore systems.
\newblock {\em IEEE Transactions on Parallel and Distributed Systems},
  29(12):2771--2784.

\bibitem[Abdulah et~al., 2018b]{abdulah2018parallel}
Abdulah, S., Ltaief, H., Sun, Y., Genton, M.~G., and Keyes, D.~E. (2018b).
\newblock Parallel approximation of the maximum likelihood estimation for the
  prediction of large-scale geostatistics simulations.
\newblock In {\em 2018 IEEE International Conference on Cluster Computing
  (CLUSTER)}, pages 98--108.

\bibitem[Abt and Welch, 1998]{abt1998fisher}
Abt, M. and Welch, W.~J. (1998).
\newblock Fisher information and maximum-likelihood estimation of covariance
  parameters in gaussian stochastic processes.
\newblock {\em Canadian Journal of Statistics}, 26(1):127--137.

\bibitem[Banerjee et~al., 2008]{banerjee2008gaussian}
Banerjee, S., Gelfand, A.~E., Finley, A.~O., and Sang, H. (2008).
\newblock Gaussian predictive process models for large spatial data sets.
\newblock {\em Journal of the Royal Statistical Society: Series B (Statistical
  Methodology)}, 70(4):825--848.

\bibitem[Cao et~al., 2022]{cao2022reshaping}
Cao, Q., Abdulah, S., Alomairy, R., Pei, Y., Nag, P., Bosilca, G., Dongarra,
  J., Genton, M.~G., Keyes, D.~E., Ltaief, H., et~al. (2022).
\newblock Reshaping geostatistical modeling and prediction for extreme-scale
  environmental applications.
\newblock In {\em 2022 SC22: International Conference for High Performance
  Computing, Networking, Storage and Analysis (SC)}, pages 13--24. IEEE
  Computer Society.

\bibitem[Cox and Reid, 1987]{cox1987parameter}
Cox, D.~R. and Reid, N. (1987).
\newblock Parameter orthogonality and approximate conditional inference.
\newblock {\em Journal of the Royal Statistical Society: Series B
  (Methodological)}, 49(1):1--18.

\bibitem[Curriero and Lele, 1999]{curriero1999composite}
Curriero, F.~C. and Lele, S. (1999).
\newblock A composite likelihood approach to semivariogram estimation.
\newblock {\em Journal of Agricultural, Biological, and Environmental
  Statistics}, 4(1):9--28.

\bibitem[De~Oliveira and Han, 2022]{de2022information}
De~Oliveira, V. and Han, Z. (2022).
\newblock On information about covariance parameters in {G}aussian {M}at{\'e}rn
  random fields.
\newblock {\em Journal of Agricultural, Biological and Environmental
  Statistics}, 27(4):690--712.

\bibitem[Geoga et~al., 2020]{geoga2020scalable}
Geoga, C.~J., Anitescu, M., and Stein, M.~L. (2020).
\newblock Scalable gaussian process computations using hierarchical matrices.
\newblock {\em Journal of Computational and Graphical Statistics},
  29(2):227--237.

\bibitem[Geoga et~al., 2023]{geoga2022fitting}
Geoga, C.~J., Marin, O., Schanen, M., and Stein, M.~L. (2023).
\newblock Fitting mat{\'e}rn smoothness parameters using automatic
  differentiation.
\newblock {\em Statistics and Computing}, 33(2):48.

\bibitem[Gonz{\'a}lez-Santander, 2018]{gonzalez2018closed}
Gonz{\'a}lez-Santander, J. (2018).
\newblock Closed-form expressions for derivatives of {B}essel functions with
  respect to the order.
\newblock {\em Journal of Mathematical Analysis and Applications},
  466(1):1060--1081.

\bibitem[Gough, 2009]{gough2009gnu}
Gough, B. (2009).
\newblock {\em GNU scientific library reference manual}.
\newblock Network Theory Ltd.

\bibitem[Handcock and Stein, 1993]{handcock1993bayesian}
Handcock, M.~S. and Stein, M.~L. (1993).
\newblock A {B}ayesian analysis of kriging.
\newblock {\em Technometrics}, 35(4):403--410.

\bibitem[Handcock and Wallis, 1994]{handcock1994approach}
Handcock, M.~S. and Wallis, J.~R. (1994).
\newblock An approach to statistical spatial-temporal modeling of
  meteorological fields.
\newblock {\em Journal of the American Statistical Association},
  89(426):368--378.

\bibitem[Hong et~al., 2021]{hong2021efficiency}
Hong, Y., Abdulah, S., Genton, M.~G., and Sun, Y. (2021).
\newblock Efficiency assessment of approximated spatial predictions for large
  datasets.
\newblock {\em Spatial Statistics}, 43:100517.

\bibitem[Huang et~al., 2022]{huang}
Huang, H., Castruccio, S., and Genton, M.~G. (2022).
\newblock Forecasting high-frequency spatio-temporal wind power with
  dimensionally reduced echo state networks.
\newblock {\em Journal of the Royal Statistical Society: Series C (Applied
  Statistics)}, 71(2):449--466.

\bibitem[Li and Vu, 2013]{li2013identification}
Li, P. and Vu, Q.~D. (2013).
\newblock Identification of parameter correlations for parameter estimation in
  dynamic biological models.
\newblock {\em BMC Systems Biology}, 7(1):1--12.

\bibitem[Loh, 2005]{loh2005fixed}
Loh, W.-L. (2005).
\newblock Fixed-domain asymptotics for a subclass of {M}at{\'e}rn type
  {G}aussian random fields.
\newblock {\em The Annals of Statistics}, 33(5):2344--2394.

\bibitem[Mardia and Marshall, 1984]{mardia1984maximum}
Mardia, K.~V. and Marshall, R.~J. (1984).
\newblock Maximum likelihood estimation of models for residual covariance in
  spatial regression.
\newblock {\em Biometrika}, 71(1):135--146.

\bibitem[Mat{\'e}rn, 1960]{Matrn1960SpatialV}
Mat{\'e}rn, B. (1960).
\newblock Spatial {V}ariation: Stochastic models and their application to some
  problems in forest surveys and other sampling investigations.
\newblock {\em Stockholm: {M}edd. Statens {S}kogsforskningsinstitut}, 49(5).

\bibitem[Myung et~al., 2005]{myung2005information}
Myung, J.~I., Navarro, D.~J., et~al. (2005).
\newblock Information matrix.
\newblock {\em Encyclopedia of Statistics in Behavioral Science}, 2:923--924.

\bibitem[Nychka et~al., 2021]{fields}
Nychka, D., Furrer, R., Paige, J., and Sain, S. (2021).
\newblock fields: Tools for spatial data.
\newblock R package version 14.1.

\bibitem[Olver et~al., 2010]{olver2010nist}
Olver, F.~W., Lozier, D.~W., Boisvert, R.~F., and Clark, C.~W. (2010).
\newblock {\em NIST {H}andbook of {M}athematical {F}unctions {H}ardback and
  CD-ROM}.
\newblock Cambridge {U}niversity {P}ress.

\bibitem[Petersen et~al., 2008]{petersen2008matrix}
Petersen, K.~B., Pedersen, M.~S., et~al. (2008).
\newblock The matrix cookbook.
\newblock {\em Technical University of Denmark}, 7(15):510.

\bibitem[Powell, 2009]{powell2009bobyqa}
Powell, M.~J. (2009).
\newblock The {BOBYQA} algorithm for bound constrained optimization without
  derivatives.
\newblock {\em Cambridge NA Report NA2009/06, University of Cambridge,
  Cambridge}, 26.

\bibitem[Ribeiro and Diggle, 2001]{geor}
Ribeiro, P.~J. and Diggle, P.~J. (2001).
\newblock geo{R}: a package for geostatistical analysis.
\newblock {\em R-NEWS}, 1(2):14--18.
\newblock ISSN 1609-3631.

\bibitem[Schlather et~al., 2019]{randomfield}
Schlather, M., Malinowski, A., Oesting, M., Boecker, D., Strokorb, K., Engelke,
  S., Martini, J., Ballani, F., Moreva, O., Auel, J., Menck, P.~J., Gross, S.,
  Ober, U., Ribeiro, P., Singleton, R., Pfaff, B., and {R Core Team} (2019).
\newblock {\em {RandomFields}: Simulation and Analysis of Random Fields}.
\newblock R package version 3.3.1.

\bibitem[Stein, 1999]{stein1999interpolation}
Stein, M.~L. (1999).
\newblock {\em Interpolation of {S}patial {D}ata: Some {T}heory for {K}riging}.
\newblock Springer Science \& Business Media.

\bibitem[Stein, 2014]{stein2014limitations}
Stein, M.~L. (2014).
\newblock Limitations on low rank approximations for covariance matrices of
  spatial data.
\newblock {\em Spatial Statistics}, 8:1--19.

\bibitem[Vecchia, 1988]{vecchia1988estimation}
Vecchia, A.~V. (1988).
\newblock Estimation and model identification for continuous spatial processes.
\newblock {\em Journal of the Royal Statistical Society: Series B
  (Methodological)}, 50(2):297--312.

\bibitem[Willmot, 1988]{willmot1988parameter}
Willmot, G.~E. (1988).
\newblock Parameter orthogonality for a family of discrete distributions.
\newblock {\em Journal of the American Statistical Association},
  83(402):517--521.

\bibitem[Zimmerman and Cressie, 1992]{zimmerman1992mean}
Zimmerman, D.~L. and Cressie, N. (1992).
\newblock Mean squared prediction error in the spatial linear model with
  estimated covariance parameters.
\newblock {\em Annals of the Institute of Statistical Mathematics},
  44(1):27--43.

\end{thebibliography}

\newpage
\setcounter{page}{1}
\section*{Supplementary Material to ``Which Parameterization of the Matérn Covariance Function?'' by Kesen Wang, Sameh Abdulah, Ying Sun, Marc G. Genton}
\setcounter{figure}{0}
\setcounter{table}{0}
\renewcommand{\thefigure}{S\arabic{figure}}
\renewcommand{\thetable}{S\arabic{table}}
\subsection*{S1. Derivatives of the Three Matérn Variants }
\begin{align}
    \frac{\partial {\cal M}_1(h;\boldsymbol{\theta}_1)}{\partial\sigma^2}=&\frac{1}{2^{\nu-1}\Gamma(\nu)}\left( \frac{h}{\beta} \right )^\nu {\cal K}_\nu \left( \frac{h}{\beta}\right)
    \label{e6} \tag{S1}
    \\
   \nonumber \\
\frac{\partial {\cal M}_1(h;\boldsymbol{\theta}_1)}{\partial\beta}=& \frac{\sigma^2}{2^{\nu-1}\Gamma(\nu)}\left\{-\frac{\nu}{\beta}\left(\frac{h}{\beta}\right)^\nu {\cal K}_\nu\left(\frac{h}{\beta}\right)-\left(\frac{h}{\beta}\right)^\nu {\cal K}^\prime_\nu\left(\frac{h}{\beta}\right)\frac{h}{\beta^2}\right\}
\label{e7} \tag{S2}\\
\nonumber\\
\frac{\partial {\cal M}_1(h;\boldsymbol{\theta}_1)}{\partial\nu}= & \sigma^2\biggl(-\log(2) 2^{1-\nu}\Gamma^{-1}(\nu)\left(\frac{h}{\beta}\right)^{\nu} {\cal K}_\nu\left(\frac{h}{\beta}\right)
\nonumber \\
&+2^{1-\nu}\biggl[-\Gamma^{-1}(\nu)\Psi(\nu)\left(\frac{h}{\beta}\right)^\nu
{\cal K}_\nu\left(\frac{h}{\beta}\right)  \nonumber\\ & +\Gamma^{-1}(\nu)\biggl\{\left(\frac{h}{\beta}\right)^\nu\log\left(\frac{h}{\beta}\right) {\cal K}_\nu\left(\frac{h}{\beta}\right)+\left(\frac{h}{\beta}\right)^\nu {\cal K}_{\nu^\prime}\left(\frac{h}{\beta}\right) \biggr\} \biggr] \biggr)
\label{e8} \tag{S3}
\\
\nonumber\\
    \frac{\partial {\cal M}_2(h;\boldsymbol{\theta}_2)}{\partial \phi}=&\frac{\pi^{1/2}}{2^{\nu-1}\Gamma(\nu+\frac{1}{2})\alpha^{2\nu}}(\alpha h)^\nu {\cal K}_\nu(\alpha h)
    \label{e9} \tag{S4}
    \\
    \nonumber \\
\frac{\partial {\cal M}_2(h;\boldsymbol{\theta}_2)}{\partial \alpha }=&\frac{\pi^{1/2}\phi h^\nu}{2^{\nu-1}\Gamma \left(\nu+\frac{1}{2} \right)}\left\{-\nu \alpha^{\nu-1}{\cal K}_\nu(\alpha h) + \alpha^{-\nu}{\cal K}^\prime_\nu(\alpha h)h\right\}
    \label{e10} \tag{S5}
\\
\nonumber\\
    \frac{\partial {\cal M}_2(h;\boldsymbol{\theta}_2)}{\partial \nu}=&2\pi^{1/2}\phi\biggl[\log\left(\frac{h}{2\alpha}\right)\left(\frac{h}{2\alpha}\right)^\nu \Gamma^{-1}\left(\nu+\frac{1}{2}\right){\cal K}_\nu\left(\alpha h\right) \nonumber \\&
    +\left(\frac{h}{2\alpha}\right)^\nu \biggl\{-\Gamma\left(\nu+\frac{1}{2}\right)^{-2}\Gamma^\prime\left(\nu+\frac{1}{2}\right){\cal K}_\nu(\alpha h)+\Gamma\left(\nu+\frac{1}{2}\right)^{-1}{\cal K}_{\nu^\prime}(\alpha h)\biggr\}\biggr]
    \label{e11} \tag{S6}\\
    \nonumber \\
    \frac{\partial {\cal M}_3(h;\boldsymbol{\theta}_3)}{\partial \sigma^2}=&\frac{1}{2^{\nu-1}\Gamma(\nu)}\left(\frac{2\nu^{1/2}h}{\rho}\right)^\nu {\cal K}_\nu\left(\frac{2\nu^{1/2}h}{\rho}\right)
    \label{e12} \tag{S7}\\
    \nonumber \\
    \frac{\partial {\cal M}_3(h;\boldsymbol{\theta}_3)}{\partial \rho}=&\frac{\sigma^2(2\nu^{1/2}h)^\nu}{2^{\nu-1}\Gamma(\nu)}\left\{-\nu\rho^{-\nu-1}{\cal K}_\nu\left(\frac{2\nu^{1/2}h}{\rho}\right)+\rho^{-\nu}{\cal K}_{\nu}^\prime\left(\frac{2\nu^{1/2}h}{\rho}\right) \frac{-2\nu^{1/2}h}{\rho^2}\right\}
    \label{e13} \tag{S8}
\end{align}
    \newpage
\begin{align}
    \frac{\partial {\cal M}_3(h;\boldsymbol{\theta}_3)}{\partial \nu}=& 2\sigma^2\biggl(-\Gamma(\nu)^{-2}\Gamma^\prime(\nu)\left(\frac{\nu^{1/2}h}{\rho}\right)^\nu {\cal K}_\nu\left(\frac{2\nu^{1/2}h}{\rho}\right) \nonumber\\
    &+\Gamma(\nu)^{-1}\biggl[\left(\frac{\nu^{1/2}h}{\rho}\right)^\nu\left\{\frac{1}{2}\log(\nu)  + \frac{1}{2} + \log\left(\frac{h}{\rho}\right)\right\}{\cal K}_\nu\left(\frac{2\nu^{1/2}h}{\rho}\right) \nonumber\\
    &+\left(\frac{\nu^{1/2}h}{\rho}\right)^\nu\biggl\{{\cal K}_\nu^\prime\left(\frac{2\nu^{1/2}h}{\rho}\right) \frac{\nu^{-1/2}h}{\rho}+{\cal K}_{\nu^\prime}\left(\frac{2\nu^{1/2}h}{\rho}\right)\biggr\}\biggr]\biggr)
    \label{e14} \tag{S9}\\\nonumber
    \\
    \frac{\partial{\cal M}(h;\btheta)}{\partial \tau^2}  = & \mathbbm{1}_{h=0}
    \label{e15} \tag{S10}
\end{align}
Equations \ref{e12}-\ref{e14} are derived from \cite{de2022information}. The rest of the equations are computed in a similar manner. 
\subsection*{S2. Jacobians of Transformations}
To transform from ${\cal M}_1$ to ${\cal M}_2$, we have  
\begin{center}
\(\textbf{J}_{1\rightarrow 2}\)=
$\begin{pmatrix}
\frac{\partial \sigma^2}{\partial \phi} & \frac{\partial \sigma^2}{\partial \alpha}  & \frac{\partial \sigma^2}{\partial \nu} & 0\\
0& \frac{\partial \beta}{\partial \alpha} & 0 & 0\\
0 & 0 & 1 & 0 \\
0 & 0 & 0 & 1
\end{pmatrix},$
\end{center}
where 
\begin{align*}
\frac{\partial \sigma^2}{\partial \phi}=&\frac{\pi^{1/2}\Gamma(\nu)}{\Gamma(\nu+\frac{1}{2})\alpha^{2\nu}},
\\
\\
\frac{\partial \sigma^2}{\partial \alpha}=&-\frac{2\pi^{1/2}\phi\Gamma(\nu)\nu}{\Gamma(\nu+\frac{1}{2})\alpha^{2\nu+1}},
\\
\\
\frac{\partial \sigma^2}{\partial \nu}=&\pi^{1/2}\phi\biggl[\Gamma^\prime(\nu)\Gamma\left(\nu+\frac{1}{2}\right)^{-1}\alpha^{-2\nu}+\Gamma(\nu)\biggl\{-\Gamma\left(\nu+\frac{1}{2}\right)^{-2}\\&
\times \Gamma^\prime\left(\nu+\frac{1}{2}\right)\alpha^{-2\nu}-2\Gamma\left(\nu+\frac{1}{2}\right)^{-1}\alpha^{-2\nu}\log(\alpha)\biggr\}\biggr],\\
\frac{\partial \beta}{\partial \alpha}=&-\frac{1}{\alpha^2}.
\end{align*}
To transform from ${\cal M}_1$ to ${\cal M}_3$, we have the following:
\begin{center}
 \(\textbf{J}_{1\rightarrow 3}\)=
$\begin{pmatrix}
1 & 0  & 0 & 0\\
0& \frac{\partial \beta}{\partial \rho} & \frac{\partial \beta}{\partial \nu} & 0\\
0 & 0 & 1 & 0\\
0 & 0 & 0 & 1
\end{pmatrix},$   
\end{center}
where 
\begin{align*}
\frac{\partial \beta}{{\redcolor\partial \rho}}=&\frac{1}{2\nu^{1/2}},\\
\\
\frac{\partial \beta}{\partial \nu}=&-\frac{\rho}{4\nu^{3/2}}.
\end{align*} 
Lastly, for transformations between ${\cal M}_1$ and ${\cal M}_3$, we can compute
\begin{center}
 \(\textbf{J}_{2\rightarrow 3}\)=
$\begin{pmatrix}
\frac{\partial \phi}{\partial \sigma^2} & \frac{\partial \phi}{\partial \rho}  & \frac{\partial \phi}{\partial \nu} & 0\\
0& \frac{\partial \alpha}{\partial \rho} & \frac{\partial \alpha}{\partial \nu} & 0\\
0 & 0 & 1 & 0 \\
0 & 0 & 0 & 1 
\end{pmatrix},$
\end{center}
where
\begin{align*}
    \frac{\partial \phi}{\partial \sigma^2}=&\frac{\Gamma(\nu+\frac{1}{2})(2\nu^{1/2})^{2\nu}}{\pi^{1/2}\Gamma(\nu)\rho^{2\nu}},\\
    \\
    \frac{\partial \phi}{\partial \rho}=&-\frac{2\sigma^2\Gamma(\nu+\frac{1}{2})(2\nu^{1/2})^{2\nu}\nu}{\pi^{1/2}\Gamma(\nu)\rho^{2\nu+1}},\\
    \\
\frac{\partial \phi}{\partial \nu}  =&\frac{\sigma^2}{\pi^{1/2}}\biggl(-\Gamma(\nu)^{-2}\Gamma^\prime(\nu)\Gamma\left(\nu+\frac{1}{2}\right)\left(\frac{2\nu^{1/2}}{\rho}\right)^{2\nu}+\Gamma(\nu)^{-1}\biggr[\Gamma^\prime\left(\nu+\frac{1}{2}\right)\left(\frac{2\nu^{1/2}}{\rho}\right)^{2\nu}\\&+\Gamma\left(\nu+\frac{1}{2}\right)\left(\frac{2\nu^{1/2}}{\rho}\right)^{2\nu}\left\{\frac{1}{2}+\log(2\nu)-2\log(\rho)\right\}\biggr]\biggr),\\
\\
\frac{\partial \alpha}{\partial \rho}=&-\frac{2\nu^{1/2}}{\rho^2},\\
\\
\frac{\partial \alpha}{\partial \nu}=&\frac{\nu^{-1/2}}{\rho}.
\end{align*}
\subsection*{S3. Additional Simulation Results}
 Due to length limitations, we provide results used in Section~4, 5, and 6. Numbers highlighted in red denote large values to take notice.
\begin{figure}[H]
\begin{subfigure}{0.33\linewidth}
  \centering
  \includegraphics[width=1\textwidth,]{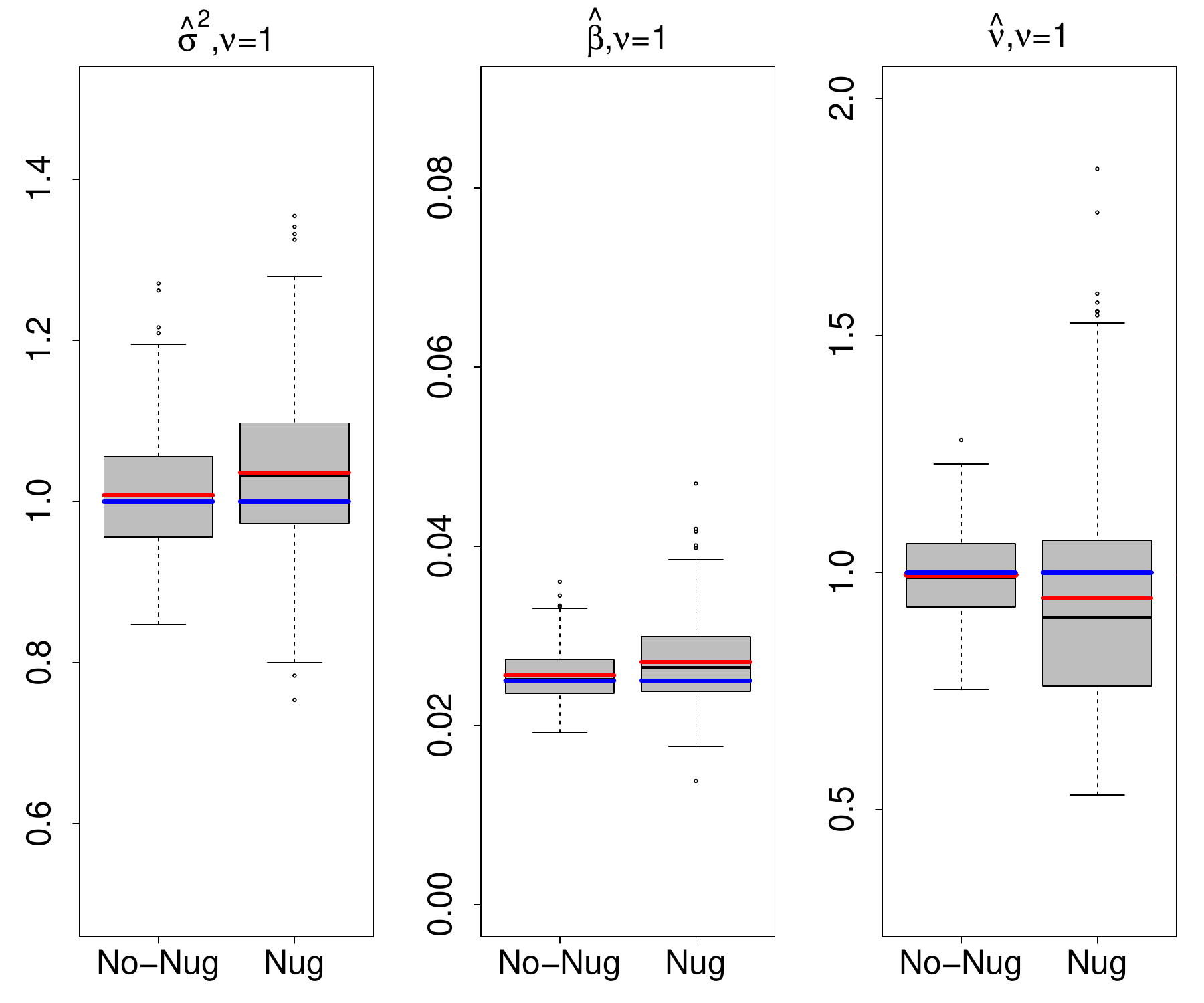}
  \caption{$\hat{\btheta}_1,{\cal M}_1$, weak correlation. }
  \label{fig9(a)}
  \vspace{5mm}
\end{subfigure}
\begin{subfigure}{0.33\linewidth}
  \centering
  \includegraphics[width=1\textwidth,]{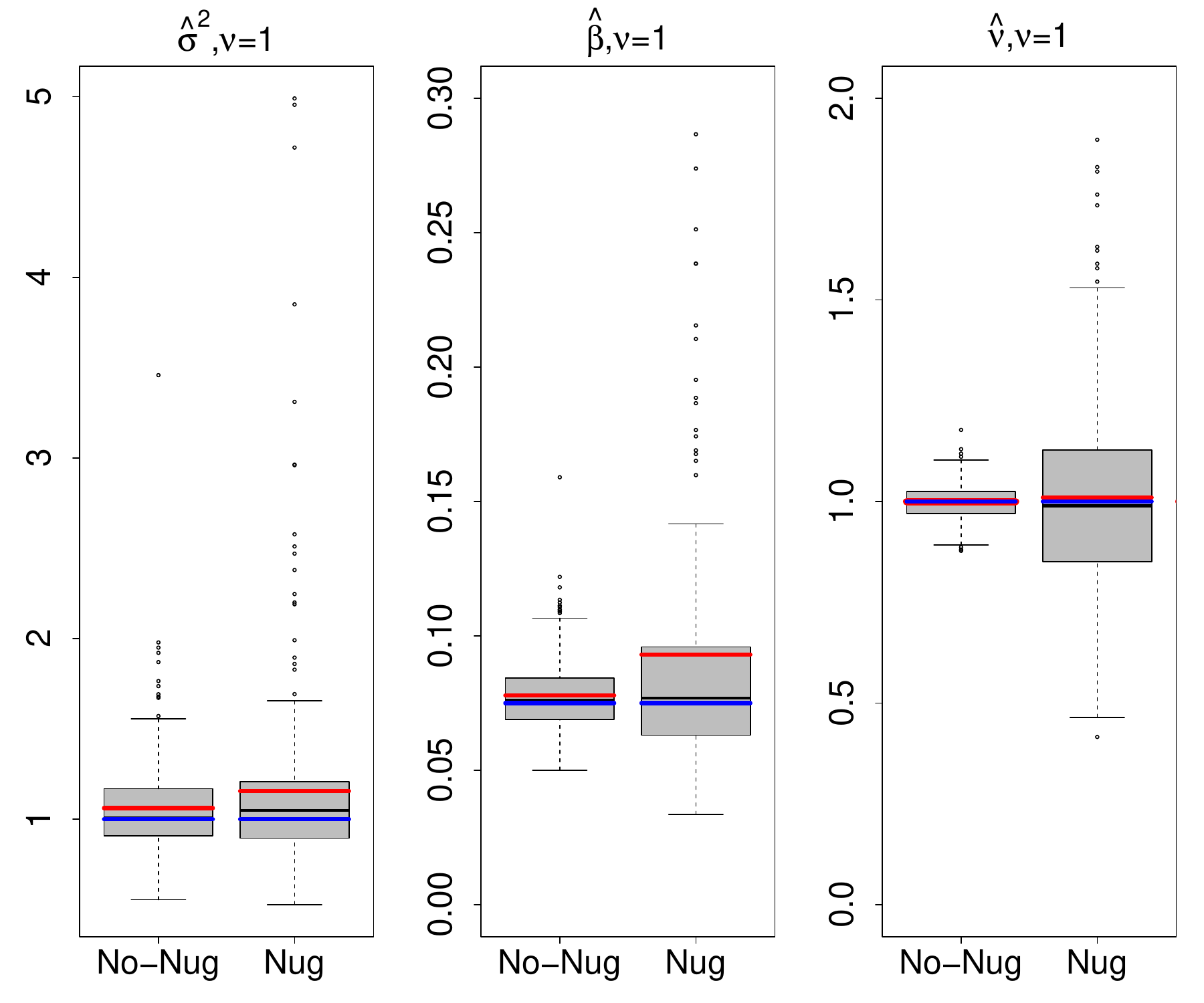}
  \caption{$\hat{\btheta}_1,{\cal M}_1$, medium correlation. }
  \label{fig9(b)}
  \vspace{5mm}
\end{subfigure}
\begin{subfigure}{0.33\linewidth}
  \centering
  \includegraphics[width=1\textwidth,]{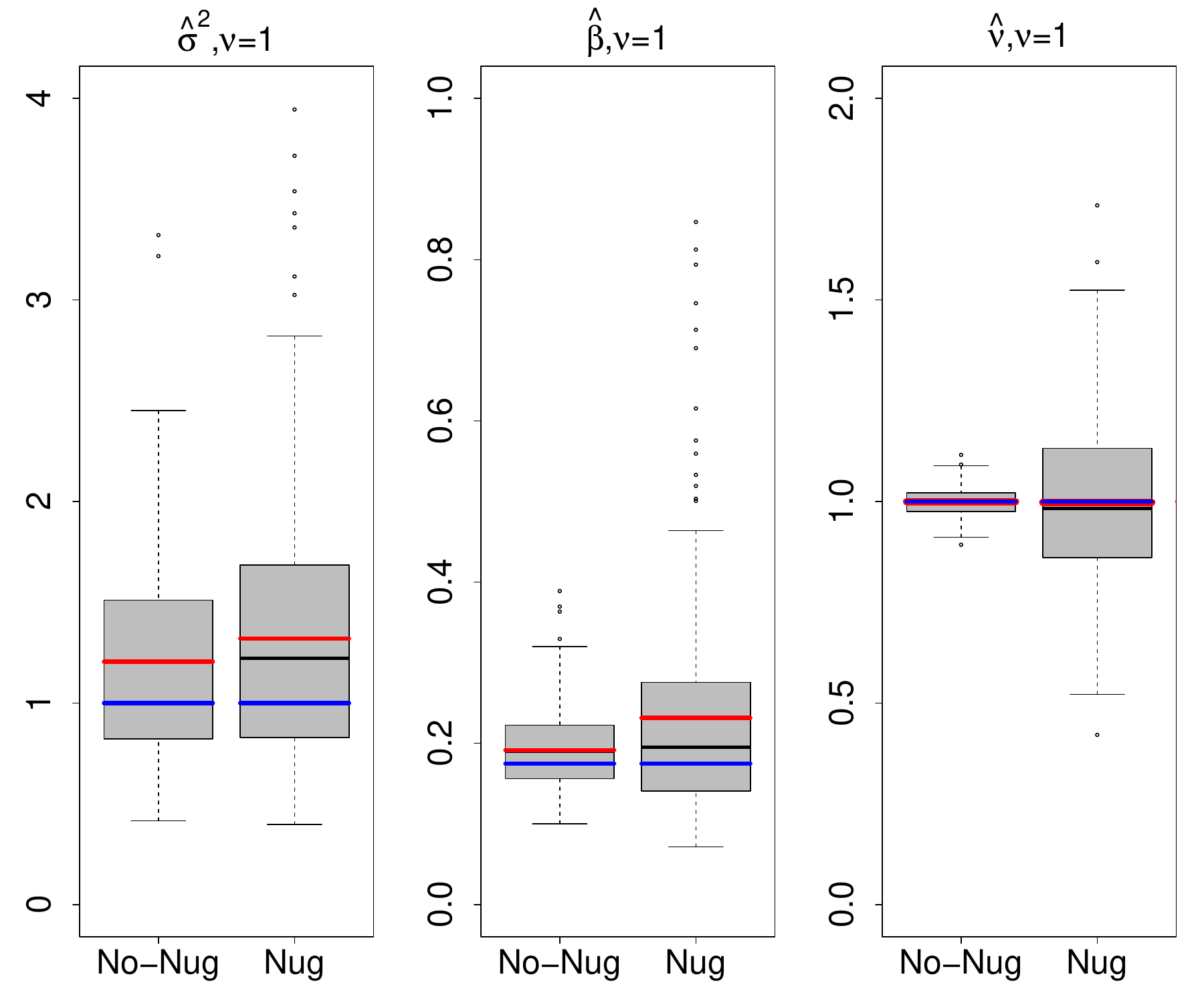}
  \caption{$\hat{\btheta}_1,{\cal M}_1$, strong correlation. }
  \label{fig9(c)}
  \vspace{5mm}
\end{subfigure}
\begin{subfigure}{0.33\linewidth}
  \centering
  \includegraphics[width=1\textwidth,]{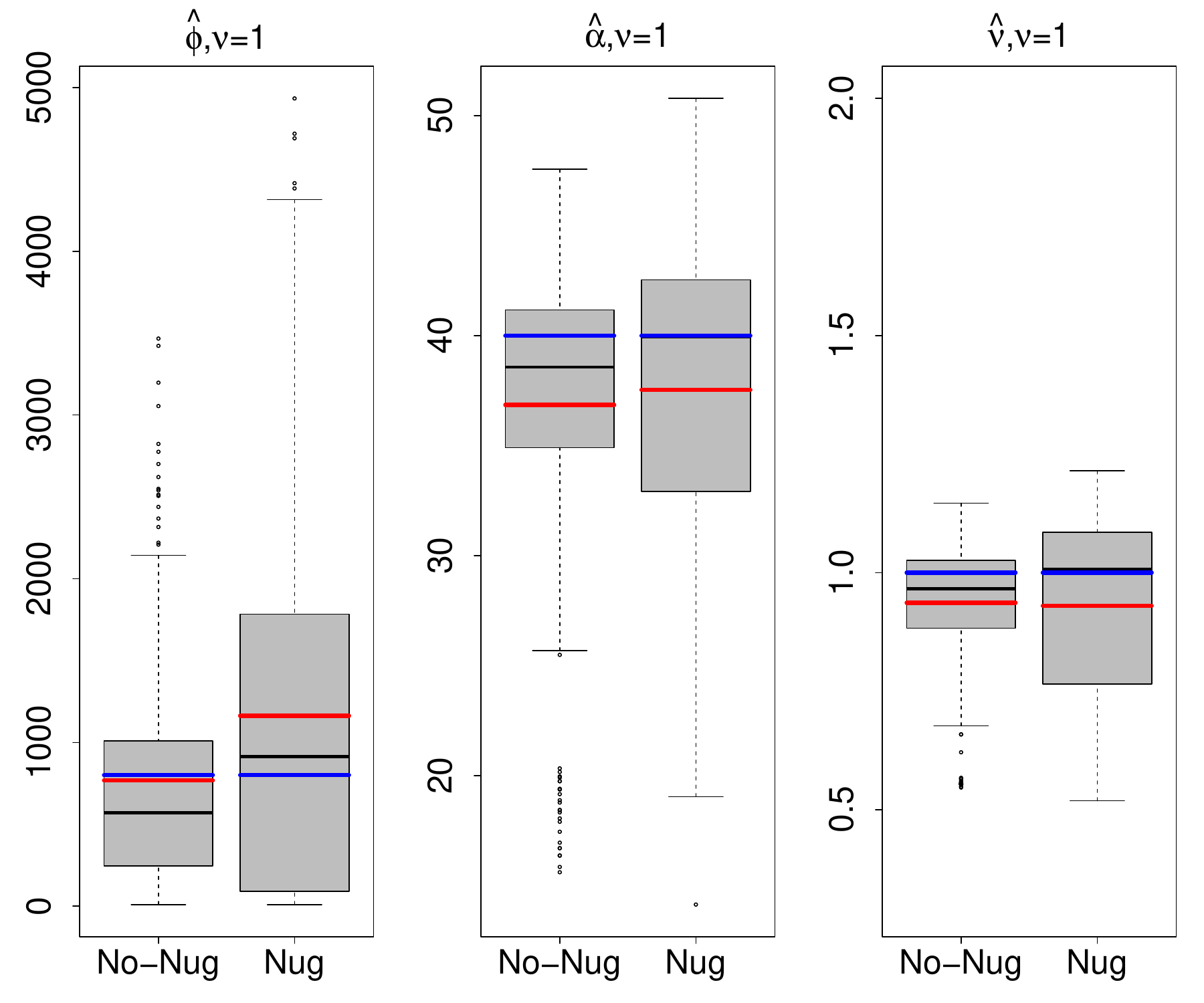}
  \caption{$\hat{\btheta}_2,{\cal M}_2$, weak correlation. }
  \label{fig9(d)}
  \vspace{5mm}
\end{subfigure}
\begin{subfigure}{0.33\linewidth}
  \centering
  \includegraphics[width=1\textwidth,]{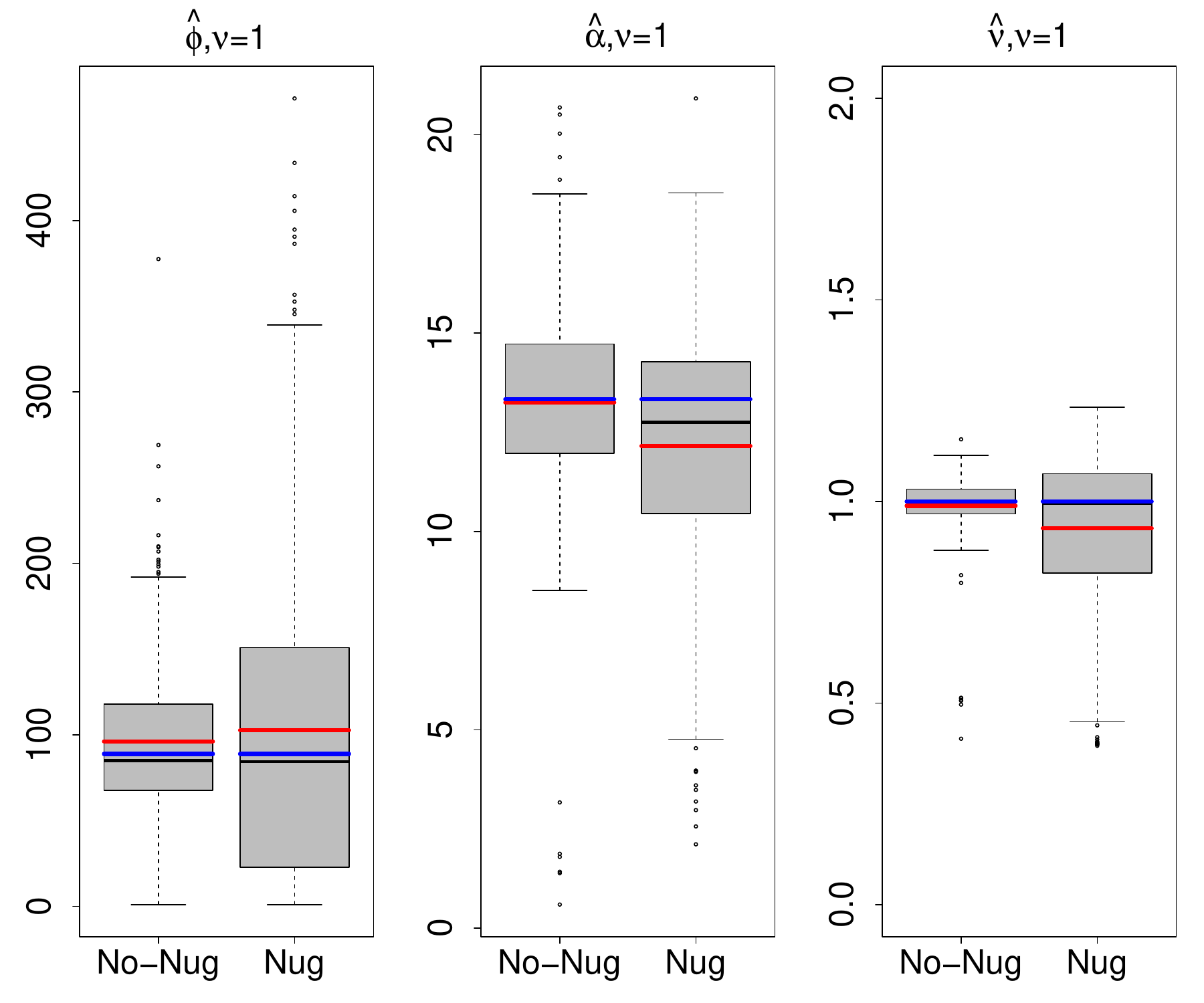}
  \caption{$\hat{\btheta}_2,{\cal M}_2$, medium correlation. }
  \label{fig9(e)}
  \vspace{5mm}
\end{subfigure}
\begin{subfigure}{0.33\linewidth}
  \centering
  \includegraphics[width=1\textwidth,]{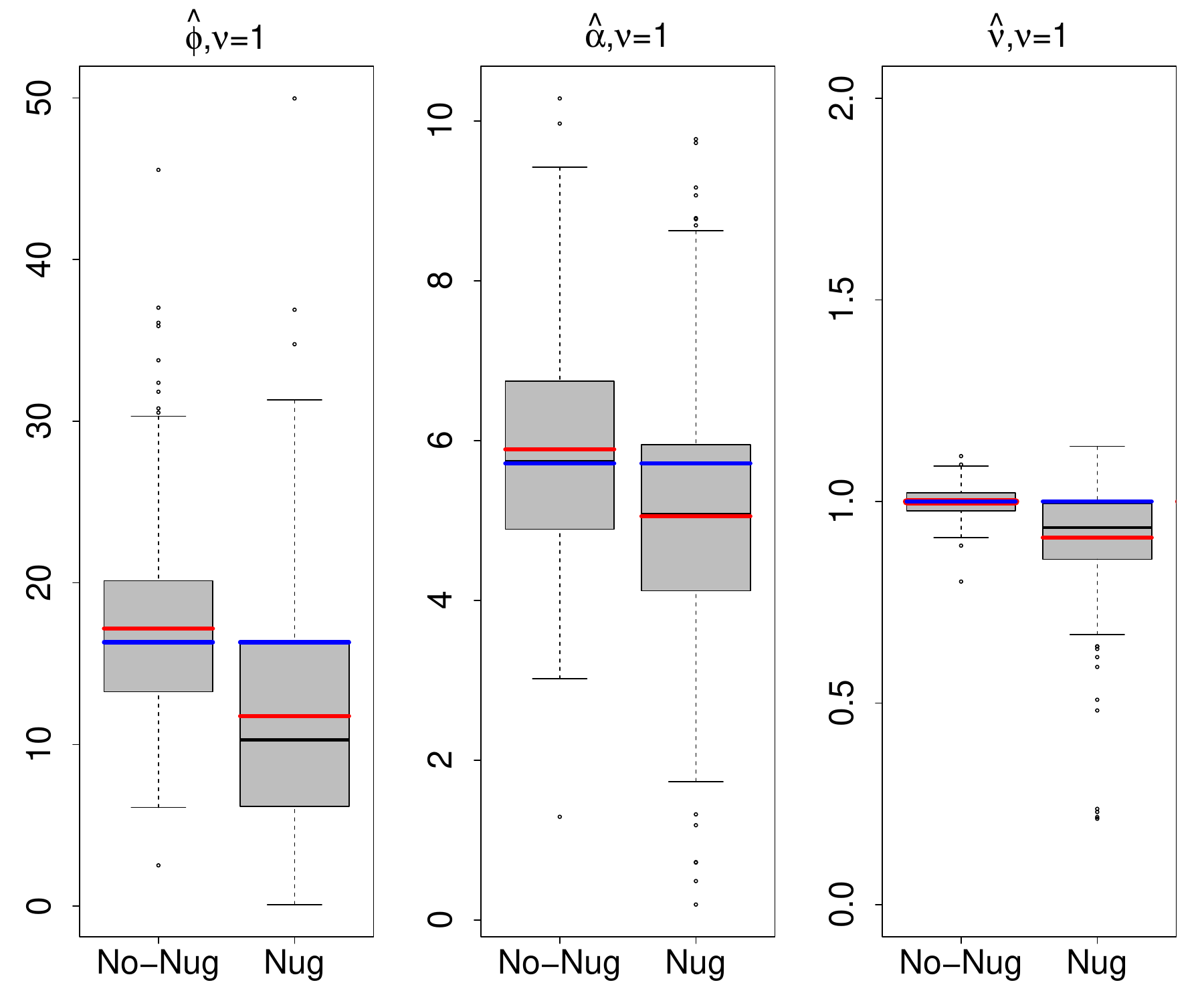}
  \caption{$\hat{\btheta}_2,{\cal M}_2$, strong correlation. }
  \label{fig9(f)}
  \vspace{5mm}
\end{subfigure}
\begin{subfigure}{0.33\linewidth}
  \centering
  \includegraphics[width=1\textwidth,]{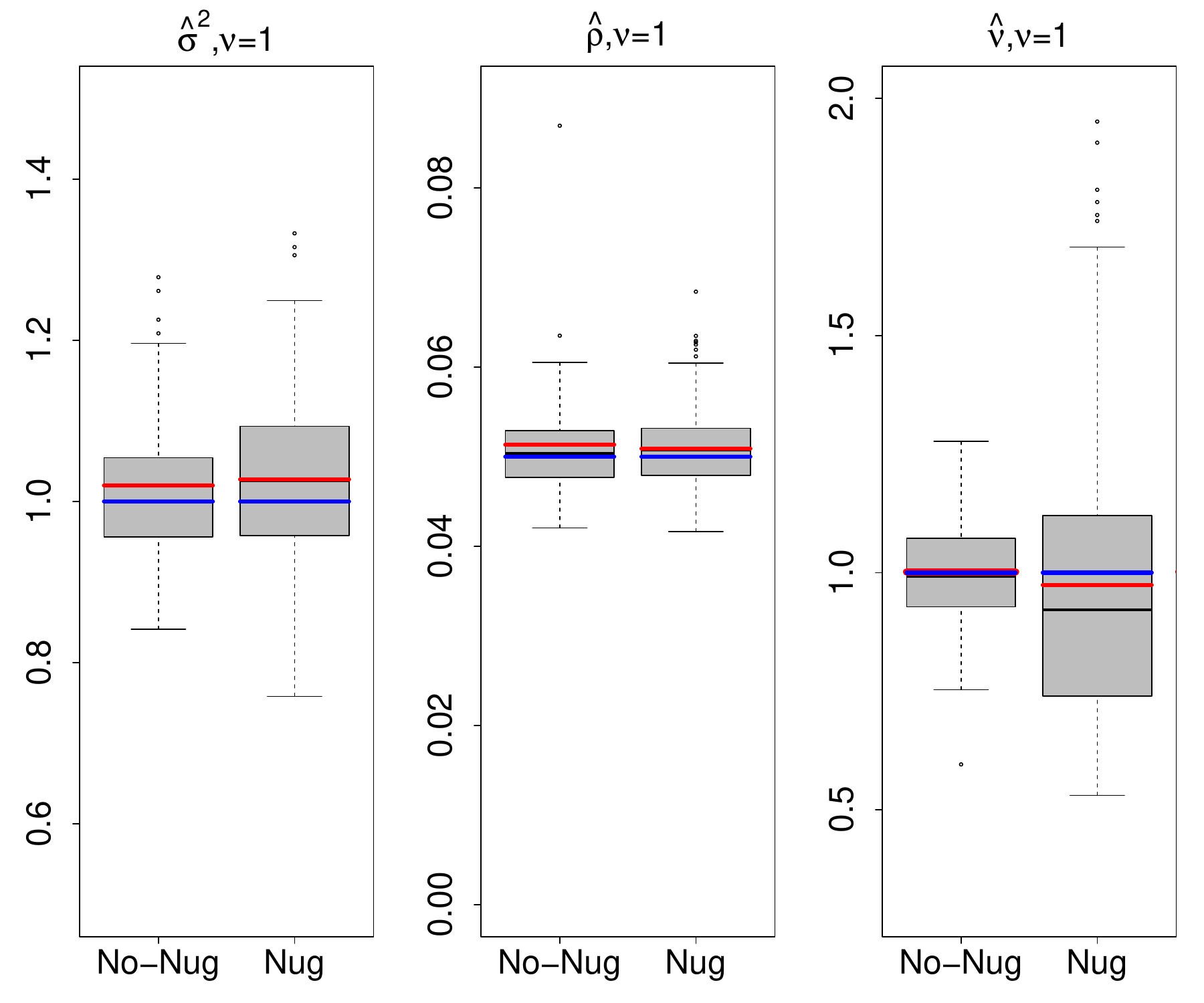}
  \caption{$\hat{\btheta}_3,{\cal M}_3$, weak correlation. }
  \label{fig9(g)}
\end{subfigure}
\begin{subfigure}{0.33\linewidth}
  \centering
  \includegraphics[width=1\textwidth,]{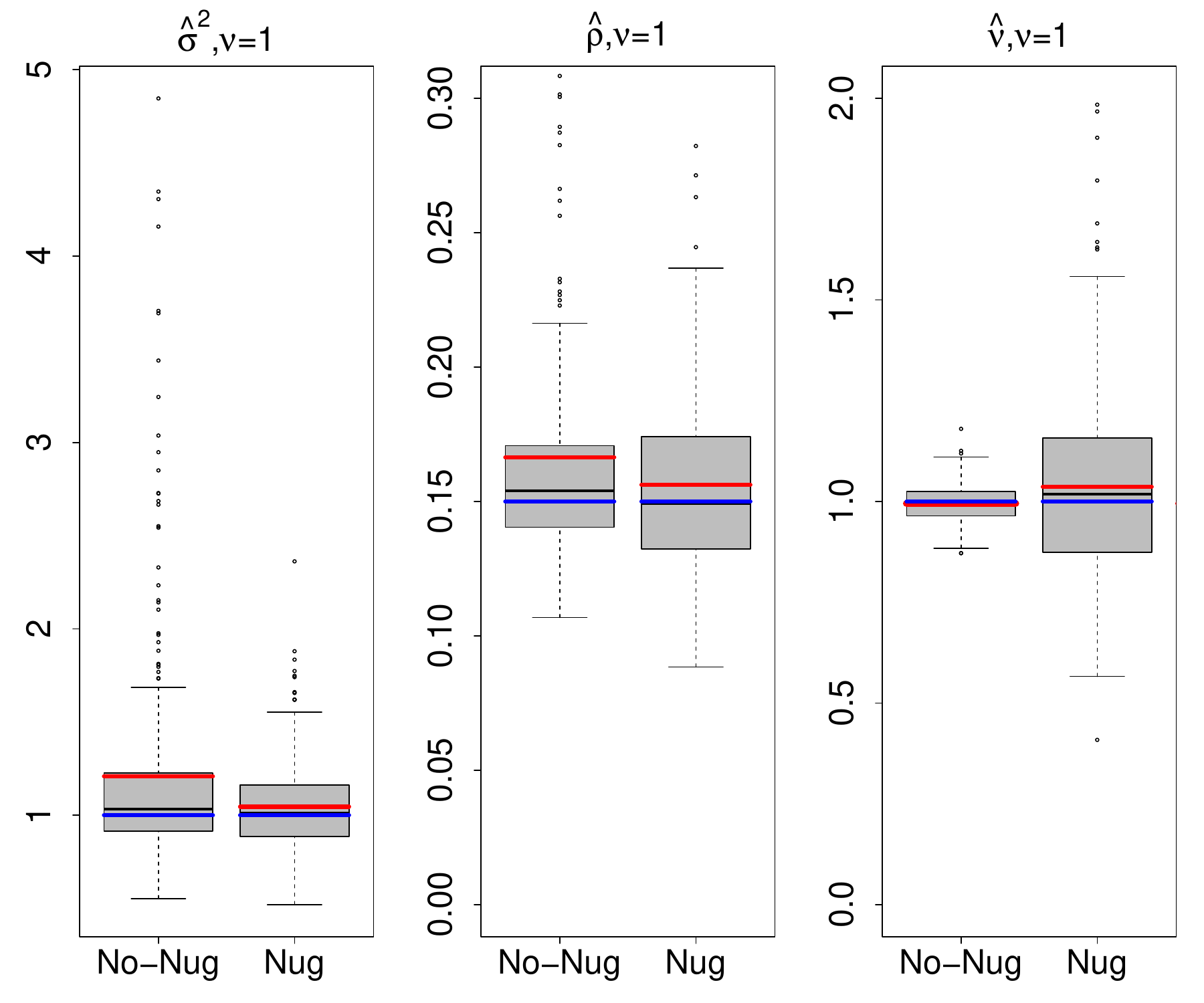}
  \caption{$\hat{\btheta}_3,{\cal M}_3$, medium correlation. }
  \label{fig9(h)}
\end{subfigure}
\begin{subfigure}{0.33\linewidth}
  \centering
  \includegraphics[width=1\textwidth,]{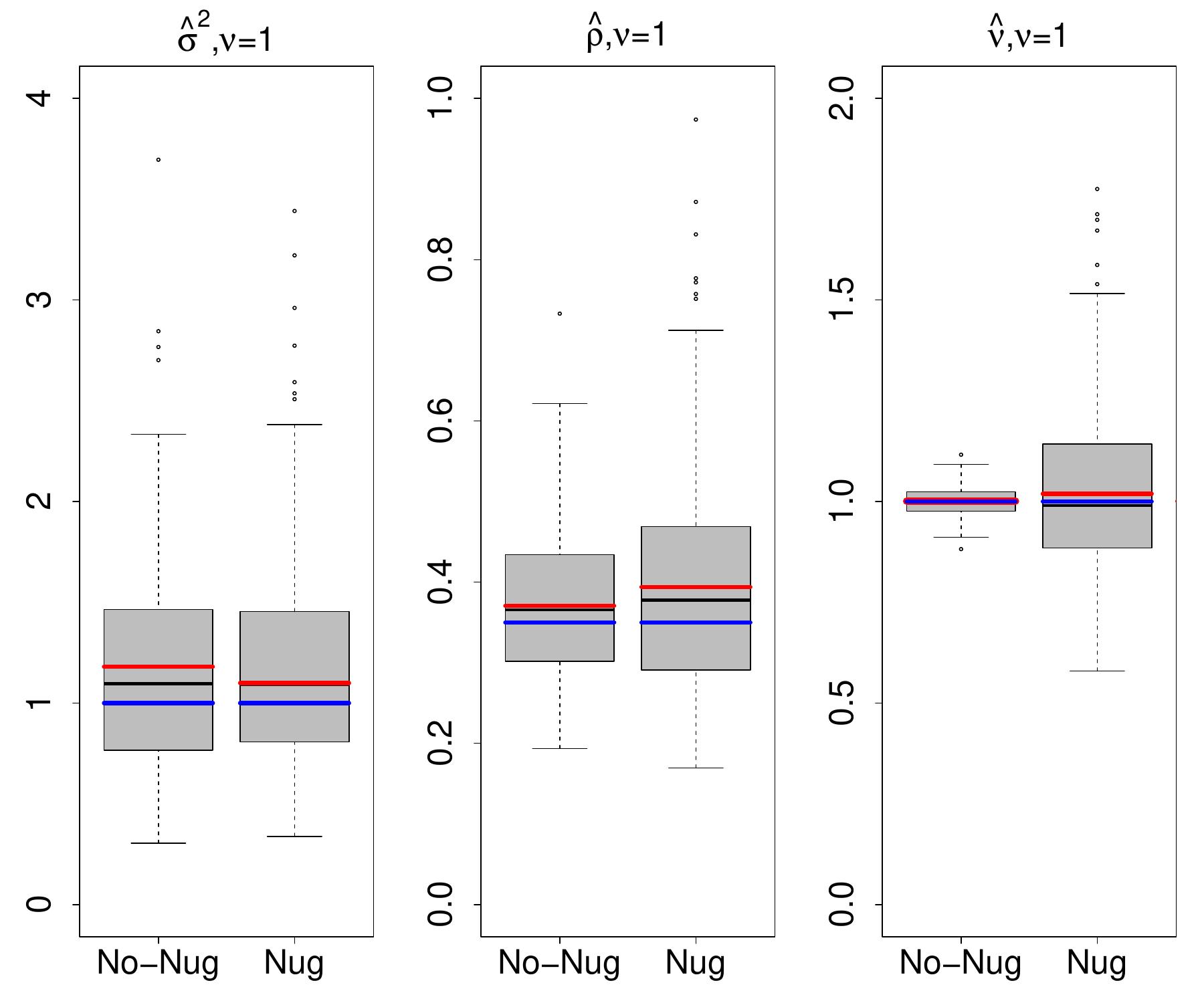}
  \caption{$\hat{\btheta}_3,{\cal M}_3$, strong correlation. }
  \label{fig9(i)}
\end{subfigure}
\caption{Boxplots of the MLEs of the Whittle Gaussian fields. Red lines represent the sample means and the blue lines denote the true values. {\redcolor No-Nug denotes the non-nugget effect models and Nug denotes the nugget effect models.}}
\label{fig9}
\end{figure}
\begin{figure}[H]
\begin{subfigure}{0.5\textwidth}
  \centering
  \includegraphics[width=0.8\textwidth]{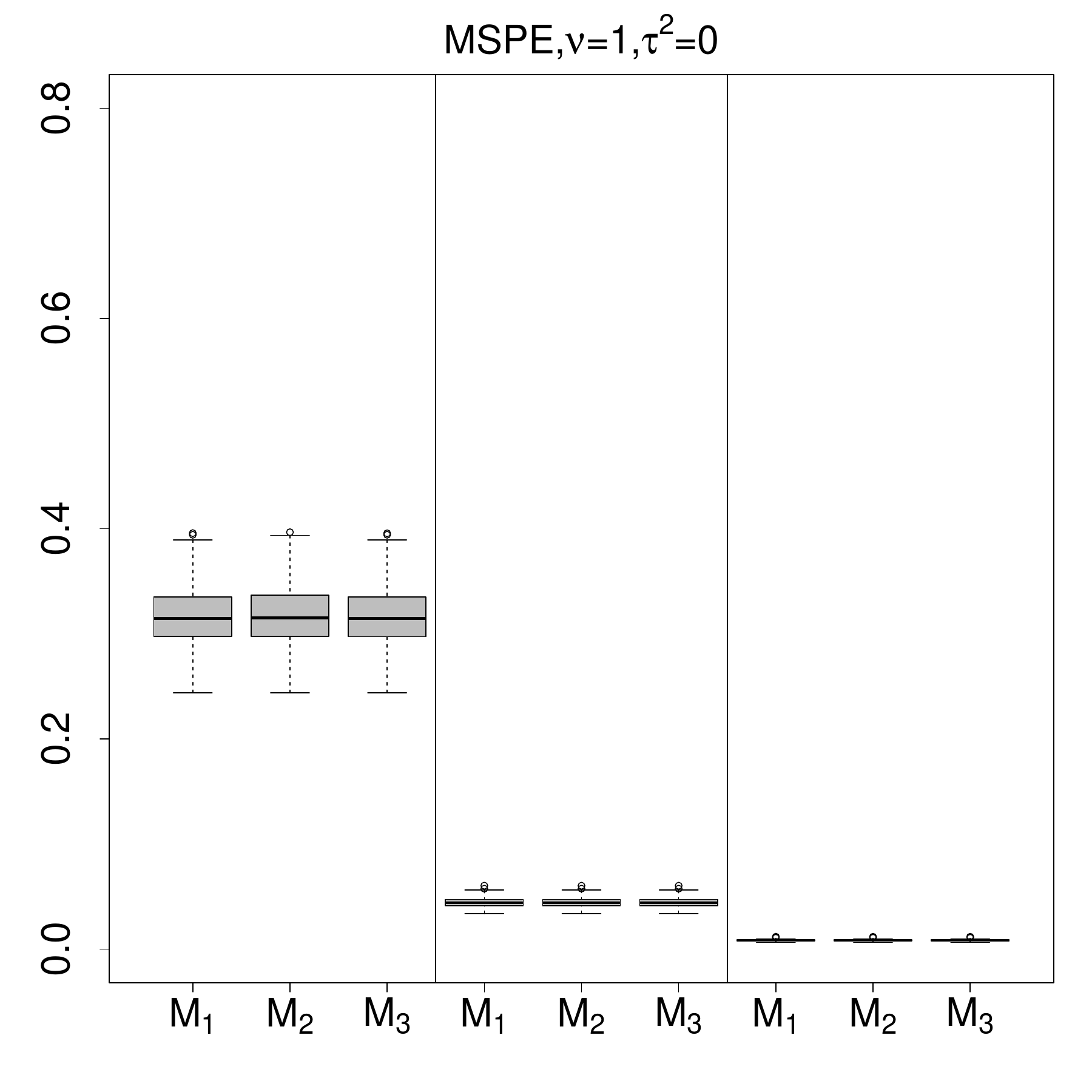}
\end{subfigure}
\begin{subfigure}{0.5\textwidth}
  \centering
  \includegraphics[width=0.8\textwidth]{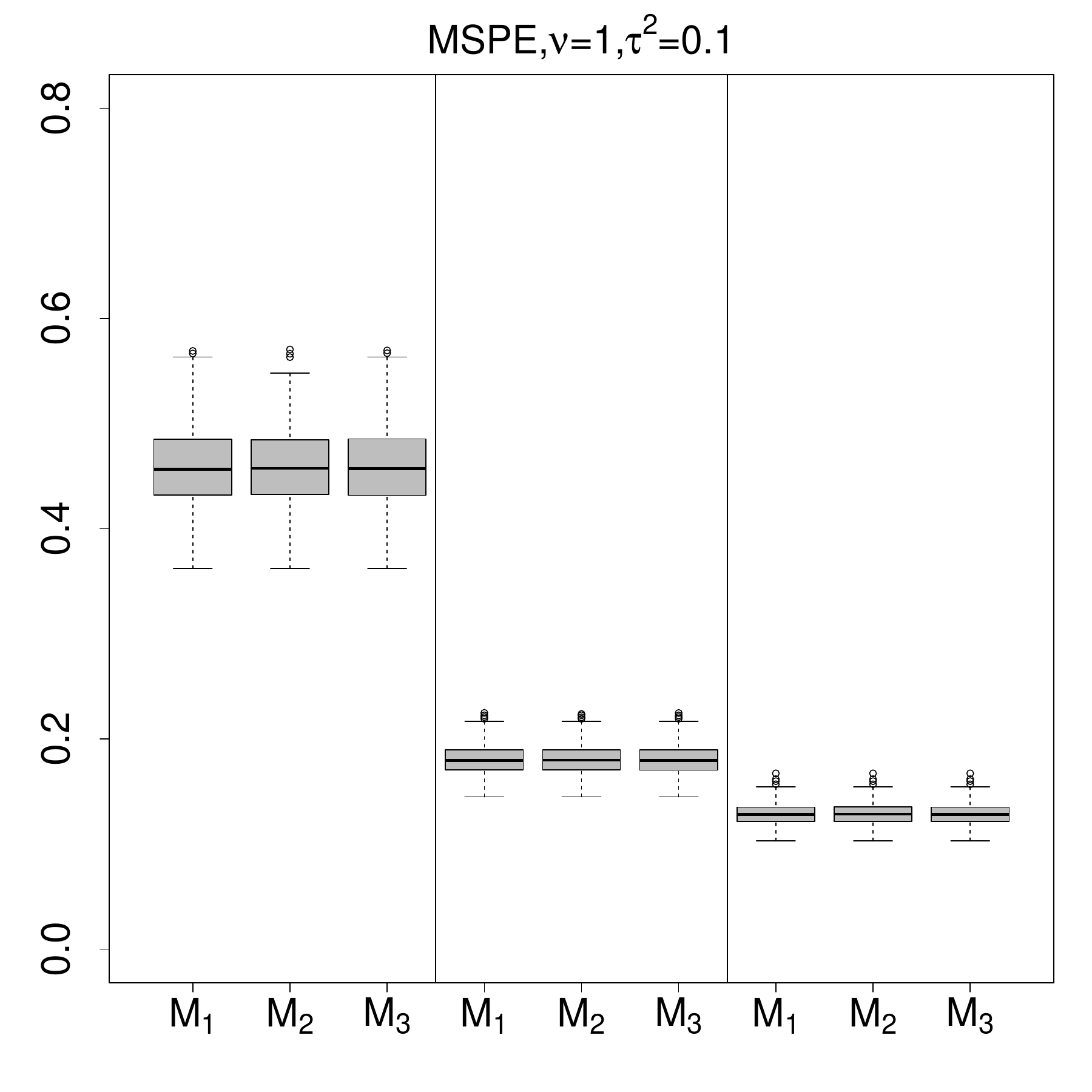}
\end{subfigure}

\caption{Boxplots of the MSPEs with and without nugget effects under weak (first block), medium (second block), and strong (third block) correlations, separated by vertical lines in each sub-figure, on the whittle fields with 300 replicates.}
\label{figmspe}
\end{figure}
\begin{figure}[H]
\begin{subfigure}{0.5\textwidth}
  \centering
  \includegraphics[width=0.8\textwidth]{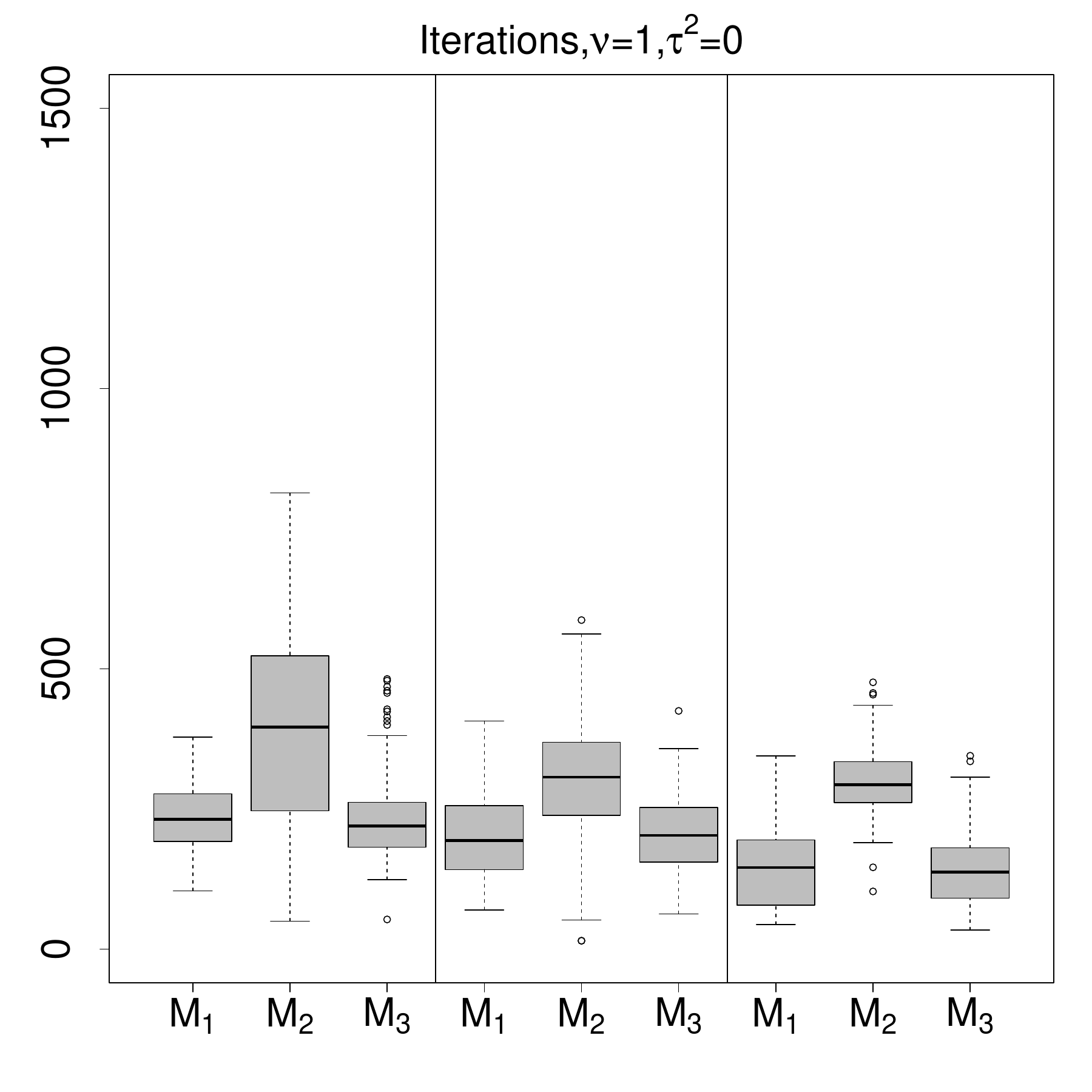}
\end{subfigure}
\begin{subfigure}{0.5\textwidth}
  \centering
  \includegraphics[width=0.8\textwidth]{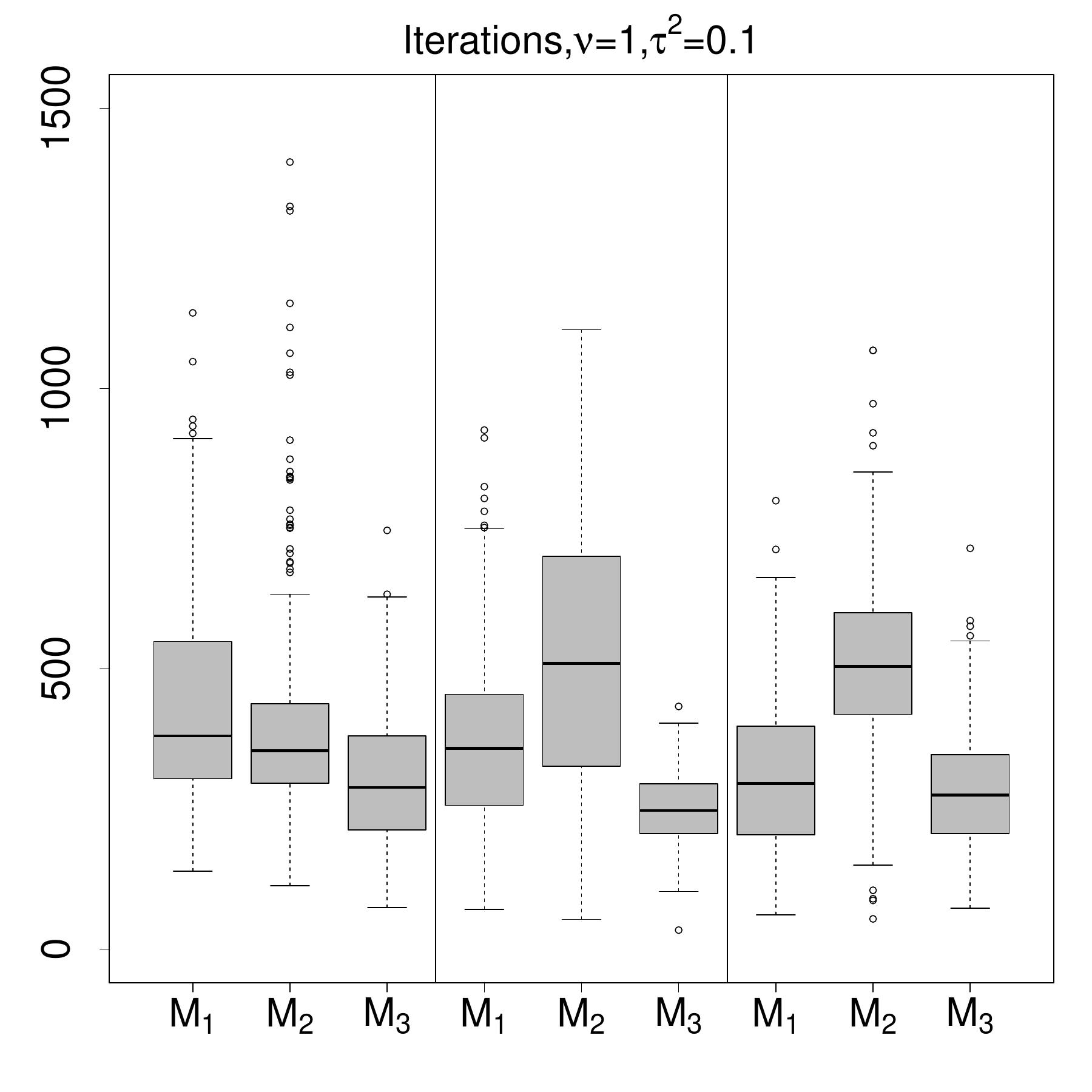}
\end{subfigure}
\caption{Boxplots of the convergent iterations with and without nugget effects under weak (first block), medium (second block), and strong
(third block) correlations, separated by vertical lines in each sub-figure, on the Whittle fields with 300 replicates.}
\label{iterw}
\end{figure}
\begin{table}[H]
    \centering
    \caption{Difference ratios between means and true values for ${\cal M}_1$, ${\cal M}_2$, ${\cal M}_3$ with and without nugget effects using 300 replicates.}
    \resizebox{1\linewidth}{!}{
    \begin{tabular}{|l|c|c|c||c|c|c||c|c|c|}
    \hline
        Model &  $\sigma^2$ & $\beta$ & $\nu$ & $\phi$ & $\alpha$ & $\nu$ &$\sigma^2$ & $\rho$ & $\nu$ \\ \hline
       Weak-exp ($h=0.1$)   &  0.00532 & 0.0208 & 0.00975 & {\redcolor0.1972} & 0.000222 & 0.00350 & 0.0259 & {\redcolor 0.1090} & 0.00678 \\   \hline
       Medium-exp ($h=0.3$) & 0.0263 & 0.0416 & 0.00528 & {\redcolor 0.1031} & 0.0217 & 0.00478 & 0.0696 & {\redcolor 0.1016} & 0.00403 \\  \hline
       Strong-exp ($h=0.7$) & 0.0584 & 0.0691 & 0.00597 & 0.0248 & 0.0122 & 0.0979 & {\redcolor0.1796} & {\redcolor0.2123} & 0.00341\\
       \hline
       \hline
        Weak-exp-nugget ($h=0.1$)   &  0.00535 & 0.00252 & 0.0996 & {\redcolor1.2609} & 0.0717 & {\redcolor0.1497} & 0.00736 & 0.0114 & 0.1121 \\   \hline
       Medium-exp-nugget ($h=0.3$) & {\redcolor0.1360} & {\redcolor0.4672} & 0.0444 & {\redcolor0.9388} & 0.0218 & 0.000744 & 0.0289 & 0.0412 & 0.0748 \\  \hline
       Strong-exp-nugget ($h=0.7$) & {\redcolor0.2435} & {\redcolor0.5356} & 0.0152 & 0.0163 & 0.0343 & 0.0538 & 0.0995 & {\redcolor0.1744} & 0.0415 \\
       \hline
       \hline
        Weak-Whittle ($h=0.1$)   &  0.00773 & 0.0236 & 0.00431 & 0.0407 & 0.0787 & 0.0635 & 0.0199 & 0.0268 & 0.00154 \\   \hline
       Medium-Whittle ($h=0.3$) & 0.0618 & 0.0377 & 0.0000581 & 0.0825 & 0.00593 & 0.0108 & {\redcolor 0.2093} & {\redcolor 0.1097} & 0.00499 \\  \hline
       Strong-Whittle ($h=0.7$) & {\redcolor0.2050} & 0.0954 & 0.000337 & 0.0521 & 0.0306 & 0.000416 & {\redcolor0.1449} & 0.0591 & 0.00102  \\
       \hline
       \hline
               Weak-Whittle-nugget ($h=0.1$)   &  0.0356 & 0.0826 & 0.0542 &  {\redcolor0.4538} & 0.0616 & 0.0698 & 0.0267 & 0.0178 & 0.0235 \\   \hline
       Medium-Whittle-nugget ($h=0.3$) & {\redcolor0.1556} & {\redcolor0.2403} & 0.00990 & {\redcolor0.1546} & 0.0883 & 0.0664 & 0.0460 & 0.0415 & 0.0363 \\  \hline
       Strong-Whittle-nugget ($h=0.7$) & {\redcolor0.3191} & {\redcolor0.3234} & 0.00199 & {\redcolor0.2795} & {\redcolor0.1159} & 0.0896 & {\redcolor0.1728} & {\redcolor 0.1257} & 0.0192 \\
       \hline
    \end{tabular}
    }
    \label{table7}
\end{table}
\begin{table}[H]
    \centering
    \caption{Difference ratios between means and true values for ${\cal M}_1$ and ${\cal M}_3$ with and without nugget effects under TLR approximated covariance using 300 replicates.}
    \resizebox{1\linewidth}{!}{
    \begin{tabular}{|l|c|c|c||c|c|c|}
    \hline
        Model &  $\sigma^2$ & $\beta$ & $\nu$  &$\sigma^2$ & $\rho$ & $\nu$ \\ \hline
       Weak-exp ($h=0.1$)   &  0.00553 & 0.0217 & 0.00975 & 0.0144 & 0.0842 & 0.00994 \\   \hline
       Medium-exp ($h=0.3$) & 0.0265 & 0.0402 & 0.00579  & 0.0933 & {\redcolor0.1399} & 0.00394 \\  \hline
       Strong-exp ($h=0.7$) & 0.0537 & 0.0598 & 0.00713 & {\redcolor0.1823} & {\redcolor0.2071} & 0.00438\\
       \hline
       \hline
        Weak-exp-nugget ($h=0.1$)   &  0.0127 & 0.0159 & 0.0851 & 0.00621 & 0.0117 & 0.1088 \\   \hline
       Medium-exp-nugget ($h=0.3$) & {\redcolor0.1439} & {\redcolor0.4254} & 0.0397  & 0.0326 & 0.0467 & 0.0728 \\  \hline
       Strong-exp-nugget ($h=0.7$) & {\redcolor0.2340} & {\redcolor0.5420} & 0.0191 & 0.0968 & {\redcolor0.1622} & 0.0473\\
       \hline
       \hline
        Weak-Whittle ($h=0.1$)   &  0.00627 & 0.0222 & 0.00295 & 0.00600 & 0.00991 & 0.00344 \\   \hline
       Medium-Whittle ($h=0.3$) & 0.0444 & 0.0239 & 0.00341 & {\redcolor 0.1175} & 0.0598 & 0.00161\\  \hline
       Strong-Whittle ($h=0.7$) & {\redcolor0.2216} & 0.0750 & 0.0116 & {\redcolor0.1572} & 0.0488 & 0.0129   \\
       \hline
       \hline
               Weak-Whittle-nugget ($h=0.1$)   &  0.0439 & {\redcolor 0.1045} & 0.0719 &  0.0278 & 0.0178 & 0.0264 \\   \hline
       Medium-Whittle-nugget ($h=0.3$) & {\redcolor0.1610} & {\redcolor0.2406} & 0.00750 & 0.0566 & 0.0533 & 0.0300  \\  \hline
       Strong-Whittle-nugget ($h=0.7$) & {\redcolor0.3603} & 0.0488 & 0.00459 & {\redcolor0.1492} & {\redcolor 0.1109} & 0.0238  \\
       \hline
    \end{tabular}
    }
    \label{tlrT}
\end{table}
\begin{table}[H]
    \centering
    \caption{Approximated effective range (AER) using the sample means of ${\cal M}_1$, ${\cal M}_2$, and ${\cal M}_3$ with nugget and without nugget effects using 300 replicates.}
    \resizebox{1\linewidth}{!}{
    \begin{tabular}{|l|c|c|c||l|c|c|c|}
    \hline
        Model/AER   & ${\cal M}_1(h;\hat{\btheta}_1)$ & ${\cal M}_2(h;\hat{\btheta}_2)$ & ${\cal M}_3(h;\hat{\btheta}_3)$ & Model/AER  & ${\cal M}_1(h;\hat{\btheta}_1)$ & ${\cal M}_2(h;\hat{\btheta}_2)$ & ${\cal M}_3(h;\hat{\btheta}_3)$\\
    \hline
       Weak-exp ($h=0.1$)  & 0.0523 & 0.0572 & 0.0668 & Weak-exp-nugget ($h=0.1$) & 0.0552 & 0.0531 & 0.0491\\
       \hline
       Medium-exp ($h=0.3$) & 0.0580 & 0.0499 & 0.0702 & Medium-exp-nugget  ($h=0.3$) & {\redcolor0.1542} & {\redcolor0.1055} & 0.0568\\
       \hline
       Strong-exp ($h=0.7$) & 0.0650 & 0.0537 & {\redcolor0.0999} & Strong-exp-nugget ($h=0.7$) & {\redcolor0.1802} & 0.0614 & 0.0858\\
       \hline
       \hline
              Weak-Whittle ($h=0.1$)  & 0.0543 & {\redcolor0.1119} & 0.0559 & Weak-Whittle-nugget ($h=0.1$) & 0.0627 & {\redcolor0.1605} & 0.0549\\
       \hline
       Medium-Whittle ($h=0.3$) & 0.0604 & 0.0585 & 0.0859 & Medium-Whittle-nugget  ($h=0.3$) & {\redcolor0.1160} & {\redcolor0.1253} & 0.0597\\
       \hline
       Strong-Whittle ($h=0.7$) & 0.0820 & 0.0444 & 0.0697 & Strong-Whittle-nugget ($h=0.7$) & {\redcolor0.1555} & 0.0864 & 0.0868\\
       \hline
    \end{tabular}
    }
    \label{AER}
\end{table}
\begin{table}[H]
    \centering
    \caption{Approximated effective range (AER) using the sample means of ${\cal M}_1$ and ${\cal M}_3$ with nugget and without nugget effects under TLR approximated covariance using 300 replicates.}
    \resizebox{1\linewidth}{!}{
    \begin{tabular}{|l|c|c|l|c|c|}
    \hline
        Model/AER   & ${\cal M}_1(h;\Tilde{\btheta}_1)$  & ${\cal M}_3(h;\Tilde{\btheta}_3)$ & Model/AER  & ${\cal M}_1(h;\Tilde{\btheta}_1)$ & ${\cal M}_3(h;\Tilde{\btheta}_3)$\\
    \hline
       Weak-exp ($h=0.1$)  & 0.0524  & 0.0620 & Weak-exp-nugget ($h=0.1$) & 0.0568 & 0.0491\\
       \hline
       Medium-exp ($h=0.3$) & 0.0578  & 0.0786 & Medium-exp-nugget  ($h=0.3$) & {\redcolor0.1456}  & 0.0579\\
       \hline
       Strong-exp ($h=0.7$) & 0.0632 & {\redcolor0.0991} & Strong-exp-nugget ($h=0.7$)  & {\redcolor0.1809} & 0.0832\\
       \hline
       \hline
              Weak-Whittle ($h=0.1$)  & 0.0541  & 0.0519 & Weak-Whittle-nugget ($h=0.1$) & 0.0657  & 0.0550\\
       \hline
       Medium-Whittle ($h=0.3$) & 0.0569  & 0.0682 & Medium-Whittle-nugget  ($h=0.3$) & {\redcolor0.1162}  & 0.0628\\
       \hline
       Strong-Whittle ($h=0.7$) & 0.0795  & 0.0680 & Strong-Whittle-nugget ($h=0.7$)  & {\redcolor0.1851} & 0.0815\\
       \hline
    \end{tabular}
    }
    \label{AER_tlr}
\end{table}
\begin{table}[H]
    \centering
    \caption{Asymptotic normality of $\hat{\tau}^2$ of ${\cal M}_1$, ${\cal M}_2$, and ${\cal M}_3$ on exponential and Whittle Gaussian fields using 300 replicates.}
    \resizebox{1\textwidth}{!}{
    \begin{tabular}{|l|c|c|c|c|c|c|}
    \hline
     Variance/field & ${\cal M}_1$-weak-exp & ${\cal M}_1$-medium-exp & ${\cal M}_1$-strong-exp & ${\cal M}_1$-weak-Whittle & ${\cal M}_1$-medium-Whittle & ${\cal M}_1$-strong-Whittle  \\
     \hline
        SV & 0.00944  & 0.00269 & 0.000801 & 0.00352 & 0.000237  & 0.0000410  \\
         \hline
        AV & 0.0486  & 0.00280 & 0.000569 & 0.00247 & 0.000140 &  0.0000418\\
        \hline
        DR & {\redcolor0.8058} & 0.0429 & {\redcolor0.4077} & {\redcolor0.4251} & {\redcolor0.6929} & 0.0195 \\
        \hline
            \hline
     Variance/field & ${\cal M}_2$-weak-exp & ${\cal M}_2$-medium-exp & ${\cal M}_2$-strong-exp & ${\cal M}_2$-weak-Whittle & ${\cal M}_2$-medium-Whittle & ${\cal M}_2$-strong-Whittle  \\
     \hline
        SV & 0.00501  & 0.00273 & 0.000676 & 0.00218 & 0.000451  & 0.000128  \\
         \hline
        AV & 0.0486  & 0.00280 & 0.000569 & 0.00247 & 0.000140 &  0.0000418\\
        \hline
        DR & {\redcolor0.8969} & 0.0250 & 0.1880 & 0.1174 & {\redcolor2.2214} & {\redcolor2.0622} \\
        \hline
        \hline
     Variance/field & ${\cal M}_3$-weak-exp & ${\cal M}_3$-medium-exp & ${\cal M}_3$-strong-exp & ${\cal M}_3$-weak-Whittle & ${\cal M}_3$-medium-Whittle & ${\cal M}_3$-strong-Whittle  \\
     \hline
        SV & 0.00999  & 0.00259 & 0.000749 & 0.00289 & 0.000183  & 0.0000335  \\
         \hline
        AV & 0.0486  & 0.00280 & 0.000569 & 0.00247 & 0.000140 &  0.0000418\\
        \hline
        DR & {\redcolor0.7865} & 0.0750 & 0.3163 & 0.1700 & 0.3071 & 0.1986 \\
        \hline
    \end{tabular}
    }
    \label{tau}
\end{table}
 \begin{table}[H]
    \centering
    \caption{SV, TAV, and DRVs, of $\hat{\btheta}_1$, $\hat{\btheta}_2$, and $\hat{\btheta}_3$ on the exponential and Whittle Gaussian fields with and without nugget effects using 300 replicates with sample size $n=1600$. }
    \resizebox{1\textwidth}{!}{
    \begin{tabular}{|l|c|c|c||c|c|c||c|c|c|}
    \hline
        Weak-exp   & $\hat{\sigma}^2$ & $\hat{\beta}$ & $\hat{\nu}$ & $\hat{\phi}$ & $\hat{\alpha}$ & $\hat{\nu}$ & $\hat{\sigma}^2$ & $\hat{\rho}$ & $\hat{\nu}$  \\
           \hline
       SV  & 0.00401 & 0.0000307 & 0.00402 & 47.3421 & 19.3994 & 0.00314 & 0.0520 & 0.00338 & 0.00435  \\
       \hline
       TAV & 0.00371 & 0.0000259 & 0.00362 & 36.2830 & 21.8370 & 0.00362 & 0.00372 & 0.0000245 & 0.00362 \\
       \hline
       DRV & 0.0809 & 0.1853 & 0.1105 & 0.3048 & 0.1116 & 0.1326 & {\redcolor12.9785} & {\redcolor136.9592} & 0.2017 \\
       \hline
       \hline
      Medium-exp   & $\hat{\sigma}^2$ & $\hat{\beta}$ & $\hat{\nu}$ & $\hat{\phi}$ & $\hat{\alpha}$ & $\hat{\nu}$ & $\hat{\sigma}^2$ & $\hat{\rho}$ & $\hat{\nu}$  \\
      \hline
        SV  & 0.0320 & 0.000683 & 0.00116 & 2.4777 & 6.7213 & 0.00162 & 0.1152 & 0.00485 & 0.00119  \\
       \hline
       TAV & 0.0244 & 0.000489 & 0.00111 & 1.0686 & 4.8899 & 0.00111 & 0.0244 & 0.000799 & 0.00111 \\
       \hline
       DRV & 0.3115 & 0.3967 & 0.0450 & {\redcolor1.3186} & 0.3746 & {\redcolor0.4596} & {\redcolor3.7213} & {\redcolor5.0701} & 0.0721 \\
       \hline
       \hline
       Strong-exp   & $\hat{\sigma}^2$ & $\hat{\beta}$ & $\hat{\nu}$ & $\hat{\phi}$ & $\hat{\alpha}$ & $\hat{\nu}$ & $\hat{\sigma}^2$ & $\hat{\rho}$ & $\hat{\nu}$  \\
      \hline
        SV  & 0.1170 & 0.00889 & 0.000741 & 0.3577 & 4.5617 & 0.0206 & 0.2805 & 0.0474 & 0.000749 \\
       \hline
       TAV & 0.0959 & 0.00718 & 0.000734 & 0.1240 & 2.3935 & 0.000734 & 0.0959 & 0.0133 & 0.000734 \\
       \hline
       DRV & 0.2200 & 0.2382 & 0.00954 & {\redcolor1.8847} & {\redcolor0.9059} & {\redcolor27.0654} & {\redcolor1.9249} & {\redcolor2.5639} & 0.0204 \\
        \hline
       \hline
        Weak-Whittle   & $\hat{\sigma}^2$ & $\hat{\beta}$ & $\hat{\nu}$ & $\hat{\phi}$ & $\hat{\alpha}$ & $\hat{\nu}$ & $\hat{\sigma}^2$ & $\hat{\rho}$ & $\hat{\nu}$   \\
           \hline
       SV  & 0.00574 & 0.00000989 & 0.00995 & 508914.0000 & 49.4198 & 0.0191 & 0.0474 & 0.000176 & 0.0115 \\
       \hline
       TAV & 0.00503 & 0.00000842 & 0.00921 & 596200.0000 & 21.5630 & 0.00921 & 0.00503 & 0.0000141 & 0.00921 \\
       \hline
       DRV & 0.1412 & 0.1746 & 0.0803 & 0.1464 & {\redcolor1.2919} & {\redcolor1.0738} & {\redcolor8.4235} & {\redcolor11.4823} & 0.2486 \\
       \hline
       \hline
      Medium-Whittle   & $\hat{\sigma}^2$ & $\hat{\beta}$ & $\hat{\nu}$ & $\hat{\phi}$ & $\hat{\alpha}$ & $\hat{\nu}$ & $\hat{\sigma}^2$ & $\hat{\rho}$ & $\hat{\nu}$  \\
      \hline
        SV  & 0.0759 & 0.000188 & 0.00230 & 2361.0720 & 7.3414 & 0.00748 & 0.3802 & 0.00239 & 0.00256 \\
       \hline
       TAV & 0.0334 & 0.000114 & 0.00208 & 1450.4000 & 3.6176 & 0.00209 & 0.0334 & 0.000359 & 0.00209 \\
       \hline
       DRV & {\redcolor1.2725} & {\redcolor0.6491} & 0.1058 & {\redcolor0.6279} & {\redcolor1.0294} & {\redcolor2.5789} & {\redcolor10.3832} & {\redcolor5.6574} & 0.2249 \\
       \hline
       \hline
       Strong-Whittle   & $\hat{\sigma}^2$ & $\hat{\beta}$ & $\hat{\nu}$ & $\hat{\phi}$ & $\hat{\alpha}$ & $\hat{\nu}$ & $\hat{\sigma}^2$ & $\hat{\rho}$ & $\hat{\nu}$  \\
      \hline
        SV  & 0.2341 & 0.00232 & 0.00129 & 34.0935 & 1.8841 & 0.00134 & 0.2253 & 0.00809 & 0.00128 \\
       \hline
       TAV & 0.1234 & 0.00143 & 0.00119 & 26.5530 & 1.5280 & 0.00120 & 0.1234 & 0.00522 & 0.00120 \\
       \hline
       DRV & {\redcolor0.8971} & {\redcolor0.6224} & 0.0840 & 0.2839 & 0.2330 & 0.1167 & {\redcolor0.8258} & {\redcolor0.5498} & 0.0667 \\
       \hline
       \hline
       Weak-exp-nugget   & $\hat{\sigma}^2$ & $\hat{\beta}$ & $\hat{\nu}$ & $\hat{\phi}$ & $\hat{\alpha}$ & $\hat{\nu}$ & $\hat{\sigma}^2$ & $\hat{\rho}$ & $\hat{\nu}$ \\
       \hline
       SV  & 0.0119 & 0.0000453 & 0.0371 & 136.9030 & 20.7309 & 0.00622 & 0.0135 & 0.0000321 & 0.0544 \\
       \hline
       TAV & 0.0545 & 0.000103 & 0.0906 & 600.5700 & 87.2620 & 0.0906 & 0.0545 & 0.0000299 & 0.0905\\
       \hline
       DRV & {\redcolor0.7720} & {\redcolor0.7624} & {\redcolor0.5905} & {\redcolor0.9456} & {\redcolor0.9076} & {\redcolor0.9313} & {\redcolor0.7523} & 0.0736 & 0.3989\\
       \hline
       \hline
       Medium-exp-nugget   & $\hat{\sigma}^2$ & $\hat{\beta}$ & $\hat{\nu}$ & $\hat{\phi}$ & $\hat{\alpha}$ & $\hat{\nu}$ & $\hat{\sigma}^2$ & $\hat{\rho}$ & $\hat{\nu}$  \\
       \hline
       SV  & 0.2055 & 0.0358 & 0.0301  & 67.8228 & 18.1832 & 0.0218 & 0.0366 & 0.00161 & 0.0288 \\
       \hline
       TAV & 0.0297 & 0.00118 & 0.0205  & 11.7100 & 11.7930 & 0.0205 & 0.0297 & 0.00113 & 0.0205\\
       \hline
       DRV & {\redcolor5.9192} & {\redcolor29.3389} & {\redcolor0.4683} & {\redcolor4.7919} & {\redcolor0.5419} & 0.0634 & 0.2323 & {\redcolor0.4248} & {\redcolor0.4049}\\
       \hline
       \hline
       Strong-exp-nugget   & $\hat{\sigma}^2$ & $\hat{\beta}$ & $\hat{\nu}$ & $\hat{\phi}$ & $\hat{\alpha}$ & $\hat{\nu}$ & $\hat{\sigma}^2$ & $\hat{\rho}$ & $\hat{\nu}$  \\
       \hline
       SV  & 0.2152 & 0.0669 & 0.0176 & 0.7974 & 3.1759 & 0.00879 & 0.1367 & 0.0349 & 0.0181 \\
       \hline
       TAV & 0.1012 & 0.0135 & 0.0132 & 1.1710 & 4.5005 & 0.0132 & 0.1012 & 0.0193 & 0.0132\\
       \hline
       DRV & {\redcolor1.1265} & {\redcolor4.1778} & 0.3333 & 0.3190 & 0.2943 & 0.3341 & 0.3508 & {\redcolor0.8083} & 0.3712\\
       \hline
       \hline
       Weak-Whittle-nugget   & $\hat{\sigma}^2$ & $\hat{\beta}$ & $\hat{\nu}$ & $\hat{\phi}$ & $\hat{\alpha}$ & $\hat{\nu}$ & $\hat{\sigma}^2$ & $\hat{\rho}$ & $\hat{\nu}$ \\
       \hline
       SV  & 0.00860 & 0.0000234 & 0.0639 & 1329457.0000  & 56.3423 & 0.0428 & 0.00869 & 0.0000197 & 0.0867 \\
       \hline
       TAV & 0.00968 & 0.0000320 & 0.1150 & 6051400.0000 & 81.8330 & 0.1150 & 0.00968 & 0.0000184 & 0.1150\\
       \hline
       DRV & 0.1116 & 0.2688 & {\redcolor0.4443} & {\redcolor0.7803} & 0.3115 & {\redcolor0.6278} & 0.1023 & 0.0707 & 0.2461\\
       \hline
       \hline
       Medium-Whittle-nugget   & $\hat{\sigma}^2$ & $\hat{\beta}$ & $\hat{\nu}$ & $\hat{\phi}$ & $\hat{\alpha}$ & $\hat{\nu}$ & $\hat{\sigma}^2$ & $\hat{\rho}$ & $\hat{\nu}$  \\
       \hline
       SV  & 0.3004 & 0.00551 & 0.0583 & 8696.0090 & 10.2706 & 0.0363 & 0.0599 & 0.00124 & 0.0551 \\
       \hline
       TAV & 0.0367 & 0.000364 & 0.0384 & 18556.0000 & 11.5170 & 0.0384 & 0.0367 & 0.000683 & 0.0384\\
       \hline
       DRV & {\redcolor7.1853} & {\redcolor14.1374} & {\redcolor0.5182} & {\redcolor0.5314} & 0.1082 & 0.0547 & {\redcolor0.6322} & {\redcolor0.8155} & {\redcolor0.4349}\\
       \hline
       \hline
       Strong-Whittle-nugget   & $\hat{\sigma}^2$ & $\hat{\beta}$ & $\hat{\nu}$ & $\hat{\phi}$ & $\hat{\alpha}$ & $\hat{\nu}$ & $\hat{\sigma}^2$ & $\hat{\rho}$ & $\hat{\nu}$  \\
       \hline
       SV  & 0.3983 & 0.0197 & 0.0391 & 57.6320 & 2.4925 & 0.0184 & 0.2466 & 0.0183 & 0.0406\\
       \hline
       TAV & 0.1377 & 0.00413 & 0.0352 & 441.2300 & 4.4028 & 0.0352 & 0.1377 & 0.0106 & 0.0352\\
       \hline
       DRV & {\redcolor 1.8925} & {\redcolor3.7699} & 0.1108 & {\redcolor0.8694} & {\redcolor0.4339} & {\redcolor0.4773} & {\redcolor0.7908} & {\redcolor0.7264} & 0.1534\\
       \hline
    \end{tabular}
    }
    \label{table13}
\end{table}
 \begin{figure}[H]
\begin{subfigure}{0.33\linewidth}
  \centering
  \includegraphics[width=1\textwidth,]{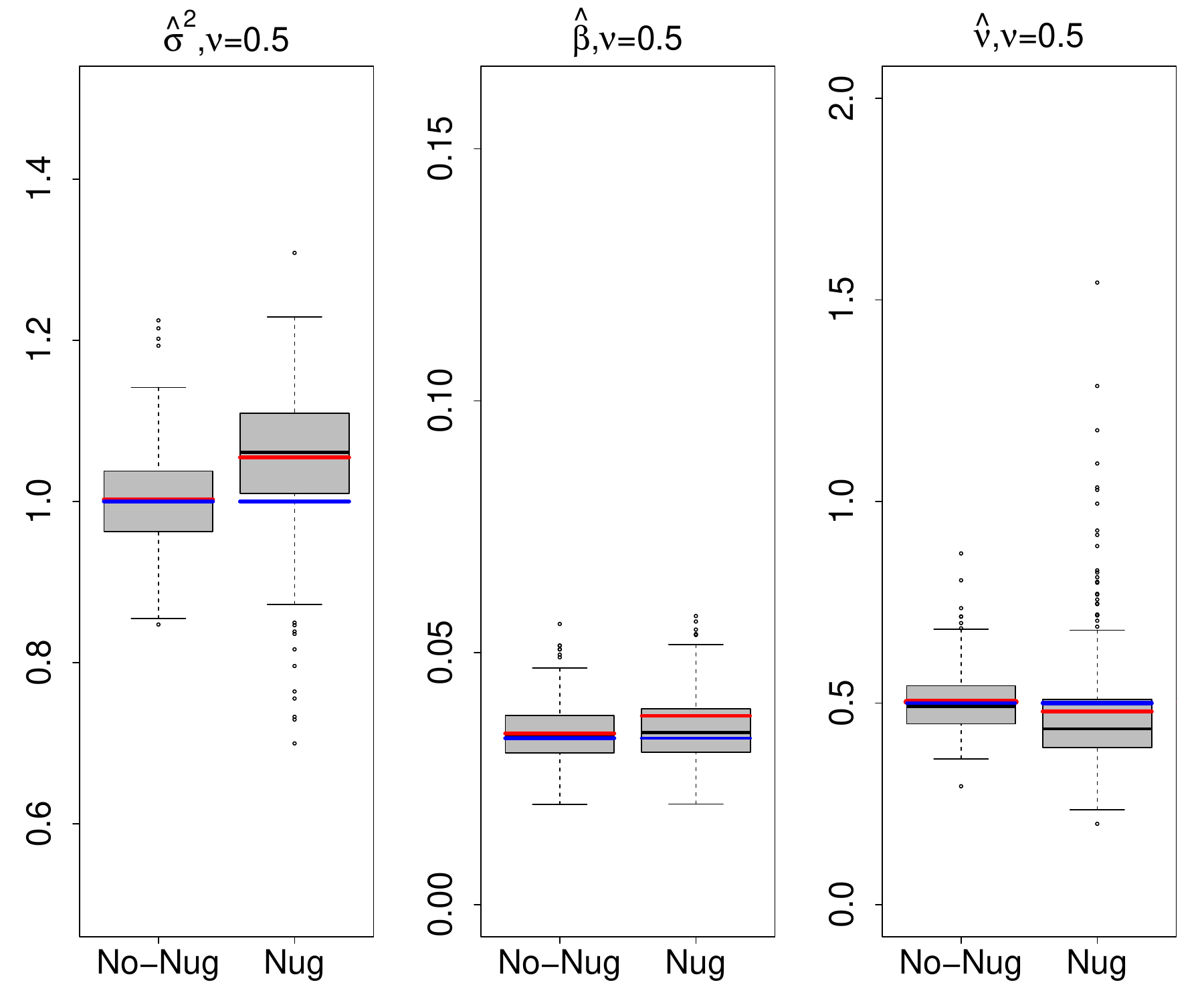}
  \caption{$\hat{\btheta}_1,{\cal M}_1$, weak correlation. }
  \label{reg(a)}
  \vspace{5mm}
\end{subfigure}
\begin{subfigure}{0.33\linewidth}
  \centering
  \includegraphics[width=1\textwidth,]{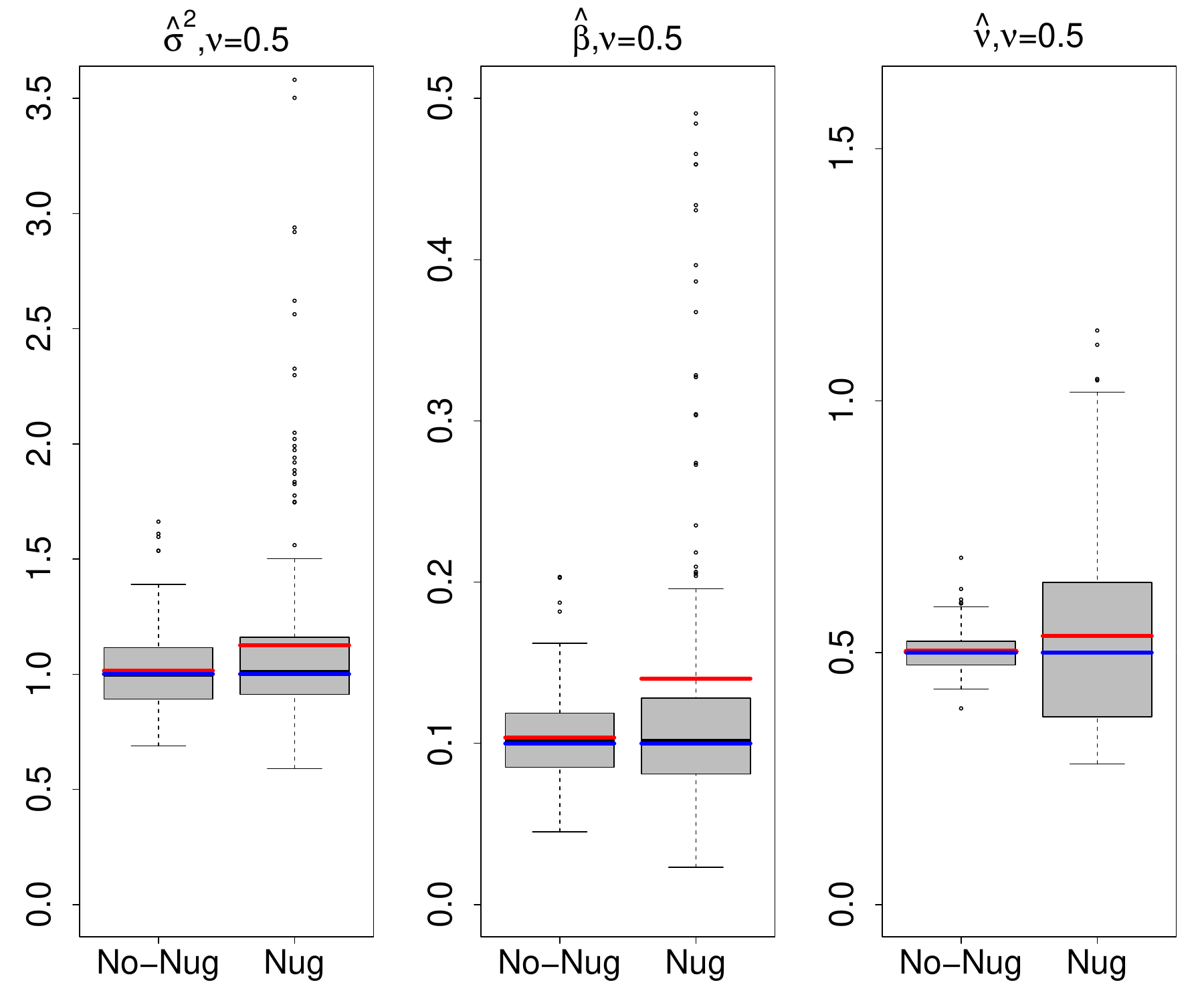}
  \caption{$\hat{\btheta}_1,{\cal M}_1$, medium correlation. }
  \label{reg(b)}
  \vspace{5mm}
\end{subfigure}
\begin{subfigure}{0.33\linewidth}
  \centering
  \includegraphics[width=1\textwidth,]{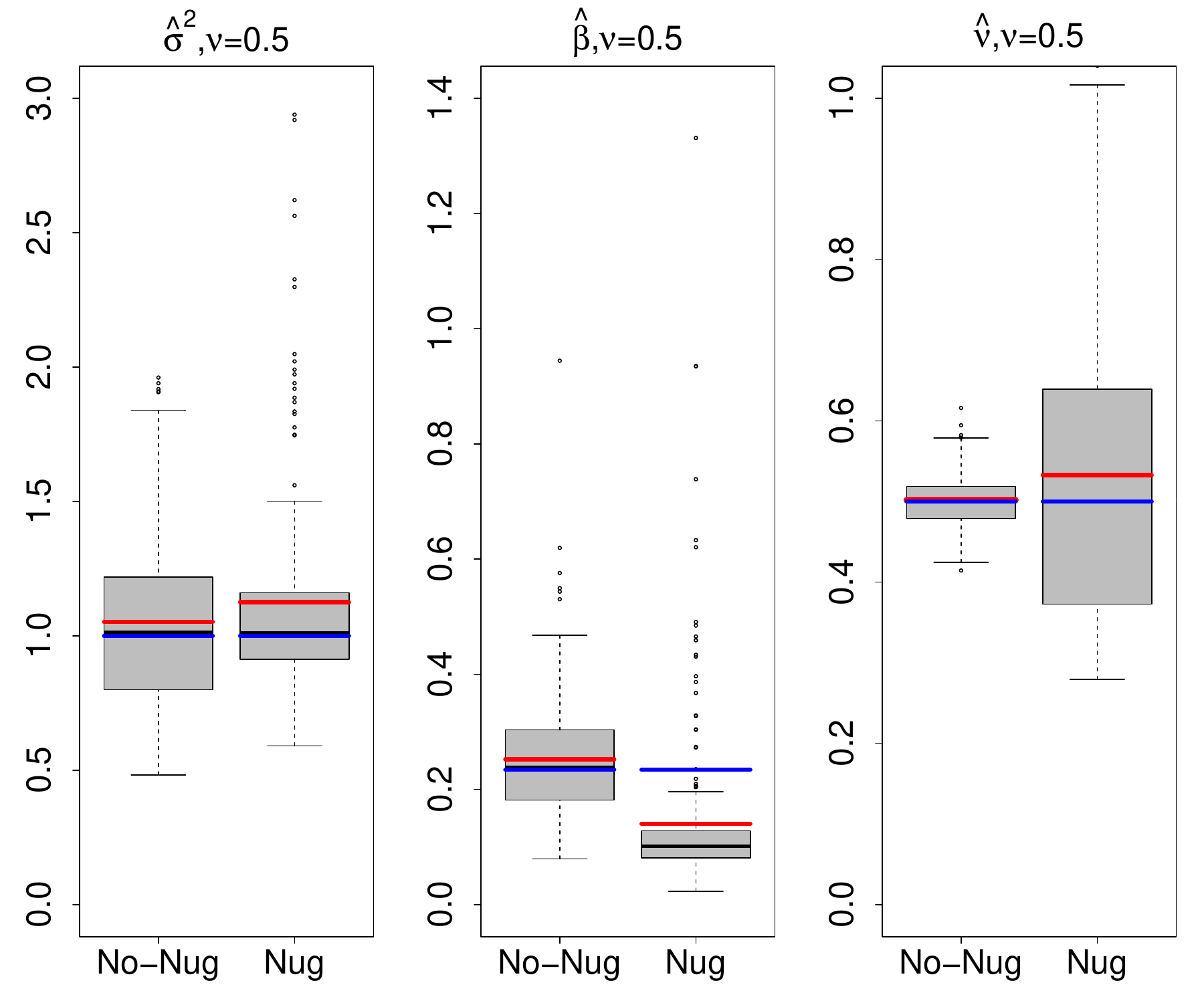}
  \caption{$\hat{\btheta}_1,{\cal M}_1$, strong correlation. }
  \label{reg(c)}
  \vspace{5mm}
\end{subfigure}
\begin{subfigure}{0.33\linewidth}
  \centering
  \includegraphics[width=1\textwidth,]{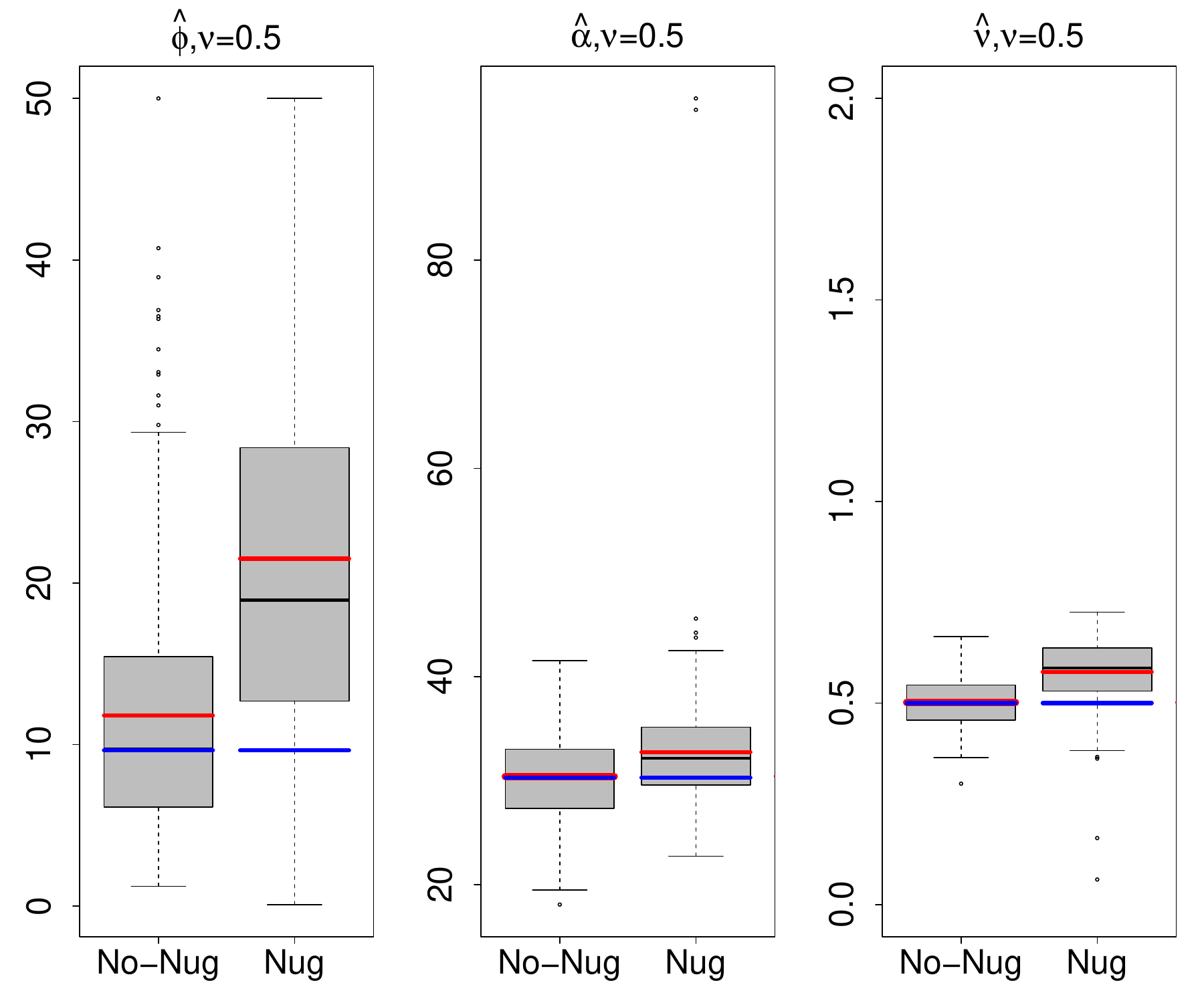}
  \caption{$\hat{\btheta}_2,{\cal M}_2$, weak correlation. }
  \label{reg(d)}
  \vspace{5mm}
\end{subfigure}
\begin{subfigure}{0.33\linewidth}
  \centering
  \includegraphics[width=1\textwidth,]{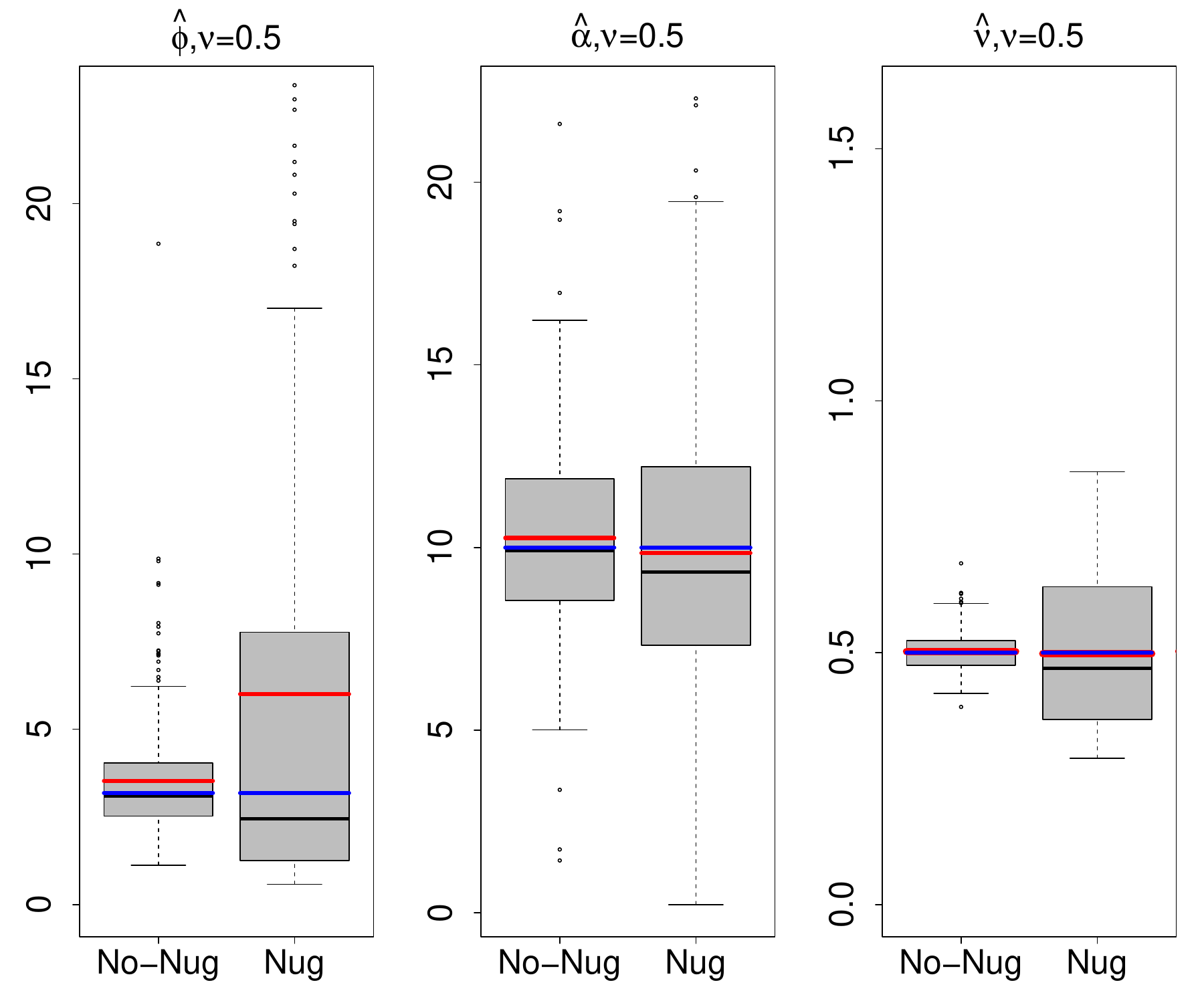}
  \caption{$\hat{\btheta}_2,{\cal M}_2$, medium correlation. }
  \label{reg(e)}
  \vspace{5mm}
\end{subfigure}
\begin{subfigure}{0.33\linewidth}
  \centering
  \includegraphics[width=1\textwidth,]{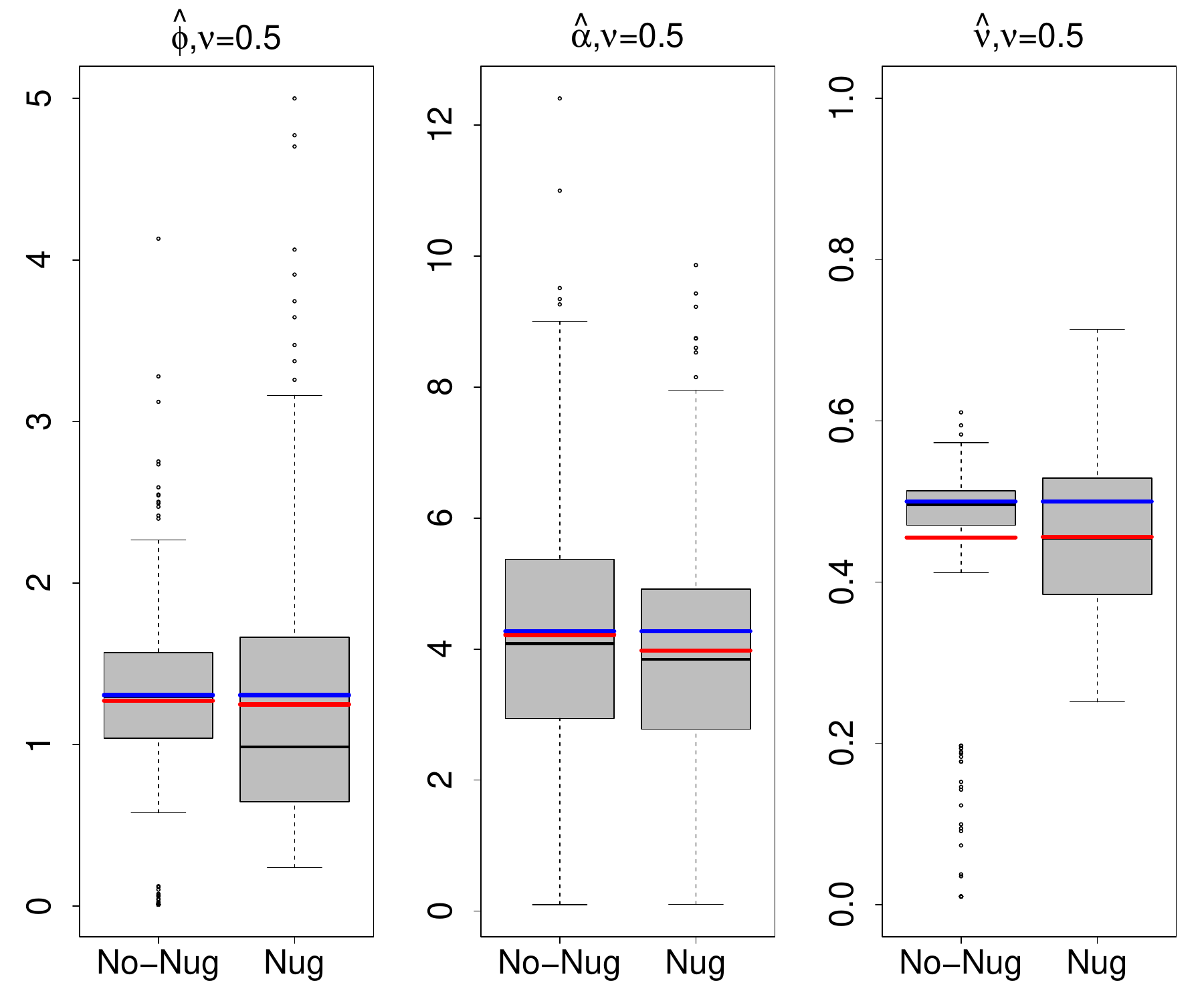}
  \caption{$\hat{\btheta}_2,{\cal M}_2$, strong correlation. }
  \label{reg(f)}
  \vspace{5mm}
\end{subfigure}
\begin{subfigure}{0.33\linewidth}
  \centering
  \includegraphics[width=1\textwidth,]{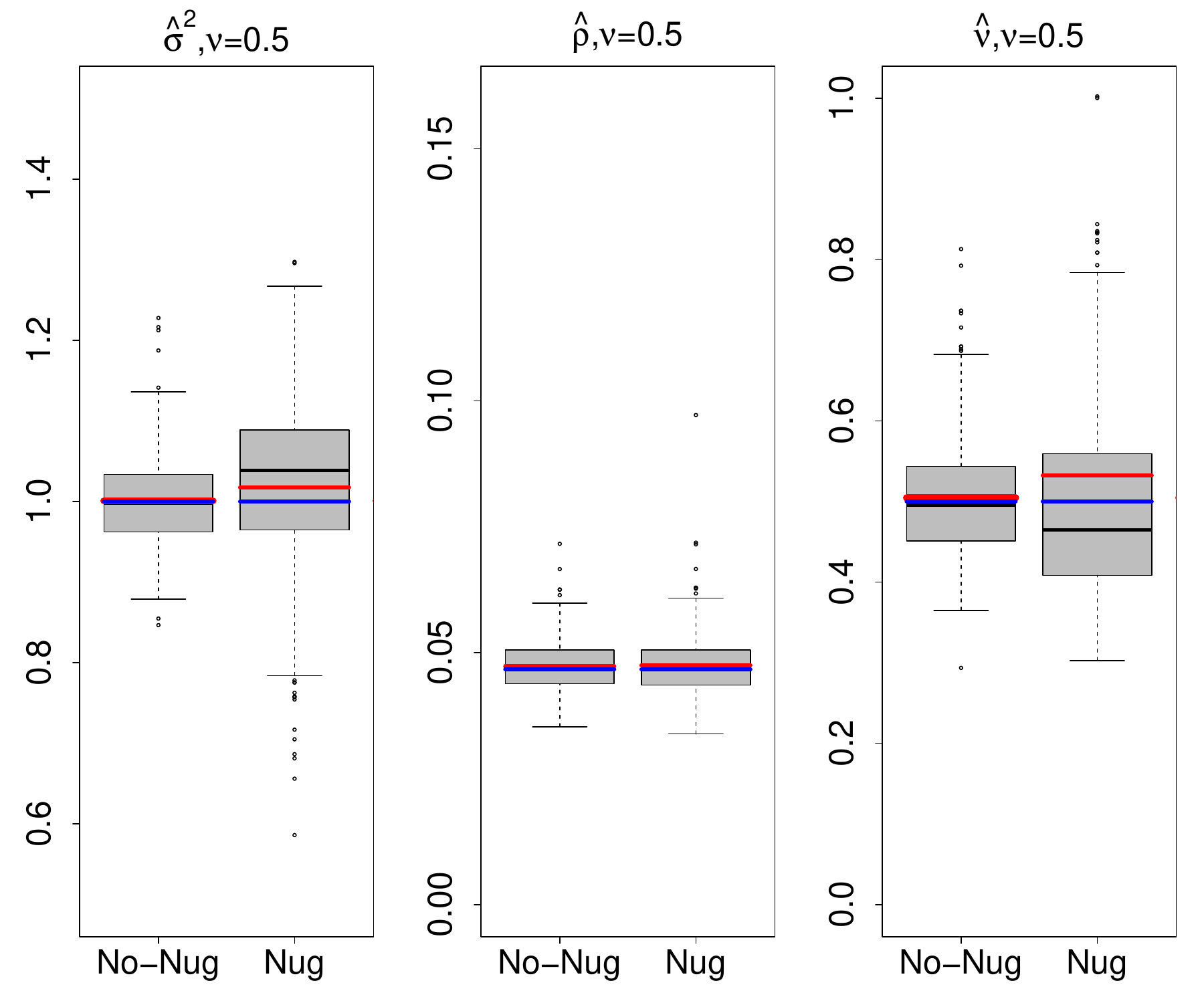}
  \caption{$\hat{\btheta}_3,{\cal M}_3$, weak correlation. }
  \label{reg(g)}
\end{subfigure}
\begin{subfigure}{0.33\linewidth}
  \centering
  \includegraphics[width=1\textwidth,]{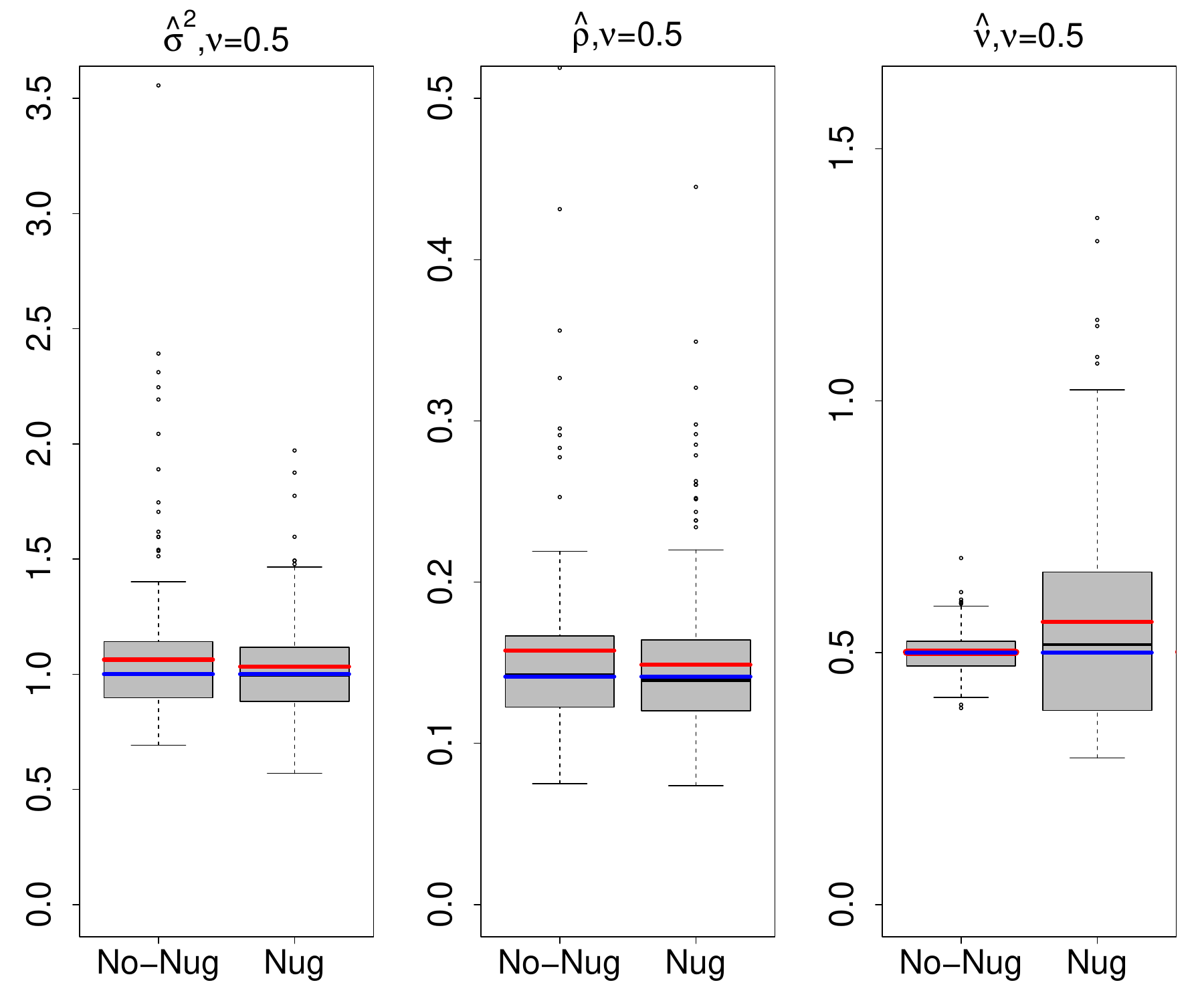}
  \caption{$\hat{\btheta}_3,{\cal M}_3$, medium correlation. }
  \label{reg(h)}
\end{subfigure}
\begin{subfigure}{0.33\linewidth}
  \centering
  \includegraphics[width=1\textwidth,]{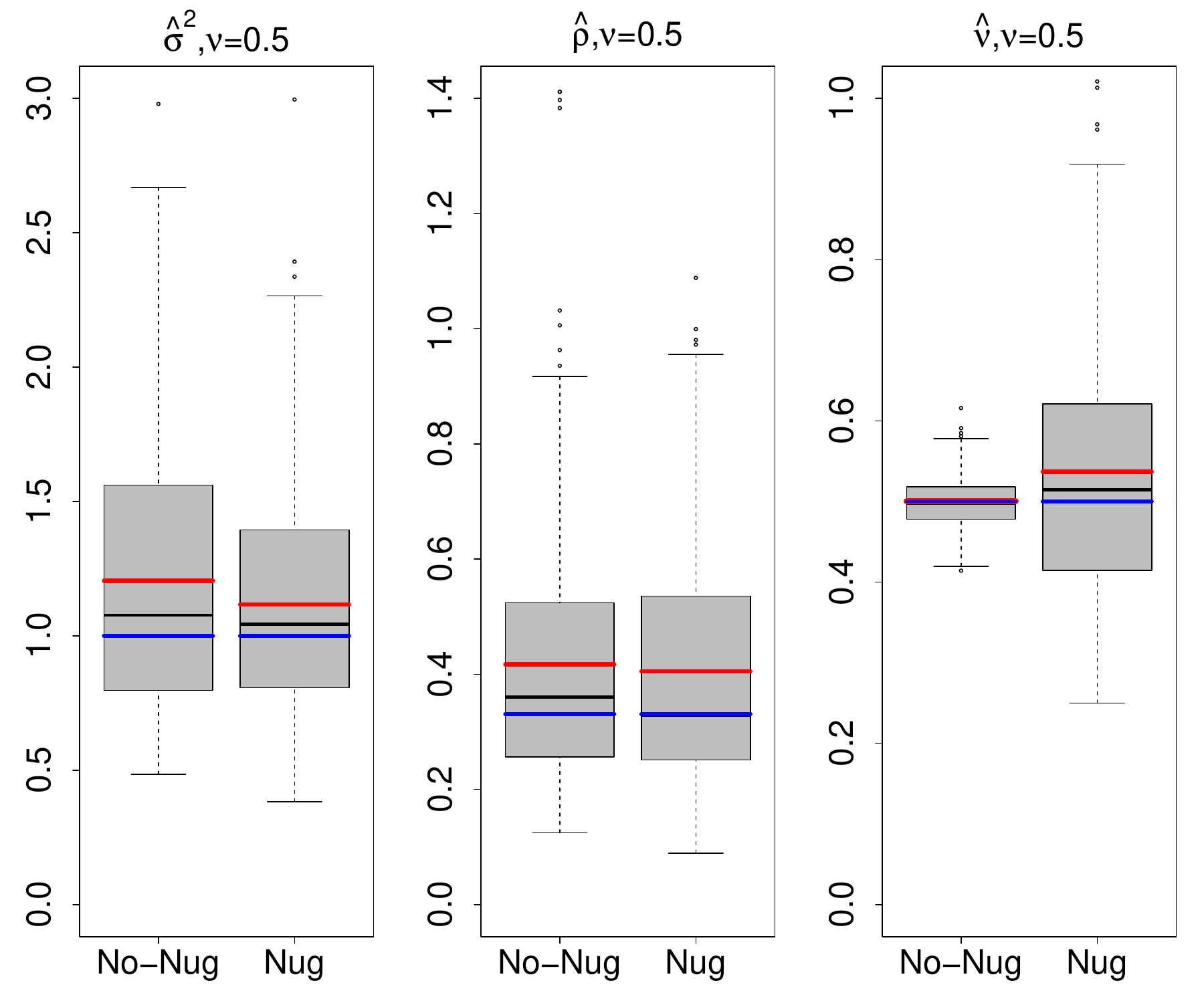}
  \caption{$\hat{\btheta}_3,{\cal M}_3$, strong correlation. }
  \label{reg(i)}
\end{subfigure}
\caption{{\redcolor Boxplots of the MLEs of the exponential Gaussian fields on a regular grid. Red lines represent the sample means and the blue lines denote the true values.  No-Nug denotes the non-nugget effect models and Nug denotes the nugget effect models.}}
\label{fig_reg}
\end{figure}
Figure \ref{fig_reg} demonstrates consistent behaviors of the parameter estimates from regular exponential Gaussian fields for ${\cal M}_1$, ${\cal M}_2$, and ${\cal M}_3$ as in Figures \ref{fig2} and \ref{fig9}.
\begin{figure}[H]
\begin{subfigure}{0.33\textwidth}
  \centering
  \includegraphics[width=1\textwidth,]{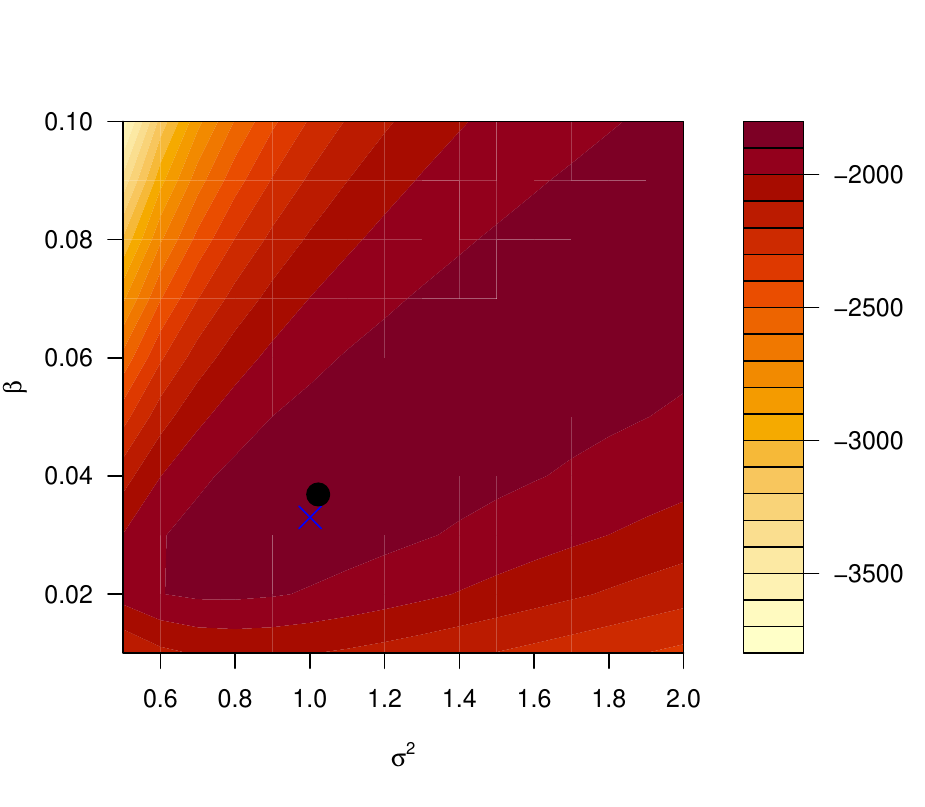}
  \caption*{${\cal M}_1$, weak}
  \label{c(a)}
\end{subfigure}
\begin{subfigure}{0.33\textwidth}
  \centering
  \includegraphics[width=1\textwidth,]{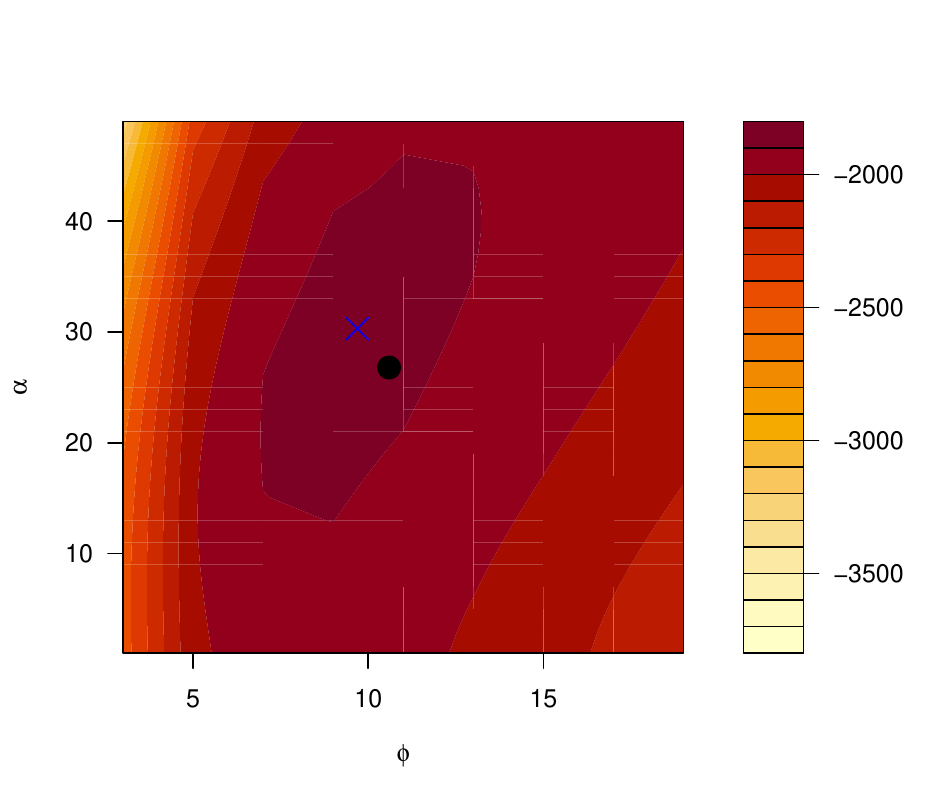}
  \caption*{${\cal M}_2$, weak}
  \label{c(b)}
\end{subfigure}
\begin{subfigure}{0.33\textwidth}
  \centering
  \includegraphics[width=1\textwidth,]{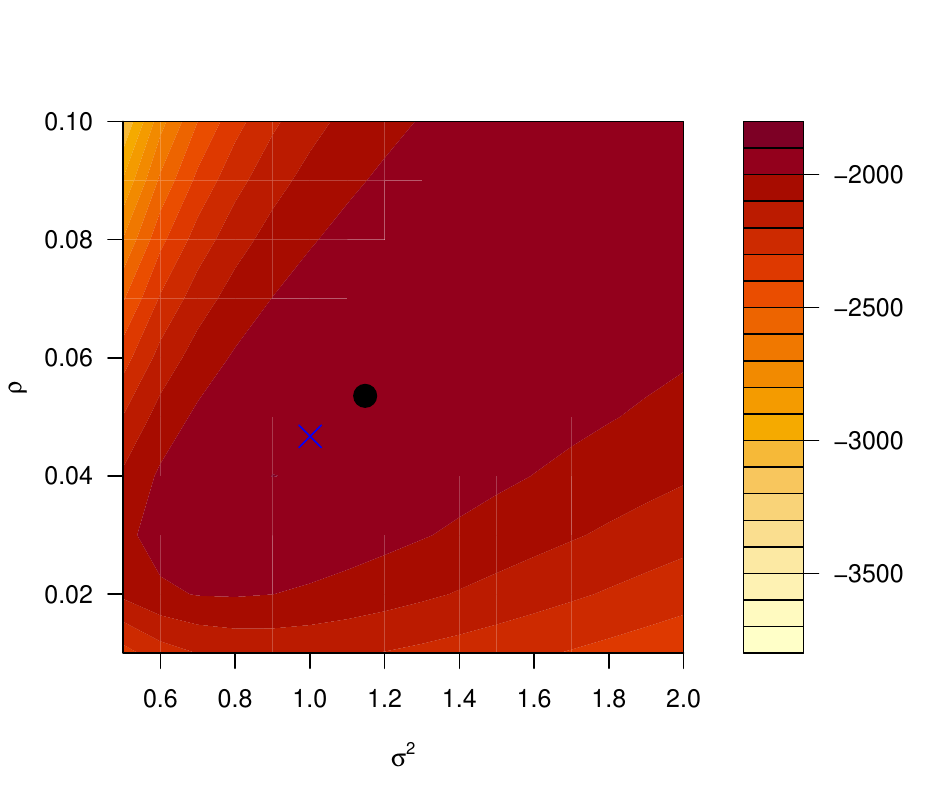}
  \caption*{${\cal M}_3$, weak}
  \label{c(c)}
\end{subfigure}
\begin{subfigure}{0.33\textwidth}
  \centering
  \includegraphics[width=1\textwidth,]{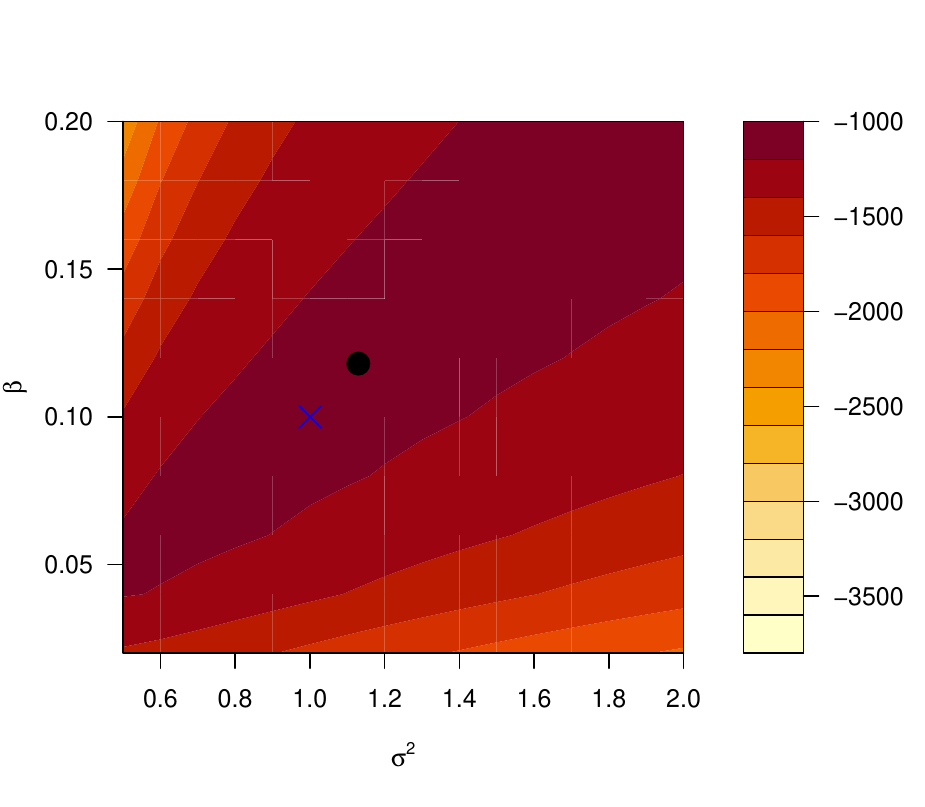}
  \caption*{${\cal M}_1$, medium}
\end{subfigure}
\begin{subfigure}{0.33\textwidth}
  \centering
  \includegraphics[width=1\textwidth,]{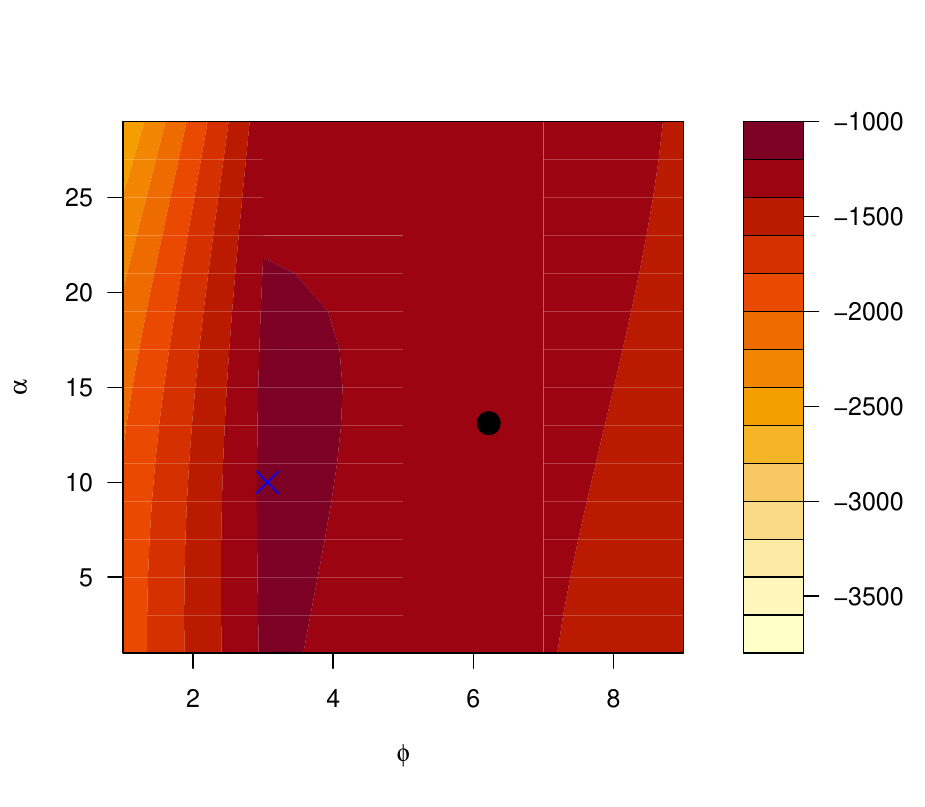}
  \caption*{${\cal M}_2$, medium}
\end{subfigure}
\begin{subfigure}{0.33\textwidth}
  \centering
  \includegraphics[width=1\textwidth,]{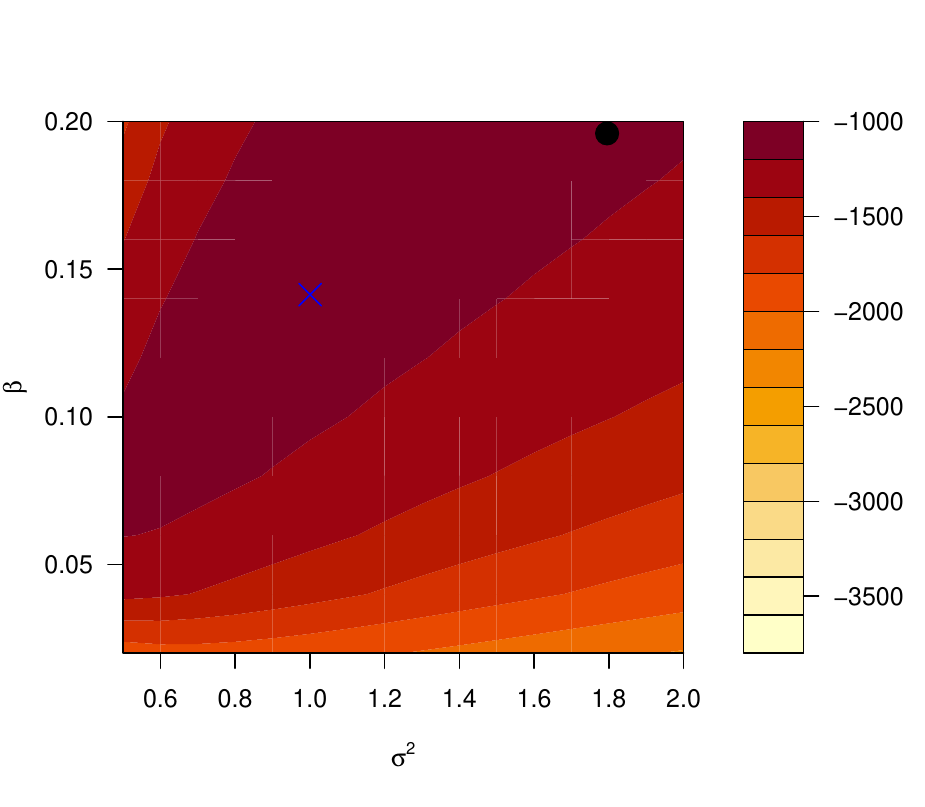}
  \caption*{${\cal M}_3$, medium}
\end{subfigure}
\caption{{\redcolor Contour plots of the log-likelihood function of 1600 samples simulated from weak and medium exponential Gaussian fields parameterized in ${\cal M}_1$, ${\cal M}_2$, and ${\cal M}_3$ with $\nu=0.5$. For the same reason as in Figure~\ref{c}, the contours for $\nu=1$ are omitted. The black dot in each figure represents the MLEs and the blue cross represents the true parameter values.}}
\label{contour}
\end{figure}
 \begin{table}[H]
    \centering
    \caption{SV, TAV, and DRVs, of $\hat{\btheta}_1$, $\hat{\btheta}_2$, and $\hat{\btheta}_3$ on the exponential without nugget effects using 300 replicates with sample size $n=3600,6400$. }
    \resizebox{1\textwidth}{!}{
    \begin{tabular}{|l|c|c|c||c|c|c||c|c|c|}
    \hline
        Strong-exp/$N=3600$   & $\hat{\sigma}^2$ & $\hat{\beta}$ & $\hat{\nu}$ & $\hat{\phi}$ & $\hat{\alpha}$ & $\hat{\nu}$ & $\hat{\sigma}^2$ & $\hat{\rho}$ & $\hat{\nu}$  \\
           \hline
       SV  & 0.0869 & 0.00615 & 0.000296 & 0.1771 & 3.3644 & 0.00906 & 0.2143 & 0.00338 & 0.000298  \\
       \hline
       TAV & 0.0936 & 0.00623 & 0.000295 & 0.0581 & 2.0770 & 0.000295 & 0.0936 & 0.0269 & 0.000295 \\
       \hline
       DRV & 0.0708 & 0.0125 & 0.00259 & {\redcolor 2.0482} & {\redcolor 0.6198} 
       & {\redcolor 29.7119} & {\redcolor 1.2900} & {\redcolor 1.2543} & 0.0102\\
       \hline
        \hline
        Strong-exp/$N=6400$   & $\hat{\sigma}^2$ & $\hat{\beta}$ & $\hat{\nu}$ & $\hat{\phi}$ & $\hat{\alpha}$ & $\hat{\nu}$ & $\hat{\sigma}^2$ & $\hat{\rho}$ & $\hat{\nu}$  \\
           \hline
       SV  & 0.0867 & 0.00585 & 0.000145 & 0.0879 & 2.4586 & 0.00906 & 0.4566 & 0.0553 & 0.000147  \\
       \hline
       TAV & 0.0927 & 0.00582 & 0.000157 & 0.0345 & 1.9420 & 0.000295 & 0.0927 & 0.0113 & 0.000157 \\
       \hline
       DRV & 0.0647 & 0.0052 & 0.0828 & {\redcolor 1.5478} &  0.2660 
       & {\redcolor 37.6624} & {\redcolor 3.9256} & {\redcolor 3.8938} & 0.0637\\
       \hline
    \end{tabular}
    }
    \label{1}
\end{table}
 \begin{table}[H]
    \centering
    \caption{SV, TAV, and DRVs, of $\hat{\btheta}_1$, $\hat{\btheta}_2$, and $\hat{\btheta}_3$ on the exponential with nugget effects using 300 replicates with sample size $n=3600,6400$. }
    \resizebox{1\textwidth}{!}{
    \begin{tabular}{|l|c|c|c||c|c|c||c|c|c|}
    \hline
        Medium-exp-Nug/$N=3600$   & $\hat{\sigma}^2$ & $\hat{\beta}$ & $\hat{\nu}$ & $\hat{\phi}$ & $\hat{\alpha}$ & $\hat{\nu}$ & $\hat{\sigma}^2$ & $\hat{\rho}$ & $\hat{\nu}$  \\
           \hline
       SV  & 0.1289 & 0.0106 & 0.00845 & 8.7519 & 12.2605 & 0.00911 & 0.0295 & 0.00132 & 0.00815  \\
       \hline
       TAV & 0.0258& 0.000771 & 0.00712 & 4.6490 & 7.7132 & 0.00712 & 0.0258 & 0.000954 & 0.00712 \\
       \hline
       DRV & {\redcolor 3.9961} & {\redcolor 12.7483} & 0.1868 & {\redcolor 0.8825} & {\redcolor 0.5895} 
       & 0.2794 & 0.1434 & 0.3836 & 0.1447\\
       \hline
        \hline
        Medium-exp-Nug/$N=6400$   & $\hat{\sigma}^2$ & $\hat{\beta}$ & $\hat{\nu}$ & $\hat{\phi}$ & $\hat{\alpha}$ & $\hat{\nu}$ & $\hat{\sigma}^2$ & $\hat{\rho}$ & $\hat{\nu}$  \\
           \hline
       SV  & 0.1852 & 0.00651 & 0.00359 & 3.2313 & 10.5353 & 0.00514 & 0.0288 & 0.000906 & 0.00298  \\
       \hline
       TAV & 0.0247& 0.000605 & 0.00352 & 2.5350 & 6.0526 & 0.00352 & 0.0247 & 0.000850 & 0.00352 \\
       \hline
       DRV & {\redcolor 6.4939} & {\redcolor 9.7568} & 0.0199 &  0.2747 & {\redcolor 0.7406} 
       & 0.4602 & 0.1659 & 0.0659 & 0.1534\\
       \hline
       \end{tabular}
    }
    \label{2}
\end{table}
\begin{table}[H]
    \centering
    \caption{Asymptotic normality of $\hat{\tau}^2$ of ${\cal M}_1$, ${\cal M}_2$, and ${\cal M}_3$ on exponential using 300 replicates with $N=3600,6400$.}
    \begin{tabular}{|l|c|c|c|}
    \hline
     Variance/field,$N=3600$ & ${\cal M}_1$-medium-exp & ${\cal M}_2$-medium-exp & ${\cal M}_3$-medium-exp \\
     \hline
        SV & 0.000783  & 0.00106 & 0.000734 \\
         \hline
        TAV & 0.000560  & 0.000560 & 0.000560 \\
        \hline
        DR & 0.3982 & 0.8928 & 0.3107 \\
        \hline
        \hline
        Variance/field,$N=6400$ & ${\cal M}_1$-medium-exp & ${\cal M}_2$-medium-exp & ${\cal M}_3$-medium-exp \\
     \hline
        SV & 0.000188  & 0.000399 & 0.000247 \\
         \hline
        TAV & 0.000185  & 0.000185 & 0.000185 \\
        \hline
        DR & 0.0162 & 1.1568 & 0.3351 \\
        \hline
    \end{tabular}
    \label{3}
\end{table}
\end{document}